\documentclass[pre,showpacs,noshowkeys,twocolumn]{revtex4}
\usepackage{graphicx}
\usepackage{bm}
\usepackage{textcomp}
\usepackage{color}
\usepackage{amsfonts}

\newcommand{\rem}[1]{}
\newcommand{\changes}{\textcolor{black}}

\begin{document}

	\title{Lattice symmetries and the topological protected transport of colloidal particles.}
	\author{Johannes Loehr$^{a1}$, Daniel de las Heras$^{a2}$, Michael Loenne$^{a3}$ Jonas Bugase$^{a1}$, 
		Adam Jarosz$^b$, Maciej Urbaniak$^b$, Feliks Stobiecki$^b$, 
		Andreea Tomita$^c$, Rico Huhnstock$^c$, Iris Koch$^c$, Arno Ehresmann$^c$, Dennis Holzinger$^c$, and Thomas M. Fischer$^{a1}$}\email{thomas.fischer@uni-bayreuth.de}
	\affiliation{$^{a1}$ Experimental Physics, $^{a2}$ Theoretical Physics, and $^{a3}$Mathematics, Institute of Physics and Mathematics, Universit\"at Bayreuth, 95440 Bayreuth (Germany).
		$^d$ Institute of Molecular Physics, Polish Academy of Sciences, 
		ul. M. Smoluchowskiego 17, 60-179 Pozna\'n (Poland).
		$^e$ Institute of Physics and Centre for Interdisciplinary Nanostructure Science and Technology (CINSaT),University of Kassel, Heinrich-Plett-Strasse 40, D-34132 Kassel (Germany)}
	\date{\today}
	
	\begin{abstract}
		The topologically protected transport of colloidal particles on top of
		magnetic patterns of all possible single lattice constant two dimensional
		magnetic point group symmetries is studied experimentally, theoretically, and with
		numerical simulations. We examine the transport of colloidal particles in
		response to modulation loops of the external field. We classify the
		modulation loops into topologically distinct classes causing different
		transport. We show that the lattice symmetry has a profound influence on
		the transport modes, the accessibility of transport networks, and the
		individual addressability of paramagnetic versus diamagnetic colloidal
		particles. We show how the transport of colloidal particles above a two
		fold symmetric stripe pattern changes from universal adiabatic transport
		at large elevations via a topologically protected ratchet motion at
		intermediate elevations toward a non-transport regime at low elevations.
		Transport above four fold symmetric patterns is closely related to the
		transport above two fold symmetric patterns. There exists a family of
		three fold symmetric patterns that vary as a function of the phase of the
		pattern. We show how this family can be divided into two topologically
		distinct classes supporting different transport modes and being protected
		by proper and improper six fold symmetries. Both classes support
		individual control over the transport of paramagnetic and diamagnetic
		particles. We discuss the topological transition when moving the phase
		from one class of pattern to the other class. The similarities and the
		differences in the lattice symmetry protected transport of classical
		over-damped colloidal particles versus the topologically protected
		transport in quantum mechanical systems are emphasized.
	\end{abstract}
	
	\maketitle

\section{Introduction}\label{Introduction}
The theoretical description of topological insulators highlighted the connection between symmetry and topology in quantum phases of matter \cite{Hasan,TI}.
Symmetries and the topology of quantum matter are deeply intertwined. The exploration of the role of symmetry in topological phases has led to a topological classification of phases of matter \cite{Chiu}. The complex quantum wave function of an excitation in a lattice can be considered as a two dimensional vector with real and imaginary part components that lives in the first Brillouin zone of the reciprocal lattice. When one identifies the borders of the first Brillouin zone it is  topologically a torus. Attaching the quantum wave function vector to this torus mathematically defines a vector bundle that can be characterized by Chern classes. These classes must be compatible with the symmetries of the Hamiltonian. Chern classes are symmetry protected against perturbations compatible with the symmetry. Amongst the most prominent symmetries protecting topological insulators are the time reversal symmetry, the particle hole symmetry, but also the point symmetry of the lattice \cite{Fu,Hsieh,Dziawa}. Different constraints 
of the lattice symmetries cause physical distinct effects on lattices of different symmetry \cite{Slager,Liu}. 
In topological nontrivial systems Dirac cones play a crucial role.  
The number of these Dirac cones in a hexagonal and a square lattice differ and their robustness against perturbations is different if they are located at a high symmetry point,  a high symmetry line or a generic location of the Brillouin zone \cite{Miert}.

The variety of phenomena enriches when considering time dependent periodically driven systems. In such systems the frequency or energy of an excitation is conserved only modulo the frequency of the driving field and the first frequency zone can be folded into a circle in the same spirit as folding the first Brillouin zone into a torus \cite{Kitagawa,Rudner}. Floquet topological insulators are one example of topologically non trivial systems arising from  periodic driving.
  
The discreteness of spectra of quantum phenomena is one ingredient shared also with spectra of bound classical waves and with the nature of topological invariants.
 The quantum Hall effect is one important example, where transport coefficients increase in discrete steps that contain only fundamental constants of nature including Planck's constant. The  discreteness of the steps are caused by topology \cite{Thouless}.
 
The topological classification of phases is not restricted to quantum systems. There are other non-quantum vector waves in lattices \cite{Kane,Paulose,Nash,Huber,Rechtsman,Mao} that can be characterized in just the same way. Hence the topological discreteness also appears in many classical wave like systems.
 The topological characterization is not restricted to classical vector bundles. It has been applied to non-equilibrium stochastic systems that describe biochemical reactions \cite{Murugan}.
We applied the concept
of topological protection to the dissipative transport of
magnetic colloidal particles on top of a modulated periodic magnetic potential \cite{Loehr,delasHeras}. There the transport of the point particle is fully characterized by the topology of the mathematical manifold on which it moves. The manifold does not carry any vector property. It can be characterized by its genus, a topological invariant somewhat more descriptive than the Chern class.  
 We have shown that the driven transport of paramagnetic or diamagnetic colloidal particles above a two dimensional lattice is topologically protected by topological invariants of the modulation loops used to drive the transport \cite{Loehr,delasHeras}.
 Non-topological transport of particles in a dissipative environment is usually vulnerable because of a spreading of the driven motion with the distribution of properties of the classical particles \cite{Olson,vortex1,vortex2,Grier,Bohlein,lab2,Arzola} as well as due to the abundance of possible hydrodynamic instabilities \cite{Loewen,Chaikin} that limit the control over their motion. Topologically protected particle transport in contrast is robust against sufficiently small continuous modifications of the external modulation. Only when the modulation loops are changed drastically they will fall into another topological class, and the direction of the transport changes in a discrete step. 

In this work we investigate how the topological classes of modulation loops are affected by the lattice
symmetry. We use experiments, theory and simulations to study transport above lattices of all possible
two dimensional magnetic point symmetry groups and examine the impact of the symmetry on the number of
transport modes, the number of topological invariants and on the type \changes{(adiabatic or ratchet)}
of transport. We show that lattice symmetry, as in topological crystalline insulators,
\cite{Fu,Hsieh,Dziawa,Slager,Liu,Miert} has a profound influence on the topologically protected transport modes. 

\changes{Applying periodic boundary conditions the unit cell of each lattice is a torus, which defines the action space. That is, the space in which the colloids move.
The colloids are driven with periodic modulation loops of an external magnetic field, the direction of which defines the control parameter space.
As a result of the interplay between the external magnetic field and the static magnetic field of the pattern, action space is divided into accessible and 
forbidden regions for the colloidal particles. For every point in an accessible region there exist a direction of the external magnetic field such that the
magnetic potential has a minimum at that point. The borders between different regions in action space are characterized by special objects in control space.
Modulation loops of the external field that wind around these special objects in control space cause colloidal transport along lattice vectors in action space.} 

\changes{In Refs.~\cite{Loehr,delasHeras} we studied the motion of colloids above hexagonal and square patterns, respectively. Here, we extend our previous studies in several ways.
We corroborate the theory developed in Refs.~\cite{Loehr,delasHeras} with experiments on four-fold symmetric patterns and prove experimentally the existence of ratchet modes in the
six-fold symmetric patterns. We also develop a theory for two- and three-fold patterns and prove their validity with experiments. 
Moreover, we find theoretically two new topological transitions, one in the non-universal stripe pattern, and one in the family of three-fold patterns.
All theoretical predictions are tested experimentally.}
 
\section{Colloidal transport system} \label{model}
In this section we introduce a soft matter system for Floquet crystalline symmetry protected driven transport of colloidal particles on top of two dimensional magnetic lattices of different symmetry.

\subsection{Magnetic colloids on magnetic lattices}

\begin{figure}
	\includegraphics[width=\columnwidth]{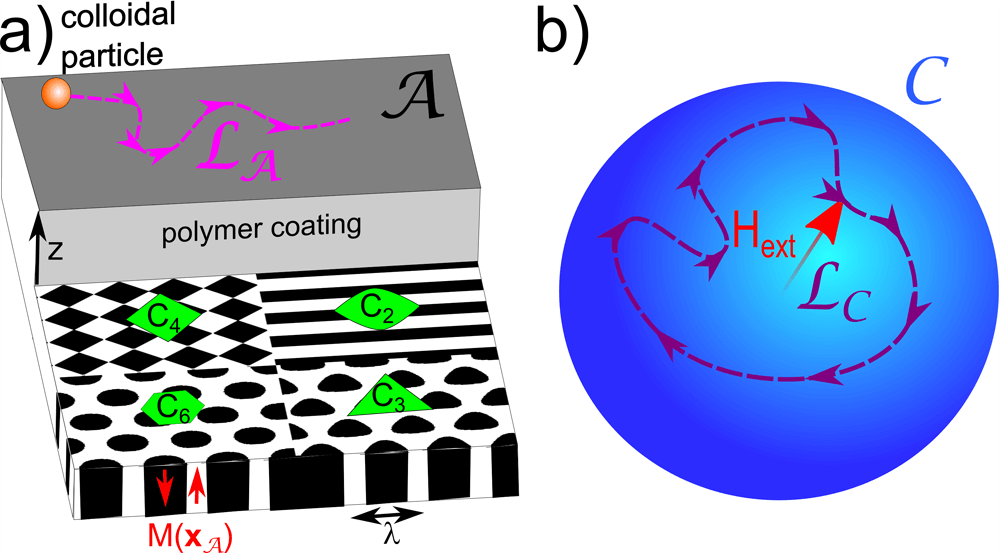}
	\caption{a) Magnetic pattern of symmetry $C_2,C_3,C_4$, and $C_6$ with wavelength $\lambda=2\pi/Q$ and magnetization $M(\mathbf x_{\cal A})$. The magnetic colloidal particles move in the two dimensional {\bf action space} $\cal A$ on top of the film at fixed elevation $z> \lambda$. b) A modulation loop $\cal L_C$ of the external magnetic field ${\mathbf H}_{ext}$ in the {\bf control space} $\cal C$ causes a transport loop $\cal L_A$ of the colloidal particle.}
	\label{figurescheme}
\end{figure}

Our system consists of a two dimensional periodic magnetic film having domains magnetized in the z-direction normal to the film (Fig. \ref{figurescheme}a). We consider a film that has as much area magnetized in the +z as in the -z direction. The magnetic field $\mathbf{H}^p$ of the pattern can be derived from a scalar magnetic potential
\begin{equation}
\mathbf{H}^p=-\nabla \psi
\end{equation}
that satisfies the Laplace equation and can be written as
\begin{equation}
\psi=\sum_\mathbf{Q}\psi_\mathbf{Q} e^{-Qz}e^{i\mathbf{Q}\cdot \mathbf{x}_{\cal A}}
\label{psisum}
\end{equation} 
where the sum is taken over the reciprocal lattice vectors $\mathbf{Q}$ ($Q=2\pi/\lambda$ for the smallest non-zero reciprocal lattice vector) of the two dimensional lattice and $\mathbf{x}_{\cal A}$ is a two dimensional vector in the lattice plane. Lower Fourier modes dominate the sum (\ref{psisum}) at higher elevation $z$. 

Magnetic colloids can be confined in a liquid at a fixed elevation $z$ that is larger than the wavelength of the pattern $\lambda$
by coating the magnetic film with a polymer film of defined thickness or by immersing the colloids into a ferrofluid that causes magnetic levitation of the colloids \cite{Loehr}. We call the two-dimensional space in which the particles move the {\bf action space} $\cal A$. \changes{We will use a number of geometric spaces and objects. Their definitions are listed in appendix \ref{definitions}.} The positions of the particles are described by the vector $\mathbf{x}_{\cal A}$.

Magnetic fields  induce magnetic moments
\begin{equation}
\mathbf {m}=\chi_{eff} V \mathbf {H}
\end{equation}

\begin{figure*}[t]\begin{center}
	\includegraphics[width=1.9\columnwidth]{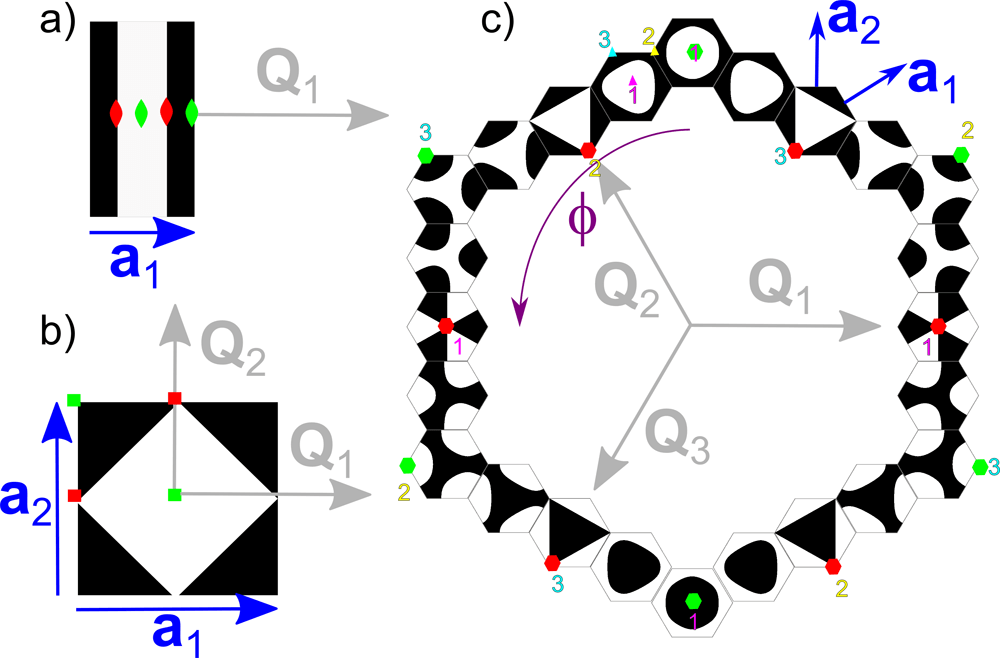}
	\caption{Wigner Seitz cells, unit vectors (blue), and reciprocal lattice vectors (gray) of all possible two dimensional single lattice constant magnetic point groups generating lattices. Black and white indicate the discrete down and up magnetized pattern according to equation (\ref{magnetizationpattern}) that creates a universal colloidal potential at an elevation $z>\lambda$ above the pattern. The ${\mathbf Q}_1$ vector is always pointing to the right in the $x$-direction.
		In a) we show
		the unit cell of the two-fold and in b) of the four-fold symmetric pattern, and in c) 24 smaller three-fold symmetric unit cells. The magnetic pattern of these three-fold symmetric unit cells continuously varies with the phase $\phi$ of equation (\ref{magnetizationpattern}). We show a sequence of such cells in steps of $\Delta\phi=\pi/12$ starting at $\phi=0$ at the top. 
		For each case there are 3 symmetry points with $C_3$ symmetry per unit cell. They are  shown in 3 different colors (pink, yellow, cyan) in the  unit cell next to $\phi=0$. For special values of $\phi$ one of  these three points acquires a proper or improper six-fold symmetry. 	$N$-fold symmetric points of all unit cells are marked in green for proper rotation symmetries $C_N$ and in red for improper rotation symmetries $S_N$.}
	\label{figpattern}\end{center}
\end{figure*}

of the colloids of effective susceptibility $\chi_{eff}$ and volume $V$. We define the colloidal potential $U=H^2$. The colloids thus acquire a potential energy $E=-\chi_{eff} V U$. This depends on the square of the total magnetic field $\mathbf{H}=\mathbf{H}^p+\mathbf{H}_{ext}$ which is the superposition of a homogeneous time dependent external field to the heterogeneous pattern field. The potential energy $E$ has a different sign for paramagnetic and diamagnetic colloids.  Hence,  paramagnetic particles move to positions that are maxima of $U$ while diamagnetic colloids move to the minima. 

 We are particularly interested in the motion of paramagnetic and diamagnetic colloids at an elevation $z>\lambda$ above the magnetic film such that only the contributions of the lowest non zero reciprocal lattice vectors to equation (\ref{psisum}) are relevant. At this elevation the response of the colloidal particles moving in action space $\cal A$ becomes universal, i.e. independent of the details of the pattern. The symmetry of the pattern becomes the only important property.
 If the lattice has a proper $C_N$ rotation symmetry or an improper $S_{N}$ symmetry there are $N$ reciprocal lattice vectors of lowest absolute value contributing to the universal scalar magnetic potential $\psi^*$ and we find
\begin{equation}\label{universalmagneticpotential}
\psi^*=\tilde\psi e^{-Qz}\sum_{n=0}^{N-1} \textrm{det}({\cal R}_N^n) e^{i[{\cal R}_N^n\mathbf{\cdot Q}]\cdot \mathbf{x}_{\cal A}}
\end{equation} 
where $\mathbf{Q}$ is one of the lowest absolute value reciprocal unit vectors and
$ {\cal R}_N$ denotes a proper rotation matrix by $2\pi/N$ ($\textrm{det}({\cal R}_N)=+1$) or an improper rotation consisting of a rotation by $2\pi/N$ and a reflection at the film plane ($\textrm{det}({\cal R}_N)=-1$). The universal scalar magnetic potential is determined only by the symmetry of the lattice and a prefactor carrying a phase $\phi$ and an amplitude, $\tilde\psi=|\tilde\psi|\exp(i\phi)$. The amplitude is irrelevant and the phase $\phi$ is only important in the $N=3$ case. 
The scalar magnetic potential will be the same for all lattices sharing the same point symmetry.

Magnetization patterns generating such  universal magnetic potentials are shown in Fig. \ref{figpattern}. The magnetization is given by

\begin{equation}\label{magnetizationpattern}
\mathbf{M}(\mathbf{x}_{\cal A})=M_s \mathbf{e}_z \textrm{sign}\left(t(\phi)+\sum_{n=0}^{N-1} \cos([{\cal R}_N^n\mathbf{\cdot Q}]\cdot \mathbf{x}_{\cal A}-\phi)\right)
\end{equation} 
with $t(\phi)\approx\frac{1}{2}\cos(3\phi)\delta_{N,3}$ chosen such that the magnetic moment
of a unit cell ($UC$) vanishes,
\begin{equation}\label{zeromagnetization}
\int_{UC}\mathbf{M}(\mathbf{x}_{\cal A})d\mathbf{x}_{\cal A}={\mathbf 0}.
\end{equation} 

The colloidal potential can now be reduced to the leading non-constant term, which is described by the universal colloidal potential:

\begin{equation}\label{universalpotential}
U^*= e^{Qz}\mathbf {H}_{ext}\cdot \mathbf{H}^{p}(\mathbf{x}_{\cal A}).
\end{equation}

Note that the prefactor $e^{Qz}$ rescales the potential such that it is independent of $z$, see Eq. (\ref{universalmagneticpotential}).

As we will see, adiabatic transport where the colloids adiabatically follow the maximum/minimum of the potential is possible along the crystallographic  directions of the lattices when the potential is modulated with external fields.
We call the space of the external field that may alter the colloidal potential the {\bf control space} $\cal C$. Following equation (\ref{universalpotential}) we see that in the universal case changing the magnitude of $\mathbf{H}_{ext}$ does not alter the position of the extrema of the colloidal potential. Control space ${\cal C}$, is thus a sphere of the external fields of constant magnitude. Each direction of the external field, which is a point in $\cal C$, produces a different colloidal potential (see Fig. \ref{figurescheme}b ).

\subsection{Lattice symmetries and topology}

 In Fig. \ref{figpattern} we depict the Wigner Seitz unit cells (with lattice vectors ${\mathbf a}_1$ and ${\mathbf a}_2$) of the periodic magnetic patterns defined by equation (\ref{magnetizationpattern}) for $N=2,3,4$ and $N=6$ and show the points of these patterns having $C_N$ (green) or $S_{2},S_{4}$ or $S_{6}$ (red) symmetry. The patterns in Fig. \ref{figpattern} exhaust all  possible single lattice constant $(a_1=a_2)$ magnetic point groups in 2d.
 White areas of the unit cell are magnetized in the positive $z$-direction and black areas in the negative $z$-direction. There are other patterns creating the same universal potential, the field of which  differs from the field of the patterns of Fig. \ref{figpattern} if experienced at lower $Qz<1$ (non-universal) elevation. Patterns having both $C_{N}$ (green) and $S_{N}$ (red) symmetries ($N=2$ or $N=4$) can be generated by using either proper or improper rotations. $N=3$ can be generated only with proper rotations. The $C_6$ and $S_6$ symmetries arise if we chose $N=3$ in equation (\ref{magnetizationpattern}) and $\phi=0$ ($\phi=\pi/6$). They  can equally well be produced with $N=6$ and using proper (improper) rotations.

Let us start with the topological characterization of action space. For a lattice with two-fold symmetry there is only one relevant reciprocal lattice vector and therefore the lattice is quasi one dimensional (see Fig. 2a). Since the lattice is periodic, we can deform the Wigner Seitz cell to merge the opposite borders. For $N=2$ the Wigner Seitz cell is a one dimensional segment, and hence action space ${\cal A}_2$ becomes topologically a circle. For all other symmetries, action space ${\cal A}_N$ , with $N>2$, is a torus.

Action space is topologically nontrivial for both $N=2$ and $N>2$ since both a circle and a torus have a hole. For $N=2$ there is one winding number around the hole, while for a torus there are two winding numbers.  The winding number of action space $\cal A$ has a very simple meaning in the underlying lattice. A winding around the circle (torus) corresponds to a translation by one unit vector in the lattice.

As we already mentioned, control space $\cal C$ is a sphere of radius $H_{ext}$.
The two-fold symmetric colloidal potential is independent of the in-plane external field component perpendicular to $\mathbf{Q}_1$.   Therefore, in the two dimensional problem we only need a reduced control space ${\cal C}^r_2$, which is the intersection of $\cal C$ with the plane spanned by $\mathbf{Q}_1$ and the vector   normal to the film $\mathbf{n}={\mathbf e}_z$. Like action space ${\cal A}_2$ the reduced control space ${\cal C}^r_2$ is a circle. 

The topology of the reduced control space ${\cal C}^r_2$ is fundamentally different from the full control space $\cal C$. 
The latter is a genus zero spherical surface that has no holes. For this reason we can continuously deform any closed loop of the external field $\cal L_C$ into any other loop  $\cal L_C'$. This is not the case if we restrict the modulation loops to lie on the reduced control space ${\cal C}^r_2$, which is a circle with a hole. Modulation loops in ${\cal C}^r_2$ can be characterized by their winding number  around the hole $w({\cal L}^r_{\cal C})$. The winding number is a topological invariant and we cannot continuously deform a modulation loop ${{\cal L_C}^r}$ with one winding number $w$ into another modulation loop  ${{\cal L_C}^r}'$ with a different winding number $w'\ne w$.

\subsection{Classification of modulation loops}
The fundamental question that we address in this work is, what are the topological requirements for a modulation loop $\cal L_C$ in  control space  to cause action loops $\cal L_A$ with different, non vanishing winding numbers in action space and hence induce transport of the colloidal particles.

For $N=2$ the answer is simple in reduced control space ${\cal C}_2^r$ but less obvious in full  control space $\cal C$. Reduced control and action space are non trivial. One might guess that the non-trivial topological classification of modulation loops in reduced control space  directly translates into the same topological classification of induced action-loops, i.e.

\begin{equation}\label{wa=wc}
w({\cal L}_{\cal A})=w({\cal L}_{{\cal C}^r}),\qquad \textrm{for  $N=2$.}
\end{equation}  

We will show that this indeed is the correct answer to the question for the universal case. But there are other, non-universal answers to this question. At low elevation the transport in the two-fold symmetric potential differs from this simple answer.

Equation (\ref{wa=wc}) does not hold in full control space, i.e there are loops with $w({\cal L}_{\cal A})\ne w({\cal L}_{\cal C})$ for any $N$. Otherwise there would not be transport since $w({\cal L}_{\cal C})=0$ for any loop. Full control space ${\cal C}$ becomes nontrivial if we puncture it at specific points or introduce even more complicated objects on it. The result is a constrained control space $\tilde{\cal C}$, for which the simple answer   

\begin{equation}\label{wat=wct}
w({\cal L}_{\cal A})=w({\cal L}_{\tilde{\cal C}})
\end{equation}  

with $w({\cal L}_{\tilde{\cal C}})$ the winding numbers around the objects of  $\tilde{\cal C}$ holds. The task is to find the objects that we need to project onto full control space and figure out how  winding around those objects allows for a classification of the modulation loops into classes that induce topologically different transport of colloids in action space.

\subsection{Computer simulations}
We use Brownian Dynamics to simulate the motion of a single point paramagnetic colloid above the different patterns. 
The motion of the particle is described by the stochastic differential Langevin equation
\begin{equation}
\gamma\frac{d{\mathbf x_{\cal A}}(t)}{dt}=-\nabla_{\cal A}E({\mathbf x}_{\cal A},{\mathbf H}_{\text{ext}}(t))+\mathbf{f}(t),\nonumber
\end{equation}
with $t$ the time, $\gamma$ the friction coefficient, and $\mathbf f$ a Gaussian random force. The variance of the random force is determined by
the fluctuation-dissipation theorem. As usual, we integrate the equation of motion in time using a standard Euler algorithm.
We always equilibrate the system before the modulation loop in control space starts, such that the colloidal particles always start in the
minimum of the potential energy $E$ at $t=0$.

The phase diagrams of the transport modes that we present in the next sections were initially obtained with computer simulations and can now also be predicted theo\-retically.

\subsection{Outline}
The rest of the paper is organized as follows. 
In section \ref{two-fold} we treat the case $N=2$. The simplicity of ${\cal C}^r_2$ allows us to visualize many concepts that cannot be visualized for $N>2$ such as the full dynamics in phase space. We also study the non-universal transport for $N=2$, and the connection to previous works \cite {Tierno2008,Pietro, Dullens,bubble}. We outline the concept of topologically protected ratchets with this very simple example. We then extend the treatment of $N=2$ to the full control space, introducing the concept of the constrained control space ${\tilde{\cal C}}$. The case $N=4$ is related to the case $N=2$ and is treated in section \ref{four-fold}. In section \ref{three-fold} we analyze the case $N=3$ that consists of a whole family of patterns continuously varying with the phase $\phi$ of the pattern. This includes the two special cases, $C_6$ symmetry ($\phi=0$) and $S_6$ symmetry ($\phi=\pi/6$). We find a new topological transition between $C_6$- and $S_6$-like three-fold symmetric lattices. Section \ref{discussion} contains a discussion of the experiments, a comparison to the theoretical and numerical predictions, and a discussion of the results in comparison to quantum systems. Finally  section \ref{conclusion} summarizes the main conclusions concerning transport.

\section{two-fold symmetry}\label{two-fold}

In this section we study the transport on top of a two-fold symmetric pattern. We start with the universal case and subsequently reduce the elevation of the colloids towards non universal cases. This allows us to first study the transition from topologically protected adiabatic motion towards ratchet motion, and then to a non transporting regime.

\subsection{Theory}
A stripe pattern is a magnetic pattern with two-fold symmetry (see Fig. \ref{figpattern}a). The magnetic field of a thick ($tQ>1$, $t$ being the thickness of the magnetic film) pattern of stripes of opposite magnetization $\pm M$ alternating along the $x$ direction reads:
\begin{eqnarray}\label{stripe}\nonumber
H^p_x+iH^p_z=\frac{2M}{\pi}\ln\left[\tan(Q(x+iz))\right]\\
=\sum_{n=0}^\infty \frac{8M}{(2n+1)^2}e^{i(2n+1)Q(x+iz)},
\end{eqnarray}
where $H^p_\alpha$ are the (real) components of the pattern magnetic field, and in the last part of equation (\ref{stripe}) we have decomposed the field into its Fourier-components. The non-universal colloidal potential valid at any height $z$ reads:
\begin{equation}
U=(H^p_x+H_{ext}\cos\varphi_{ext})^2+(H^p_z+H_{ext}\sin\varphi_{ext})^2,
\end{equation}
where 
\begin{equation}
\mathbf H_{ext}=H_{ext}\left(\begin{array}{c}
\sin\varphi_{ext}\\\cos\varphi_{ext}
\end{array}\right),\qquad \varphi_{ext}\in[0,2\pi]
\end{equation}
denotes the external magnetic field lying in the reduced control space ${\cal C}^r_2$. In the limit $Qz>1$ the pattern field is well described by 
\begin{equation}
\mathbf H^p(Qz>1)=8Me^{-Qz}\left(\begin{array}{c}
\sin Qx\\\cos Qx
\end{array}\right),\qquad Qx\in[0,2\pi]
\end{equation}
and the universal potential reads, c.f. (\ref{universalpotential})
\begin{equation}
U^*=8MH_{ext}\cos(Qx-\varphi_{ext}).
\end{equation}

The over-damped Brownian motion  of a colloidal particle in the $x$-direction is given by
\begin{equation}
\gamma\dot x=\chi_{eff}V\frac{\partial U(x,\varphi_{ext})}{\partial x}+f_{Brown}
\end{equation}
where $f_{Brown}$ is a zero average random force fulfilling the fluctuation dissipation theorem, $\gamma\propto \eta$ the friction coefficient of the colloid in the liquid of viscosity $\eta$, and the effective magnetic susceptibility $\chi_{eff}$ has a different sign for the paramagnets and diamagnets. Since our colloidal potential is sufficiently strong we can neglect the random force.

There are two kinds of colloidal dynamics that occur on separate time scales, when we adiabatically modulate the direction of the external field, which is described by $\varphi_{ext}(t)$. One is the intrinsic dynamics of the colloids on an intrinsic short time scale $t_{int}$
\begin{equation}
\gamma\dot x(t_{int})=\chi_{eff}V\frac{\partial U(x(t_{int}),\varphi_{ext}(t_{fixed}))}{\partial x}
\end{equation}
with which the colloids follow the path of steepest descent along the slope of the colloidal potential along the $x$-direction towards an extremum in $U$. The typical angular speed of this intrinsic motion is of the order $\omega_{int}=Q\dot x\propto e^{-Qz}\chi_{eff}\mu_0 MH_{ext} (QV^{1/3})^2/\eta$; (the intrinsic angular frequency renormalizes by an additional factor $tQ<1$ for thin magnetic films). Since the external modulation frequency $\omega_{ext}\ll\omega_{int}$ is significantly slower this happens at fixed external field ($\varphi_{ext}(t)=\varphi_{ext}(t_{fixed})$). The other timescale is an adiabatic creeping of the colloid with the maximum/minimum of the colloidal potential,
\begin{equation}
0=\pm\frac{\partial U(x(t_{ext}),\varphi_{ext}(t_{ext}))}{\partial x},
\end{equation}
with a small velocity dictated by the much slower time scale $t_{ext}$ of the external field modulation. Making use of the periodicity of the pattern we wrap the $Qx$-coordinate into a circle of circumsphere $2\pi$ such that action space $\cal A$ is a circle. Reduced control space ${\cal C}^r_2$ is also a circle with radius $H_{ext}$ and  coordinate $\varphi_{ext}$. The full dynamics occurs in phase space ${\cal C}^r_2\otimes{\cal A}$, which is the product space of the reduced control and action space and thus a torus.

\begin{figure*}\begin{center}
	\includegraphics[width=1.9\columnwidth]{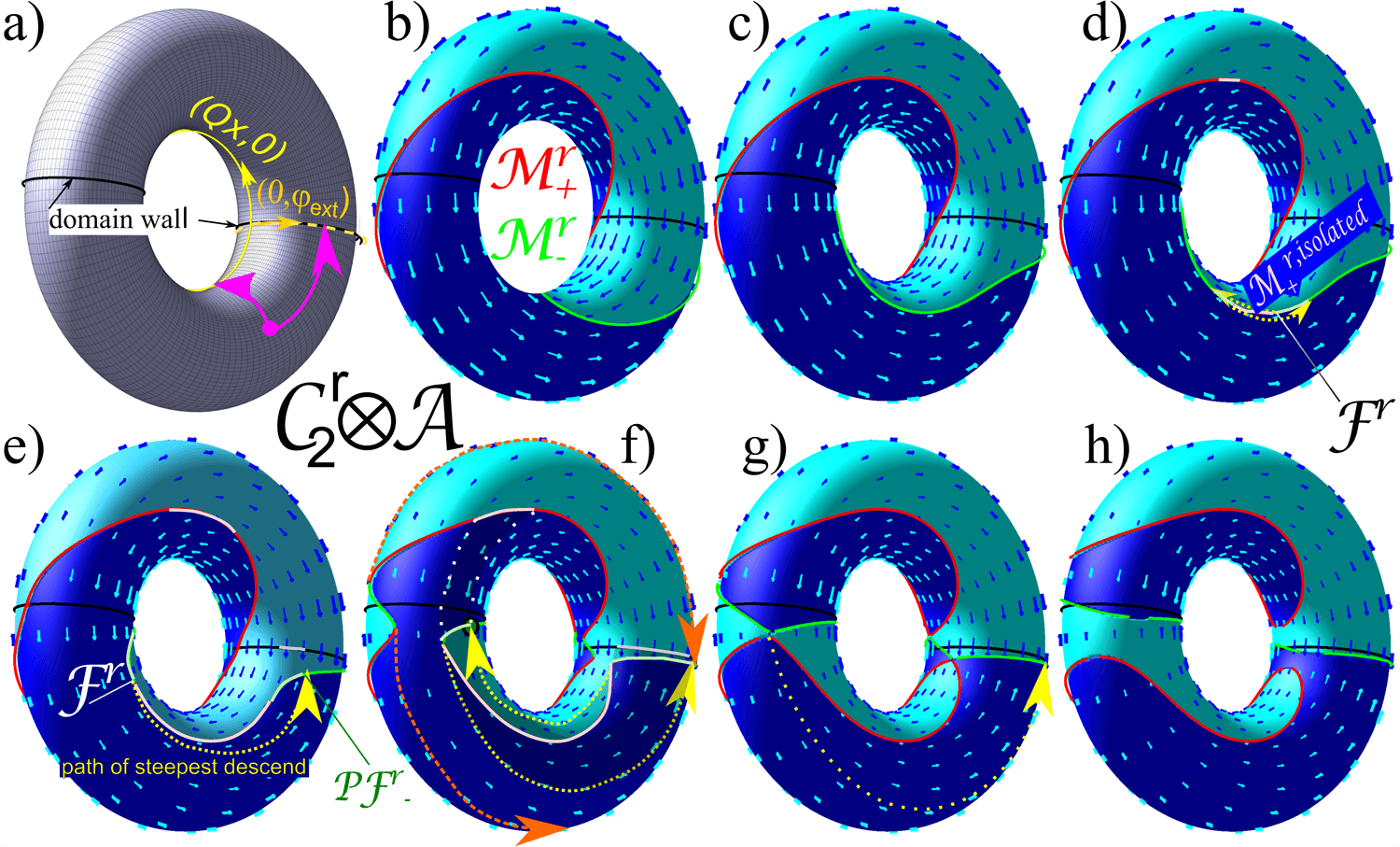}
	\caption{Reduced phase space of the two-fold symmetric system: a)The black lines depict the locations of the domain walls separating regions of opposite magnetization in phase space, which are two copies of control space at $x=0, \pi$. We may use the first one as reduced control space ${\cal C}^r$ (orange). Equally a level curve at fixed angle $\varphi$ (yellow) is a copy of action space. A point in phase space may be projected onto either control or action space, see an example in panel a) pink arrows. b) Reduced phase space and intrinsic dynamics of paramagnetic colloids for the universal potential in the limit $Qz>1$. The stationary manifolds ${\cal M}_{-}^r$ and ${\cal M}_{+}^r$ are depicted in green and red. The intrinsic dynamics is shown as a vector field of generalized velocities with cyan arrows (pointing in positive $x$-direction) and blue arrows (pointing in negative $-x$-direction). Adiabatic motion of colloidal particles occurs on the stable stationary manifold via the external modulation. c) At a lower non universal elevation $Qz=0.4$. the topology is still the same  as in the universal case. As in all the following cases we choose $H_{ext}=M$. d) Development of fences in ${\cal M}_{-}^r$ at $Qz=0.34$ and the transition towards topological protected ratchet jumps (yellow) from the fence (border between pink and green color on ${\cal M}^r$) toward the pseudo fence (border between green and light green color on ${\cal M}^r$) for paramagnetic particles \changes{(see appendix \ref{definitions} for a concise definition of fence and pseudo fence)}. Both fence and pseudo fences on $\cal M$ are projected into the same fence points in control space (border between gray and black on the domain wall). Preimages of the gray ($m=4$) excess line of control space are the two pink and two bright green lines. Preimages of the black ($m=2$) part of control space are the full red and full green colored lines.  e) Dynamics at an elevation $Qz=0.2$. f) At $Qz=0.1$ fences also start to develop in ${\cal M}_{+}^r$ causing ratchet jumps for the diamagnets (not shown) and additional feeder ratchet jumps (orange) starting from ${\cal M}_{-}^{r,isolated}$ for the paramagnets. g) Dynamics at the transition elevation $Qz=0.09$ toward a non transporting regime. h) Phase space and dynamics at low $Qz=0.07$ elevation. There are now four disconnected stationary manifolds (two of each kind) which all have zero winding number in action space.}
	\label{CA2theory}\end{center}
\end{figure*}

In Fig. \ref{CA2theory}a we depict the reduced phase space ${\cal C}^r_2\otimes{\cal A}$, together with the directions $Qx$ of action space and ${\varphi_{ext}}$ of the reduced control space. 
As indicated by the pink arrows in Fig. \ref{CA2theory}a each point $(x,\varphi_{ext})$ can be projected into the copy $x=0$ of reduced control space $(0,\varphi_{ext})=P_{{\cal C}^r_2}(Qx,\varphi_{ext})$ as well as into the copy $\varphi_{ext}=0$ of action space $(Qx,0)=P_{{\cal A}}(Qx,\varphi_{ext})$.

Figs \ref{CA2theory}b-h are plots of  the phase space ${\cal C}^r_2\otimes {\cal A}$ at different elevations $Qz$ above the pattern and for an external field-strength of $H_{ext}=M$. As we will see in the non-universal case the magnitude of $H_{ext}$ matters. The intrinsic dynamics, see Eq. (16),  is shown as a vector field on the torus. According to equation (16) the trajectories move along lines of constant external field direction $\varphi_{ext}(t_{fixed})=const$, either in $Qx$ or $-Qx$ direction. Regions of phase space with one sense of motion are colored in blue, regions of phase space with opposite sense in cyan. Both regions are separated from each other by the reduced stationary manifold ${\cal M}^r$, a line consisting of all points for which the potential is stationary $\partial_x U=0$. A stationary point is either  a minimum $(Qx,\varphi_{ext})\in{\cal M}_+^r$ (red) or a maximum $(Qx,\varphi_{ext})\in{\cal M}_-^r$(green). The intrinsic dynamics of the paramagnetic colloids starts at the red minimum line ${\cal M}_+^r$  and ends at the green maximum line ${\cal M}_-^r$.

 The reduced stationary manifold ${\cal M}^r$ of the universal potential (Fig. \ref{CA2theory}b) consists of two lines: The line $\varphi_{ext}=Qx$ (red) is  the set of minima ${\cal M}_+^r$ and the line $\varphi_{ext}=Qx+\pi$ (green) is the set of maxima ${\cal M}_-^r$. 
Following equation (17) the adiabatic creeping of the particles has to happen along the stationary manifolds. Paramagnetic colloids will adiabatically follow the green ${\cal M}^r_-$ line while diamagnetic ones will follow ${\cal M}^r_+$ (red). The simplicity of the universal stationary manifold  (Fig \ref{CA2theory}b) thereby converts any motion in control space into similar motion in action space. If we loop around the control circle we also loop around the action circle and thus induce transport by one unit vector. Both, paramagnetic and diamagnetic particles move at a fixed distance $\lambda/2$. A general  modulation loop ${\cal L}_{\cal C}^r$ in reduced control space causes an action loop $\cal L_{\cal A}$ in action space with similar winding number $w_{\cal A}=w_{\cal C}^r$. The particles can stay on the corresponding manifold during the entire modulation. 
Therefore the dynamics is completely adiabatic and thus dominated by the external modulation.

When we lower the colloidal plane to $Qz=0.4$ the manifold ${\cal M}^r$ deforms (Fig. \ref{CA2theory}c). Eventually at $Qz=0.34$, ${\cal M}^r_-$ becomes parallel to the tangent vector of action space ${\mathbf e}_x$  in one critical  point of ${\cal M}^r_-$. At this critical point $\partial_x U=\partial^2_x U=\partial ^3_x U=0$ and therefore the point is no longer a maximum. As one further lowers $Qz$ an isolated section ${\cal M}_+^{r,isolated}$ (pink) interrupts ${\cal M}_-^r$.

Two \changes {{\bf fence points}} ${\cal F}^r=\{(x,\varphi_{ext})| \partial_x U=\partial^2_x U=0\}$ as common borders between  ${\cal M}_+^{r,isolated}$ (pink) and ${\cal M}_-^r$ (bright green) develop from the formerly closed ${\cal M}_-^r$ loop (Fig. \ref{CA2theory}d). When a paramagnetic colloid adiabatically creeps along ${\cal M}_-^r$ via the externally induced dynamics and reaches the fence ${\cal F}^r$ it must leave the stationary manifold, follows the intrinsic dynamics and jumps (yellow arrow) toward a new maximum that we call the pseudo fence ${\cal PF}^r_-$ (border between the bright and full green in Fig. \ref{CA2theory}e). A pseudo fence is a point on $\cal M$ different from the fence that has the same projection onto reduced control space (border between the black and gray line) as the fence but different projections onto action space. 

 The intrinsic dynamics is irreversible, i.e. one can move along the path of steepest descent only in one direction. When we are at the critical elevation the 
${\cal M}_+^{r,isolated}$ interruption has zero length, fence and pseudo fence fall on top of each other. Like this the path of steepest descent  has zero length.
When we decrease the elevation $Qz$ the path of steepest descent continuously grows.
 Although it is no longer on ${\cal M}^r$ it falls into the same homotopy class as the section  of ${\cal M}^r$ that it bypasses. That is, both are topologically equivalent and transport by one unit vector can still be achieved by winding around the control space.
   The dynamics of the colloids, however, undergoes a phase transition from adiabatic toward a ratchet motion \cite{lab1,Reimann,many particles,Kohler,disorder,noise,Sinitsyn}. The ratchet jumps occur along the path of steepest descend with jump times short compared to the external modulation dynamics.   
The result of a ratchet transport is the same as the adiabatic motion at higher elevations because of the homotopy between the avoided section of ${\cal M}^r$ and the path of steepest descent. Like this the transport is topologically protected at the adiabatic to ratchet transition. 

If we further decrease the elevation to $Qz=0.1$ the same thing happens to the other sub-manifold ${\cal M}^r_+$. It is now interrupted by a ${\cal M}_-^{r,isolated}$ section resulting in irreversible jumps for the diamagnetic colloids (Fig. \ref{CA2theory}f). This section also opens up a new possible ratchet jump of paramagnetic particles initially located on ${\cal M}_-^{r,isolated}$ onto the disconnected other parts of ${\cal M}_-^{r}$. The special thing about these feeder jumps is, that once a colloidal particle leaves the isolated section it will never return due to the absence of pseudo fences in ${\cal M}_-^{r,isolated}$. 

\changes{The projection of a point in ${\cal C}^r\otimes {\cal A}$ onto a point in ${\cal C}^r$ defines a mapping from ${\cal M}^r$ onto ${\cal C}^r$. The inverse of this map is not a map because the projection maps several points of ${\cal M}^r$ onto the same point in ${\cal C}^r$. We call the number of preimages of the projection on ${\cal M}^r$ the {\bf multiplicity}.} 
Note that, the two (bright green) sections between pseudo fence and fence, the (pink) ${\cal M}_{+}^{r,isolated}$ insertion as well as a non isolated section (pink) of  ${\cal M}_{+}^{r}$ are projected onto the same (gray) excess segment of control space. Consequently the (gray) excess segment has multiplicity $m=4$ (it has four preimages on the manifold ${\cal M}^r$.) The rest of ${\cal M}^r$ is projected twice on the remaining (black, multiplicity $m=2$) section of ${\cal C}^r_2$. Like this there are sections of control space with different multiplicity. When we move from the $m=2$-region of control space to the $m=4$ region a maximum minimum pair is created in ${\cal M}^r$.

The topology of ${\cal M}^r$ does not change at the adiabatic to ratchet transition. It is only the distribution of points on ${\cal M}^r$ into the subsets  ${\cal M}^r_-$ and ${\cal M}^r_+$ that changes.
A transition of the topology of ${\cal M}^r$ occurs at $Qz=0.09$ when the formerly disconnected parts of ${\cal M}^r$ touch each other in four fence points (Fig. \ref{CA2theory}g) and then separate into four disconnected parts (Fig. \ref{CA2theory}h). Two of the new disconnected parts after the disjoining are entirely of type ${\cal M}_-^r$ and two are of type   ${\cal M}_+^r$. The  ${\cal M}_-^r$ parts are localized near the domain walls, while the  ${\cal M}_+^r$ parts lie on top of a domain. All four parts of ${\cal M}^r$ have non vanishing winding number around the reduced control space but vanishing winding numbers around action space. Any control loop will thus only create periodic motion in action space that is associated with no net transport over a period.  

We have given a description of the dynamics of paramagnets. The dynamics of diamagnets is the reversed intrinsic dynamics coupled with the external dynamics on ${\cal M}_+^r$. For the universal case at high elevations both types of particles move exactly the same way however they are separated by half the wavelength $\Delta Q x=\pi$ of the pattern. At lower elevation the transitions to a ratchet motion occurs for different elevations $Qz=0.34$ (Fig. \ref{CA2theory}d) for the paramagnets and $Qz=0.1$ (Fig. \ref{CA2theory}f) for the diamagnets. The transition from transport to no transport happens for both particles simultaneously at an elevation of $Qz=0.09$ (Fig. \ref{CA2theory}g). Paramagnets are then confined to the domain walls and diamagnets to the domains.

\begin{figure*}[t]\begin{center}
	\includegraphics[width=1.9\columnwidth]{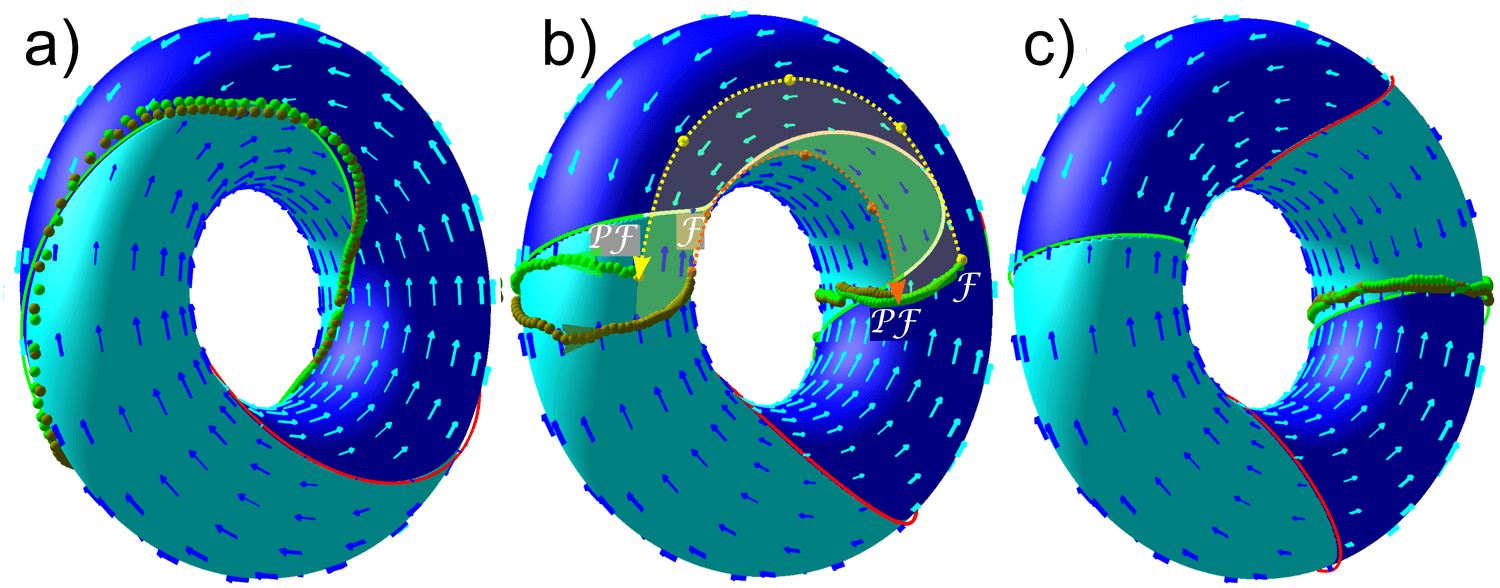}
	\caption{Reduced phase space (torus), intrinsic dynamics (vector field), stationary manifolds (green and red solid lines), and experimental trajectories (green, olive, yellow and orange) for three different non-universal elevations. a) adiabatic motion in a nearly universal potential  $Qz= 4.34$, $H_{ext}=0.2M$. b) Ratchet motion at an elevation $Qz=0.43$, $H_{ext}=0.2M$. c) No motion at a elevation $Qz=0.43$,$H_{ext}=0.1M$ below the topological transition. Experimentally measured data for a forward (backward) modulation loop with $w_{\cal C}=1(-1)$ is shown as green (olive) spheres for adiabatic, i.e. slow, motion and in yellow (orange) for the fast ratchet jumps. The ratchet motion in b) exhibits hysteresis between forward and backward motion (yellow shaded area).
		The experimental data does not perfectly match the theory (solid green line) since the changes of the stripe pattern of the garnet film with the external field (relevant at non-universal elevation) have not been included into the theory. Both experimental data and theory however fall into the same homotopy class. A video clip of the adiabatic motion of the paramagnetic colloidal particle in a) is provided in \cite{CLIPS}.}
	\label{Phase behaviour 2 experiment}\end{center}
\end{figure*}
\subsection{Experiments}\label{two-foldexperiments}
We have performed experiments with paramagnetic colloids above the stripe pattern  of wavelength $\lambda=7.2\, \mu\textrm{m}$, and magnetization $M \approx 20 \textrm{kA/m}$ of a magnetic garnet film \cite{Bobeck,0022-3727-38-12-R01}. We covered the garnet film with a ferrofluid of defined thickness $d$. Magnetic levitation lifts the colloids to the mid plane of the film at a fixed elevation $z$. Since we were limited in the variation of the thickness $d$ we used the amplitude $H_{ext}$ of the external field as a second control parameter. Both, decreasing the field or decreasing the elevation renders the transport behavior non-universal.
The modulation of the external magnetic field that drove the dynamics was generated by three coils arranged along the x,y, and z axes \cite{Martin}.
We applied a \changes{\bf palindrome modulation loop}
${{\cal{L}}_{{\cal C}^r}}={\tilde{\cal{L}}_{{\cal C}^r}}{\tilde{\cal{L}}_{{\cal C}^r}}^{-1}$, i.e 
 a combination of a forward loop ${\tilde{\cal{L}}_{{\cal C}^r}}$ of winding number $w({\tilde{\cal{L}}_{{\cal C}^r}})=1$ followed by the time reversed backward loop ${\tilde{\cal{L}}_{{\cal C}^r}}^{-1}$ with winding number  $w({\tilde{\cal{L}}_{{\cal C}^r}}^{-1})=-1$, each subloop has a  duration of $\Delta t=5\,\textrm{s}$.

We measured the corresponding trajectories in reduced phase space ${\cal C}_2^r\otimes{\cal A}$ at different heights. By video tracking we obtained the coordinate $x_{\cal A}(t)$ of the trajectory in action space. 
Simultaneously we determine $\varphi_{ext}(t)$  by measuring the width of a stripe magnetized that periodically varies with the external field and is visualized in the same video (see \cite{CLIPS}) via the polar Faraday effect.

At the universal elevation (Fig. \ref{Phase behaviour 2 experiment}a) the colloids creep adiabatically along the stationary manifold ${\cal M}_-^r$. Forward (green) and backward (olive) trajectories fall almost on top of each other. If we lower the elevation we can observe ratchet motion (Fig. \ref{Phase behaviour 2 experiment}b). There we can identify the sections of the trajectories that lie on ${\cal M}_{-}^r$ as those where the speed of the colloids on the trajectories is slow (adiabatic) (see green data in Fig. \ref{Phase behaviour 2 experiment}b). The paths of steepest descent are  the regions where the velocity is high (intrinsic dynamics). 
In the forward loop the adiabatic motion passes the pseudo fence and the particle jumps when it reaches the fence. The path of steepest descent reunites with the backward trajectory at the pseudo fence. The two sections on ${\cal M}^r$ between fence and pseudo fence together with the paths of steepest descend connecting fence and pseudo fence define the hysteresis between forward and backward ratchet loops. A fully adiabatic motion has negligible hysteresis.

At even lower elevations, below the topological transition height, we no longer observe transport. The paramagnetic particles are attached to the domain walls (Fig. \ref{Phase behaviour 2 experiment}c).

In a ratchet motion the path of steepest descent, and therefore the hysteresis, develops continuously from the critical point. The winding number $w(\tilde{\cal L}_{\cal A})=-w(\tilde{\cal L}_{\cal A}^{-1})$ of the forward loop does not change across this continuous transition. In contrast, the topological transition towards the non transporting regime is discontinuous.

\begin{figure}
	\includegraphics[width=\columnwidth]{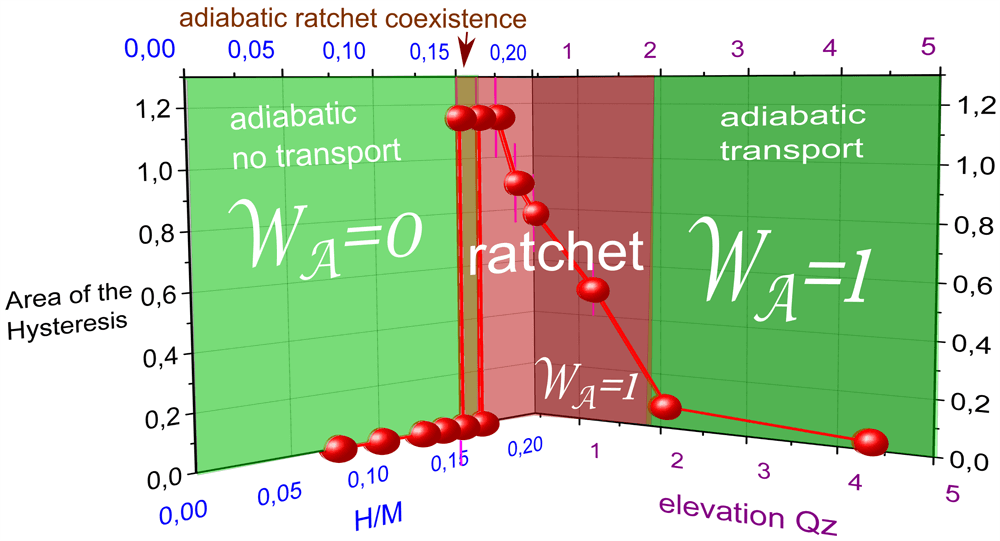}
	\caption{Experimentally measured area of the hysteresis of the transport. 
		The area of the hysteresis is measured on the surface of the torus ${\cal C}^r_2\otimes{\cal A}$. The total area of a torus is $(2\pi)^2\approx 40$.
		On the right we lowered the elevation $Qz$. This reveals the continuous transition from adiabatic transport toward ratchet motion. On the left side we decrease the external field amplitude at constant elevation. This reveals the discontinuous topological transition towards no transport.}
	\label{fighysteresis}
\end{figure}

 In Fig. \ref{fighysteresis} we plot the area of the hysteresis versus the non-universality parameters (external field $H_{ext}$ and elevation $Qz$). Both the continuous adiabatic toward ratchet transition as well as the discontinuous ratchet to adiabatic non-transport transition can be clearly identified from the figure.     

\subsection{Constrained control space}
In section \ref{four-fold} we will discuss the universal potential of a four-fold symmetric pattern. It is useful to first reiterate the universal case of the two-fold symmetric problem, using full control space $\cal C$.

In the section \ref{two-foldexperiments} we reduced the control space of the stripe system to fields that are lying in the plane spanned by the normal vector $\mathbf n$ to the pattern and by the unique reciprocal unit vector $\mathbf Q$. We just dropped the physically possible external field
component along the indifferent $\pm \mathbf n \times\mathbf Q$ direction. 
Here we do not ignore this component. Hence, since the magnitude of the external field $H_{ext}$ does not play a role for the universal case, full control space is a sphere.
 The constrained control space $\tilde{\cal C}_2$ of the stripe pattern is a two punctured sphere. The two points along the $\pm \mathbf n \times\mathbf Q$ direction are removed from the sphere of the full control space $\cal C$ since these points produce an indifferent constant potential in action space.

\begin{figure}
	\includegraphics[width=1\columnwidth]{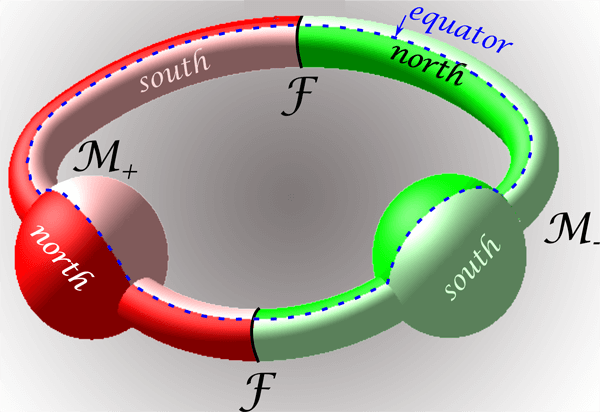}
	\caption{The stationary manifold for the universal potential of the stripe pattern for the full control space $\cal C$. ${\cal M}_{+}$ is depicted in red and ${\cal M}_{-}$ in green. Both are connected by two circular fences $\cal F$. Copies of the northern hemispheres of $\cal C$ are shown in full colors, while the southern ones are shown in light colors. \changes{(see appendix \ref{definitions} for a concise definition of the hemispheres and the equator)}}
	\label{figM2}
\end{figure}

 Topologically, the two punctured sphere $\tilde{\cal C}_2$ and the circle  ${\cal C}_2^r$ are equivalent. Since only the topology of control space is important we may expand ${\cal C}_2^r$ to the constrained control space $\tilde{\cal C}_2$. Note that the winding number of a modulation loop in ${\cal C}_2^r$ becomes the winding number of a modulation loop around the indifferent $\pm \mathbf n \times\mathbf Q$ axis through the two removed points of the punctured sphere in $\tilde{\cal C}_2$. The reduced control space is just the grand circle on the sphere around this axis. We can predict the result of modulation loops in the constrained control space $\tilde{\cal C}_2$: winding around the punctured points induces transport in action space.
 
 \begin{figure*}\begin{center}
 	\includegraphics[width=1.95\columnwidth]{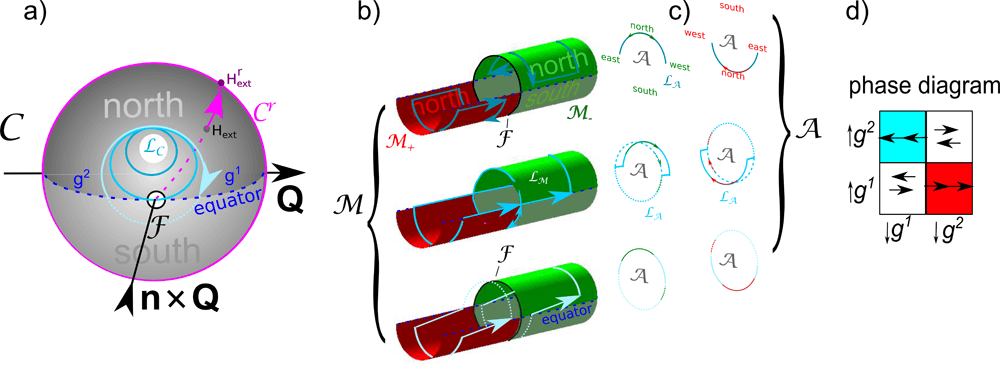}
 	\caption{a) Full control space of the stripe pattern. b) section of the stationary manifold $\cal M$ and 
 		c) its projection into action space.	
 		Several modulation loops $\cal L_C$ in $\cal C$ and their preimages $\cal L_M$ on $\cal M$ and the further projections $\cal L_A$ into $\cal A $ are shown.  In a) the reduced control space is shown in pink together with a projection of a full external field $\mathbf H_{ext}$ into the reduced external field $\mathbf H_{ext}^r$ is also shown. d) Phase diagram of the transport modes for the fundamental loops ${\cal L}_{\cal C}={  \downarrow \kern -4px g^i} { \uparrow \kern -4px g^j}$. Colored squares indicate transport, white squares indicate no transport. 	
 	}
 	\label{figM2loops}\end{center}
 \end{figure*}
 
To make the connection to the topologically trivial full control spaces of lattices with higher point symmetries, we can reinsert the removed points into the punctured sphere $\tilde{\cal C}_2$. That is, we recover the topologically trivial full control space ${\cal C}$ allowing fields pointing into the indifferent direction. This enables us to continuously deform a modulation loop with one winding number around the axis into a modulation loop with different winding number. The transition in winding number occurs when the modulation loop passes through the reinserted point.
 
Note that the indifferent direction satisfies 
\begin{equation}\label{stationary}
\nabla_{\cal A}U^*=\mathbf 0, 
\end{equation}
and 
\begin{equation}\label{fence}
\det(\nabla_{\cal A}\nabla_{\cal A}U^*)=0,
\end{equation} 
for any point $\mathbf x_{\cal A}\in \cal A$. We call points in ${\cal C}\otimes{\cal A}$ that fulfill equations (\ref{stationary}) and (\ref{fence}) the \changes{{\bf fences}} $\cal F$ on $\cal M$. For the stripe pattern and the universal case fence points only exist in ${\cal C}\otimes{\cal A}$, not in ${\cal C}_r^2\otimes{\cal A}$. In the stationary manifold  of the reduced control space ${\cal M}^r$ the sub-manifolds are two disconnected lines (maximum and minimum) without fences (Fig. \ref{CA2theory}b).
On the full stationary manifold $\cal M$ the fence consists of two copies (one for each of the opposite indifferent points in $\cal C$) of the one dimensional action space and thus consists of two disconnected circles.

The fences separate the maxima of the stationary manifold from the minima (Fig. \ref{figM2}). Hence using the constrained control space the stationary manifold $\cal M$ is a two dimensional manifold that is not disconnected. ${\cal M}_{-}$ and ${\cal M}_{+}$ are both copies of the punctured sphere, with the puncture point enlarged to a circular fence and there joined to one closed surface.
Fig. \ref{figM2} shows the topology of the universal stationary manifold $\cal M$ for the full control space. ${\cal M}_{+}$ is depicted in red and ${\cal M}_{-}$ in green.

 The constrained control space $\tilde{\cal C}_2$ can be subdivided into two hemispheres, the northern hemisphere for which $H_{ext,z}>0$ and the southern hemisphere ($H_{ext,z}<0$). Both hemispheres are simply connected areas, i.e. areas where every loop is zero homotopic. The areas are glued together at the two sections $g^1$ and $g^2$ of the equator between the puncture points. In Fig. \ref{figM2} we show the simply connected areas of the stationary manifold that are projected into both hemispheres of $\tilde{\cal C}_2$.

  Two lines circle the stationary manifold, see Fig. \ref{figM2}. We call these lines the equator since they are projected onto the equator of $\tilde{\cal C}_2$, see Fig. \ref{figM2loops}a.  When the equator hits the puncture point in $\tilde{\cal C}_2$ the two equators of the stationary manifold cross the fences in $\cal M$. Topologically $\cal M$ is a genus one surface with two winding numbers. The winding numbers of the fences are different from the winding numbers of the equator. 

Fig. \ref{figM2loops} shows the topological transition of the transport modes on $\cal M$ and $\cal A$ due to the continuous deformation of a control loop in $\cal C$.
We start with a control loop (dark blue loop) that is entirely in the north and hence does not wind around the indifferent point. The loop has two preimages on $\cal M$, one on ${\cal M}_-$ and one on ${\cal M}_+$. Both are zero homotopic. Now we further deform the modulation loop such that it crosses the fence point (blue loop). The preimage on $\cal M$ is the union of the two formerly disconnected loops and the fence itself. Mathematically the preimage is not a loop but a \changes{\bf lemniscate} \cite{Bauer}. When we slightly enlarge the loop (cyan), such that it is now winding around the fence point in $\cal C$, the  lemniscate on $\cal M$ disjoins again into two loops on ${\cal M}_-$ and ${\cal M}_+$. Now, both loops have non vanishing winding numbers. The projection of the loop in ${\cal M}_{-}$ (${\cal M}_{+}$) corresponds to a maximum (minimum) of the potential in $\cal A$ that adiabatically moves around with a winding number  similar to the winding number around the indifferent axis in $\cal C$, $w_{\cal A}=w_{\cal C}$. 

\begin{figure*}[t]\begin{center}
	\includegraphics[width=1.9\columnwidth]{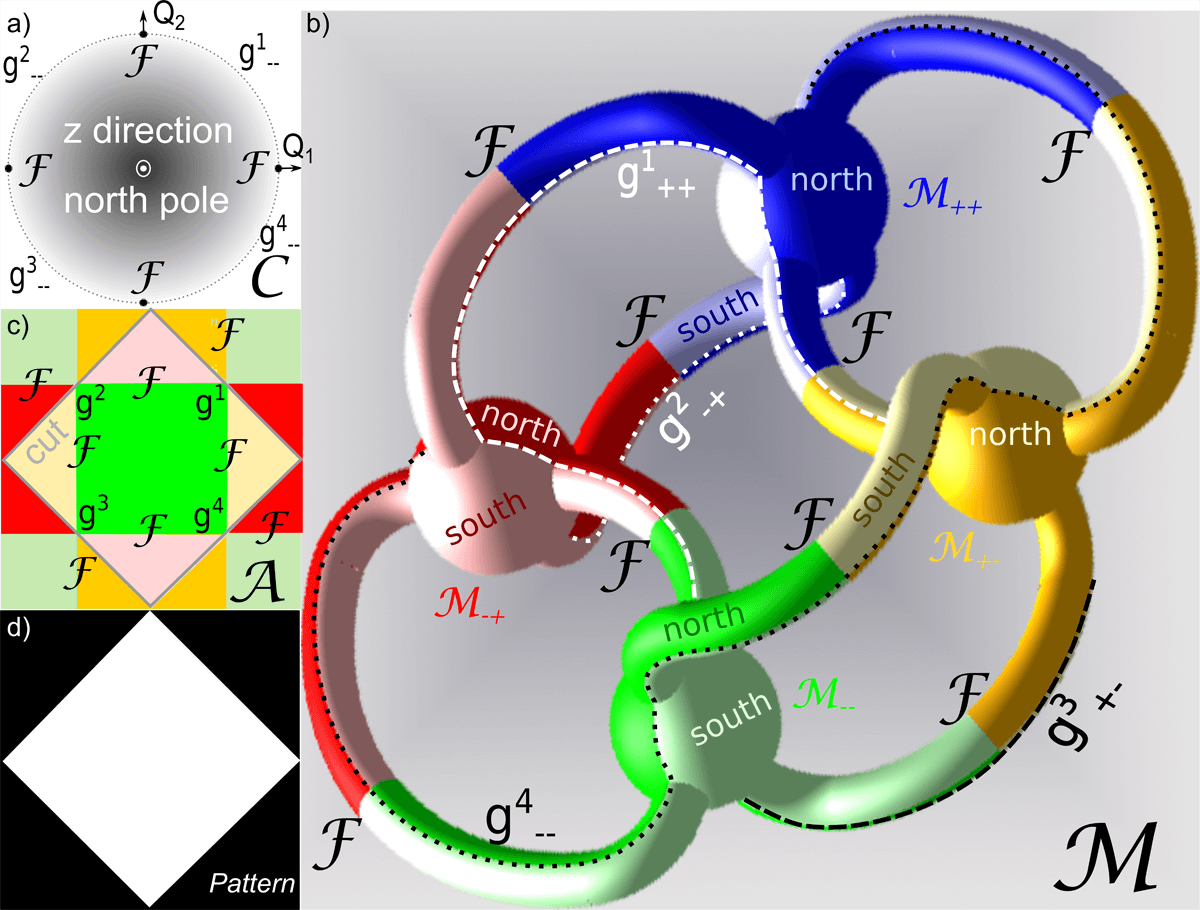}
	\caption{  a) Top view of the four-fold symmetric control space including the fence points and the maximum segments $g_{--}^i$, $i=1,2,3,4$ of the four gates.
		b) Genus five stationary manifold $\cal M$. Blue colors correspond to minima (${\cal M}_{++}$), green to maxima (${\cal M}_{--}$), red and yellow to saddle points.
		c) Projection of the half of $\cal M$ lying closer to ${\cal M}_{--}$ into action space. The cut in $\cal A$ is the projection of the points in $\cal M$ separating both halves.  d) Magnetic pattern generating the four-fold symmetric universal potential. \changes{(see appendix \ref{definitions} for a concise definition of the gates)} }
	\label{figM4loops}\end{center}
\end{figure*}

We now understand how to produce a topological transition of the transport modes by continuously deforming the loop in control space. The transport direction in action space is topologically protected for any deformation of the modulation loop that does not alter the winding number around the fence points. A topological transition occurs when we move the loop across one of the fence points.

We can characterize the simplest modulation loops by the two segments of the equator that they cross. We define $  \downarrow \kern -4px g^i$, $i=1,2$ as a \changes{\bf south traveling path} that passes the equator segment $g^i$ between the two fence points. We complete the loop with an analogous  \changes{\bf north traveling path}, $ \uparrow \kern -4px g^j$.
In Fig. \ref{figM2loops}d we depict a phase diagram of the transport for the fundamental loops ${\cal L}_{\cal C}={  \downarrow \kern -4px g^i}  {\uparrow \kern -4px g^j}$. Modulation loops that do not cross the equator, as well as those passing the same equator segment south and north, cause no transport. Modulation loops passing one segment south and the other one north induce transport.

\section{Four fold symmetry}\label{four-fold}
In Ref. \cite{delasHeras} we study in detail theoretically and with computer simulations four-fold symmetric patterns.
Here we summarize the theoretical results, present experimental data, and show the connection to the two-fold symmetric system.
\subsection{Theory}
The four-fold symmetric magnetic potential 
\begin{equation}
\psi_4(z,x,y)=\psi_2(z,x)+\psi_2(z,y)
\end{equation}
is closely related to the two-fold symmetric potential $\psi_2$,
where ${\mathbf e}_x$ points along $\mathbf{Q}_1$ and ${\mathbf e}_y$  points along ${\mathbf Q}_2$.
Action space ${\cal A}_4={\cal A}_2\otimes{\cal A}_2$ is the product space of two circles and thus a torus with both $Qx$ and $Qy$ varying from $0$ to $2\pi$. 
There is no indifferent direction and hence it is simpler to use full control space $\cal C$. However there exist fence-points satisfying equations (\ref{stationary}) and (\ref{fence}). These fence points play the same role as in $N=2$-case in generating transport.

The universal scalar magnetic potential is the superposition of two stripe potentials that separate the variables $x$ and $y$ in action space. Therefore, we have four fence points on the equator of the control space sitting in the $\pm {\mathbf e}_x$ and $\pm {\mathbf e}_y$ directions (Fig. \ref{figM4loops}a). 

We define the unit vectors 
\begin{equation}\label{vectorse}\begin{array}{ccc}
\mathbf{e}_{1}({\mathbf x}_{\cal A})=\frac{\partial_{1}{\mathbf H}^p}{|\partial_{1}{\mathbf H}^p|},&\qquad&\mathbf{e}_{2}({\mathbf x}_{\cal A})=\frac{\partial_{2}{\mathbf H}^p}{|\partial_{2}{\mathbf H}^p|}
\end{array},
\end{equation}
where $\partial_{1,2}$ denote the partial derivatives with respect to the two coordinates in $\cal A$. Points in $\cal A$ with $ \mathbf{e}_{1}\times\mathbf{e}_{2}\ne \mathbf{0}$ are made stationary by two opposite external fields \cite{Loehr,delasHeras}
\begin{equation} \label{Hexts}
\mathbf{H}^{(s)}_{ext}=\pm\frac{\mathbf{e}_{1}\times\mathbf{e}_{2}}{|\mathbf{e}_{1}\times\mathbf{e}_{2}|}.
\end{equation}
The two signs in (\ref{Hexts}) cause  opposite curvature of $U^*$ and thus each point in $\cal A$ can be made either an extremum (maximum or minimum) or a saddle point.
Hence, we can split action space into forbidden and accessible regions (see Fig. \ref{figM4loops}c). Allowed regions are regions of extrema and they are colored green, while forbidden regions are regions of saddle points and are colored red and yellow.

Each field in control space renders 4 points in action space stationary, a maximum a minimum and two saddle-points. Hence our stationary manifold consists of four copies of control space (instead of two for the case $N=2$). The indices of the four sub-manifolds ${\cal M}_{++}$,${\cal M}_{+-}$, ${\cal M}_{-+}$, and  ${\cal M}_{--}$ correspond to a minimum (index $+$) or  a maximum (index $-$) along the $x$ (first index) and $y$ (second index) coordinates.  The four fence points in control space deform into circular fences in $\cal M$. The four sub-manifolds are glued together at eight fences to form the full stationary manifold, see Fig. \ref{figM4loops}b.
The stationary manifold is a genus five surface. 

The fences in ${\cal M}$ are projected onto lines in action space that are the borders between the \changes{{\bf forbidden and allowed regions}}. The fences do not intersect on $\cal M$ but they do in $\cal A$. 
This is possible because the fences meet at special points in  $\cal A$ with $ \mathbf{e}_{1}\times\mathbf{e}_{2}= \mathbf{0}$, that we call \changes{{\bf the gates}}. As we will show below, the gates are the only points that connect two consecutive allowed regions.  From equations (\ref{universalpotential}), (\ref{vectorse}), and (\ref{Hexts}) we conclude that the gates are rendered stationary by the whole grand circle on $\cal C$ around $\mathbf{e}_{1}=\mathbf{e}_{2}$. For the four-fold symmetric pattern there are four coinciding gates $g^{i}, i=1,2,3,4$ in $\cal C$ that run across the equator right through the four fence points, see Fig. \ref{figM4loops}a. In ${\cal C}\otimes{\cal A}$ each gate is a line on $\cal M$ that lies in a single copy of the equator of control space and that is projected into the gate 
 in $\cal A$. Since one gate in $\cal C$ cuts through all four fences the gate in $\cal A$ must be the same as the intersection of fences in $\cal A$.

\begin{figure}[t]
	\includegraphics[width=0.9\columnwidth]{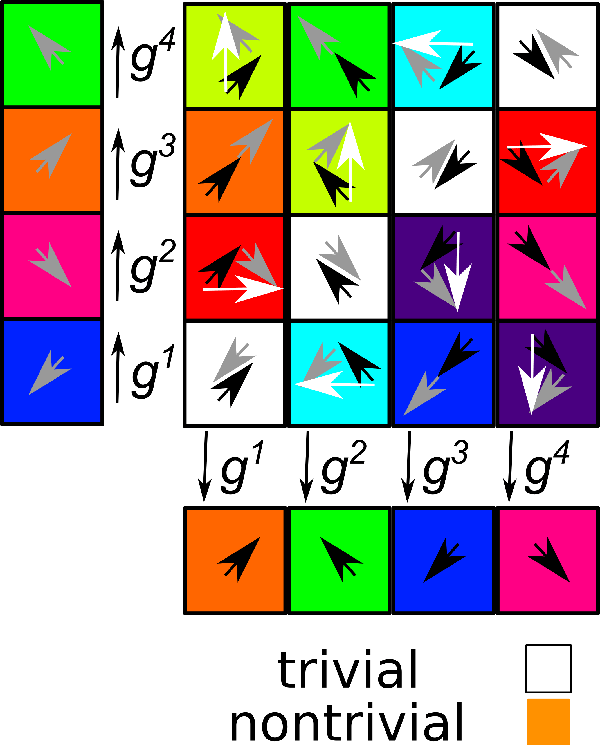}
	\caption{Phase diagram of the transport modes in a four-fold symmetric system. Black arrows denote the traveling direction in the first, south heading part of the modulation, gray ones describe the transport direction of the second part, and white arrows describe the travel direction of the full loop. The colors of the squares indicate the traveling direction. Loops passing through the same gate twice do not induce transport (white). All other combinations induce transport in one of the eight neighboring unit cells.}
	\label{figPhasediagram4}
\end{figure}

In $\cal C$ the fence points cut each gate into 4 segments $g^{i}_{++},g^{i}_{+-},g^{i}_{-+},g^{i}_{--},$ that are projections of the gates in the corresponding sub-manifolds of $\cal M$. Each gate crosses four of the eight fences in $\cal M$ and passes over all four sub-manifolds.  Each fence crosses two of the four gates. The gate $g^{i+1}_{\alpha\beta}$ in $\cal C$ coincides with the gate $g^{i}_{\alpha\beta}$ rotated by $\pi/2$. Therefore the maximum segments of the gates $g^{i}_{--},i=1,2,3,4$ fill the whole equator and subdivide $\cal C$ as well as all sub-manifolds ${\cal M}_{\alpha,\beta}$ and their projections on $\cal A$ into simply connected northern and southern hemispheres. Northern and southern allowed regions touch each other in $\cal A$  only at the gates. Nontrivial adiabatic transport therefore must pass these singular points.

In the following we will first deal with the transport of paramagnetic particles. Since these reside on the maxima of $U^*$, we are only interested in loops on ${\cal M}_{--}$. Modulation loops that remain in one hemisphere of control space are zero homotopic loops of the four punctured sphere and have zero homotopic preimage loops on $\cal M$. The simplest
non trivial modulation loop must cross the equator twice. Such loop ${\cal L}_{\cal C}={  \downarrow \kern -4px {g}^i}  {\uparrow \kern -4px {g}^j}$  
consists of two paths $  \downarrow \kern -4px g^i$ and $  \uparrow \kern -4px g^j$. $  \downarrow \kern -4px g^i$ is a path from north to south passing the gate $g_{--}^{i}$ and $  \uparrow \kern -4px g^j$ is the reverse path passing through gate $g_{--}^{j}$ from south to north.
The winding numbers in control space around the fences  cause  similar winding in action space. Fig. 
\ref{figPhasediagram4} shows the phase diagram of the transport directions of the simplest gate crossing modulation loops.
The topological transition between different transport modes is similar to the two-fold case. Modulation loops passing a fence cause topological transitions. \\
Diamagnetic particles move synchronously with the paramagnetic ones at a fixed distance $\mathbf{d}=1/2(\mathbf{a}_1+\mathbf{a}_2)$, to the paramagnets.

\begin{figure}
	\includegraphics[width=\columnwidth]{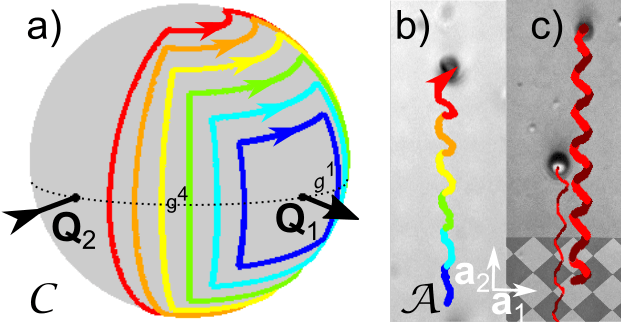}
	\caption{a) various modulation loops in control space of the type ${\cal L}_{\cal C}={  \downarrow \kern -4px {g}^1} { \uparrow \kern -4px {g}^4}$. b) Resulting trajectories of paramagnetic colloids. All modulation loops induce transport into the same ${\mathbf a}_2$-direction. c) Trajectories of a paramagnetic (thick line) and a diamagnetic colloid (thin line) subjected to the large (red) modulation loop.  Trajectories are colored in dark red for the ${  \downarrow \kern -4px {g}^1}$ segment and in bright red for the $ { \uparrow \kern -4px {g}^4}$ segment of the loop. Both types of particles are synchronously transported into the same direction. The trajectories however are shifted by $\mathbf{d}=1/2(\mathbf{a}_1+\mathbf{a}_2)$. The background in b) and c) are reflection microscopy images of the four-fold symmetric pattern. We have added the theoretical pattern to the lower part of c) for clarity. The length of the arrows indicating the lattice vectors is equivalent to the lattice constant $a=7\,\mu\textrm{m}$. A video clip of the motion of the paramagnetic and the diamagnetic colloidal particle in c) is provided in \cite{CLIPS}. 
	}
	\label{four-foldexperimentalamplitude}
\end{figure}

\begin{figure}[t]
	\includegraphics[width=\columnwidth]{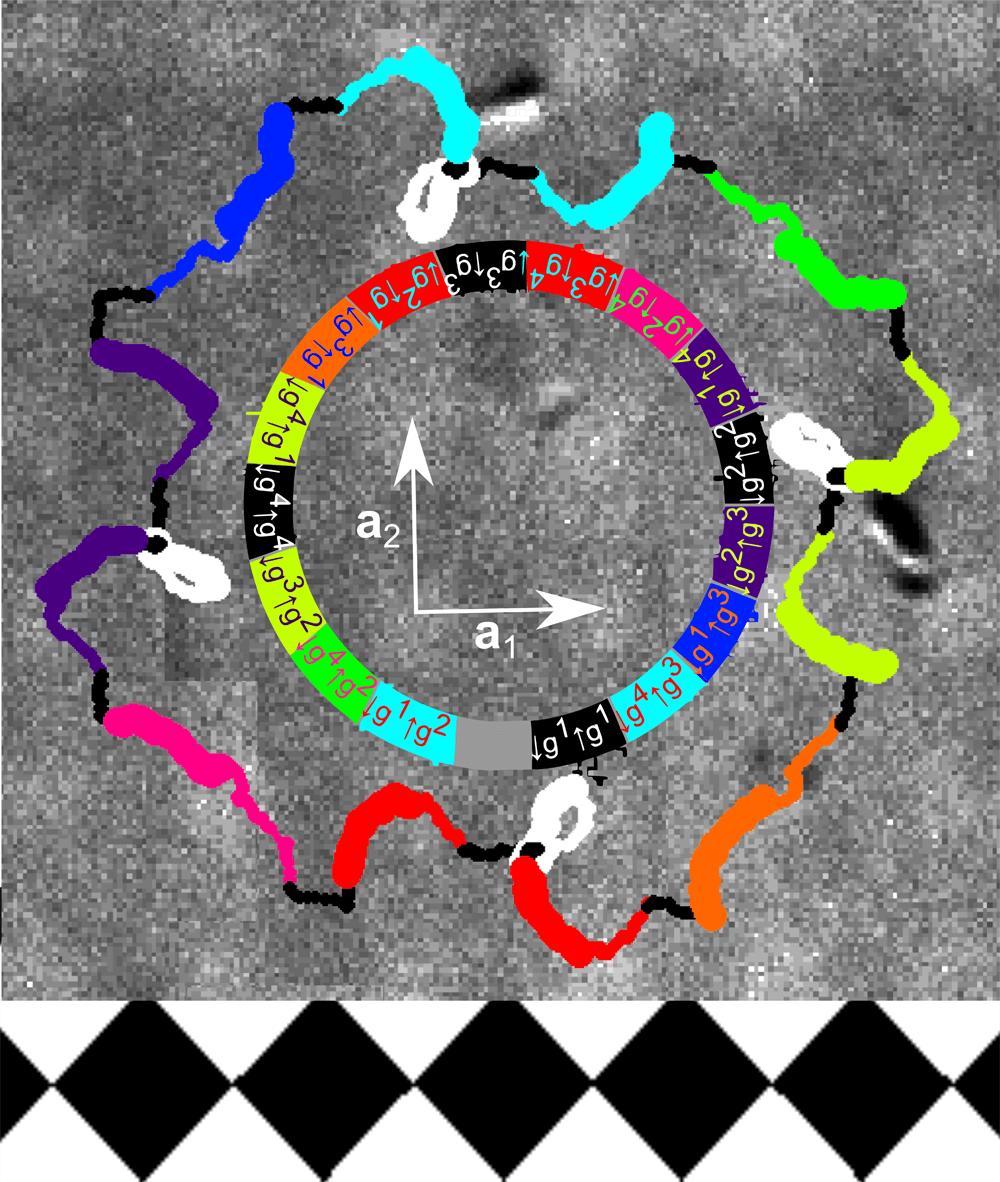}
	\caption{Experimental trajectory of a paramagnetic colloidal particle in action space $\cal A$ caused by a modulation poly-loop in $\cal C$. The poly-loop consists of a sequence of all fundamental modulation loops in the phase diagram of Fig. \ref{figPhasediagram4}. The single fundamental loops are colored according to the color in the phase diagram in Fig. \ref{figPhasediagram4}. South traveling segments are marked as thick lines while north traveling segments are marked as thin lines. Consecutive  loops are connected by trivial constant latitude connections that remain in the north  of $\cal C$ (black trajectories). The type of the single loops is indicated inside the region surrounded by the trajectory. The background is the reflection microscopy image of the underlying square magnetic pattern. At the bottom we show a scheme of the theoretical pattern aligned and oriented to the weakly visible experimental pattern on the top.  A video clip of the motion of the paramagnetic colloidal particle is provided in \cite{CLIPS}.} 
	\label{figsuperloop4}
\end{figure}
\subsection{Experiments}

Four fold symmetric patterns have been created by lithography \cite{Jarosz,KET2010,UKK2010,CBF1998}. 
The lithographic magnetic patterns are designed to have the four-fold symmetric pattern of Fig. \ref{figpattern}b with a period $a=7\mu m$. The strength of the pattern field directly on top of the surface of the thin $Qt<1$ lithographic film is $H^p\approx 3 \,\textrm{kA/m}$. Details on the production process are given in the appendix \ref{appendixpatterning}.

Lithographic edge effects of the pattern production process render white regions larger than the black regions such that the average magnetization of the film is non-zero. This breaks the $S_4$-symmetry of the pattern, but it does not affect the $S_4$-symmetry of the universal limit $Qz>1$ and the $C_4$ symmetry is preserved for the pattern and the universal limit.  We coat the patterned magnetic film with a photo-resist of thickness $1.6\,\mu \textrm{m}$. The thickness is a compromise of achieving universality and keeping the magnetic field of the pattern sufficiently strong. Paramagnetic colloids (diameter $d=2.7\,\mu \textrm{m}$) immersed into deionized water are placed on top of the coating.

In Fig. \ref{four-foldexperimentalamplitude}a we apply fundamental modulation loops. They all fall in the class ${\cal L}_{\cal C}={  \downarrow \kern -4px {g}^1} { \uparrow \kern -4px {g}^4}$, but have different proximity to the fence point in the ${\mathbf Q}_1$  direction in $\cal C$. In Fig. \ref{four-foldexperimentalamplitude}b we plot the corresponding experimental trajectories of paramagnetic particles. No matter which particular modulation loop within the same homotopy class we choose, the global result after completing the loop is the transport of the particle by one unit vector ${\mathbf a}_2$. Modulation loops closer to the encircled fence point have a straighter trajectory than loops passing the equator far from it (see Fig. \ref{four-foldexperimentalamplitude}b). 

In Fig. \ref{four-foldexperimentalamplitude}c we repeat the experiment with {paramagnetic and diamagnetic colloids} using the largest modulation loop (red). 
We immerse paramagnetic and non magnetic (polystyrene $d=4 \,\mu\textrm{m}$, susceptibility $\sim -10^{-5}$) particles in ferrofluid which renders the non magnetic particles effectively diamagnetic. The direction of the magnetic field inside the ferrofluid is used for the direction in control space. It has a higher tilt angle to the film normal then the tilt of the external field applied by the coils, because of refraction at the glass ferrofluid interface. All loops with colloids immersed in ferrofluids are corrected for this effect.  
Both particles are transported in $\mathbf{a}_2$ direction by the red loop and the predicted shift of both trajectories by $1/2(\mathbf{a}_1+\mathbf{a}_2)$
is clearly visible.

 In Fig. \ref{figsuperloop4}   we show the motion of a paramagnetic particle subject to a modulation poly-loop
 that consists of all sixteen fundamental loops ${  \downarrow \kern -4px {g}^i}  {\uparrow \kern -4px {g}^j}$ of the phase diagram of Fig. \ref{figPhasediagram4}.  We plot the fundamental sections of the trajectory of the particles in the colors of the corresponding fundamental loops in the phase diagram (Fig. \ref{figPhasediagram4}). 
It can easily be seen that all fundamental loops induce the theoretically  predicted transport. 
Due to the lack of $S_4$-symmetry the lemniscates of the zero homotopic loops in $\cal A$ (white) lose their inversion symmetry with respect to the gate in $\cal A$ (the crossing point of the lemniscate) resulting in a big and a tiny white loop.  We conclude that the experimental response of the particles to all modulation loops is in perfect agreement with the theoretical predictions.

\section{three-fold symmetry}\label{three-fold}

In Ref. \cite{Loehr} we studied the motion on a $C_6$-symmetric pattern theoretically and provided experiments of the adiabatic motion on this pattern. The $C_6$-symmetric pattern is part of the family of three-fold symmetric patterns. Here, we extend the theory to this entire family, explain a new topological transition within the family and corroborate the theory with experiments on adiabatic and ratchet transport for all family members. We also confirm experimentally the new topological transition from $C_6$-like toward $S_6$-like topology. 
 
\subsection{Control Space, stationary Manifold and Action Space}
\begin{figure*}\begin{center}
	\includegraphics[width=1.75\columnwidth]{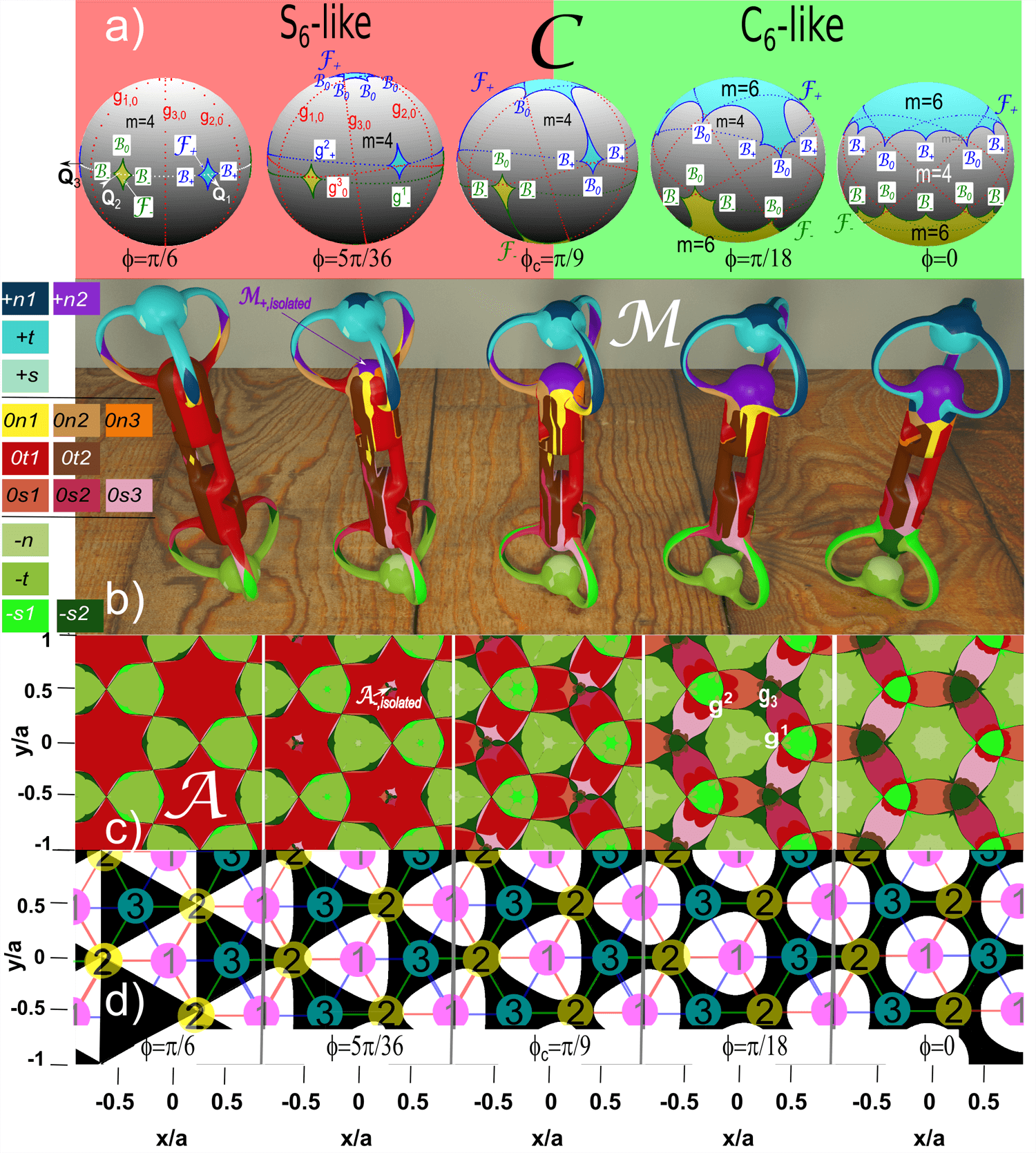}
	\caption{Topology  of the three-fold symmetric case  as a function of the phase $\phi$: a) control spaces with areas of different multiplicity $m=4$ (gray), $m=6$ extra maximum areas (yellow) which are surrounded by the southern fence ${\cal F}_-$ (green lines) and $m=6$ extra minimum areas (cyan) surrounded by the northern fence ${\cal F}_+$ (blue lines). The gates $g$ are colored according to their segments. b) Genus seven stationary manifolds $\cal M$. Blue colors correspond to minima (${\cal M}_+$), red colors to saddle points (${\cal M}_0$) and green colors to maxima (${\cal M}_-$). Fences are the boundaries between the color families and pseudo fences are the boundaries between the colors of one family. Areas with labeled with a prefix (n) are projected into the northern area  or the northern satellites in $\cal C$,  with a prefix (s) to the south , with a prefix (t) to the tropical $m=4$ area of $\cal C$. 
	c) Projection of the lower half of the stationary manifold  into action space $\cal A$. The projection of the upper half exactly matches the lower projection, however, with the colors of the upper half replacing those of the lower half. The areas $0t1$ and $0t2$ contain cuts (not shown) that connect the shown projection of the southern half of ${\cal M}_0$ to its similar twin projection of the northern half. d) Magnetic patterns corresponding to the different phases. Up magnetized regions shown in white and down magnetized regions in black. The pink, yellow and cyan circles mark the three high symmetry points of the lattice and the high symmetry lines connecting the points form the \changes{\bf $12$-, $23$-, and $31$-network}. Higher resolution images of $\cal A$, $\cal C$, and $\cal M$ for each of the phases with further details can be found in the appendix \ref{appendixpicturesM} \changes{, definitions of the various geometrical objects in appendix \ref{definitions}.}}
	\label{figM3s}\end{center}
\end{figure*}

The transport on the three-fold symmetric pattern is more complex than on the two-fold and four-fold patterns. The increased complexity
is related to the fact that the three reciprocal lattice vectors $\mathbf{Q}_1$, $\mathbf{Q}_2$ and $\mathbf{Q}_3$ are linearly dependent. In Fig. \ref{figM3s} we show the control spaces, the stationary manifolds, and the action spaces of the three-fold symmetric system for various values of the phase $\phi$ of the pattern. The phase $\phi$ varies in an interval $0\leq \phi \leq \pi/6$ which covers all possible three-fold symmetries including $C_6$ ($\phi=0$) and $S_6$ ($\phi=\pi/6$). We call the range 
 $\pi/9<\phi\leq\pi/6$ the $S_6$-like case and the range  $0\leq \phi<\pi/9$ the $C_6$-like case. 
The range $\pi/6<\phi<2\pi$ repeats those patterns, however, centered around one of the other two three-fold symmetric points and/or interchanging up and down magnetized regions, see Fig. \ref{figpattern}c. For each value of the phase $\phi$ of the pattern the stationary manifold $\cal M$ in Fig. \ref{figM3s}  is a genus seven surface. As in the two and four-fold cases there are fences of $\cal M$ separating different sub-manifolds. 
We distinguish two different fences: i) the maximum fence ${\cal F}_-={\cal M}_-\cap{\cal M}_0$, which is the border between the regions of maxima of the colloidal potential (green colors) and the saddle point regions (red colors), and ii) the minimum fence ${\cal F}_+={\cal M}_+\cap{\cal M}_0$, which is the border between saddle points and minima (blue colors).

Due to the separability of the two-fold and four-fold problem the fences were projected onto single points in control space. For $N=3$ the fences in control space $\cal C$ are not points but  closed lines. In Fig. \ref{figM3s}a the maximum fences ${\cal F}_-$ are shown as green lines and the minimum fences ${\cal F}_+$
as blue lines in control space.
The fences in $\cal C$ separate regions of different multiplicity of preimages in $\cal M$. For any value of $\phi$ there is one multiply connected area (gray) that we call the tropics. This area has multiplicity $m=4$, that is, one external field renders 4 points in action space $\cal A$ stationary: one maximum, one minimum and two saddle points. In addition there are concave excess regions of multiplicity $m=6$. In the yellow regions surrounded by ${\cal F}_-$ there is an extra maximum and an extra saddle point, while in the cyan regions (surrounded by ${\cal F}_+$) there is an additional minimum and also a saddle point. The control space always shows the $C_3$ symmetry and the inversion symmetry $U^*({\mathbf H}_{ext})= -U^*( -{\mathbf H}_{ext})$, see equation (7).
For this reason the cyan regions are the inverted yellow regions on the opposite side of control space. A rotation of control space by $2\pi/3$ leaves the control space invariant. Not all excess regions are visible in Fig. \ref{figM3s}a. We can infer the location of hidden excess regions from the visible excess regions using these two symmetry operations.

The stationary manifold is formed from multiple copies (according to the multiplicity) of the areas in $\cal C$. As already mentioned the two fences separate the three sub-manifolds of $\cal M$. But on $\cal M$ there are additional preimages of the fences in $\cal C$ that are different from the fences in $\cal M$. As in the two-fold case we call these pseudo fences. The pseudo fences in $\cal M$ and in $\cal A$ (Fig. \ref{figM3s}b and c) are the borders between the areas with different colors belonging to the same color family (red, green or blue). 

In the three-fold case we have an additional type of point that we did not have in the two- and four- fold cases. They are \changes{\bf bifurcation points } \cite{bifurcation}, located on the fences on $\cal M$. These are the only points where more than two areas of different colors meet. We have ${\cal  B}_-$ (${\cal  B}_+$) bifurcation points where three areas on ${\cal M}_-$ (${\cal M}_+$) and one area on ${\cal M}_0$ meet, and ${\cal  B}_0$ bifurcation points where three areas on ${\cal M}_0$ and one area in either ${\cal M}_-$ or ${\cal M}_+$ meet. Both types of bifurcation points split the fences on $\cal M$, as well as their projection onto $\cal C$ and onto $\cal A$, into single segments (Fig. \ref{figM3s}a).

We now consider a control loop ${\cal L}_{\cal C}$ that passes through a multiplicity $m=6$ excess region. When the loop crosses the fence towards this region the multiplicity increases by two. This happens via the creation of an extremum- saddle point pair at the fence on $\cal M$. At the same time the other preexisting stationary points pass a pseudo fence. When the modulation loop leaves the excess region the multiplicity returns to $m=4$. Now a extremum- saddle point pair is annihilated at the fence. When the loop transports a paramagnetic colloidal particle, the particle is now either adiabatically transported through the pseudo fence or the colloid carrying maximum is annihilated at the fence resulting in ratchet motion.

The type of transport is directly related to the number of bifurcation points of each excess area enclosed by the modulation loop. When the modulation loop in $\cal C$ encircles an even number of ${\cal B}^-$ (${\cal B}^+$) bifurcation points of one excess area,
then the exit of the excess area corresponds to a pseudo fence on ${\cal M}_{-}$  (${\cal M}_{+}$) and the transport is adiabatic. If the number of encircled ${\cal B}^-$ (${\cal B}^+$) bifurcation points in an excess area is odd the exit of the excess area corresponds to the fence of ${\cal M}_{-}$ and  the loop induces a ratchet. This ratchet is time reversible if the number of encircled ${\cal B}^0$ bifurcation points is a multiple of 2 (3) for each excess area in the $S_6$ ($C_6$)-like case, and non-time reversible otherwise. A \changes{\bf time reversal ratchet} is a ratchet where the reversed modulation results in the reversed transport direction.

\subsection{$S_6$-$C_6$-Topological Transition}
The topology of the $S_6$-like ($C_6$-like) systems is the same as the $S_6$- ($C_6$) symmetric system. A topological transition between $S_6$-like and $C_6$-like occurs at a critical phase $\phi_c=\pi/9$ of the pattern. 
The topological transition can be easily seen in control space. Control space consists of areas with different multiplicity. The shape and location of the areas vary with the phase $\phi$.  
The topology of these areas, however, only differs for the two situations  $\pi/9<\phi\leq\pi/6$ ($S_6$-like) and $|\phi|<\pi/9$ ($C_6$-like). Fig. \ref{figM3s}a shows examples of the control spaces $\cal C$ for these two cases as well as for the critical transition value $\phi_c=\pi/9$. 

For any value of the phase $\phi$ there is one multiply connected area in control space $\cal C$, the tropics (gray) having four preimages ($m=4$). In the $S_6$-like case there are four areas (yellow) surrounded by a maximum fence ${\cal F}_-$ (green) with multiplicity $m=6$ housing an extra maximum- saddle point pair. One area is a (hidden) southern area (opposite to the visible cyan northern area) surrounded by a maximum fence ${\cal F}_-$ with 6 segments joined at  six ${\cal B}_{0}$ bifurcation points. The other three are southern \changes{{\bf satellites}} surrounded by a maximum fence ${\cal F}_-$ with four segments joined at two ${\cal B}_{0}$ and two ${\cal B}_{-}$ bifurcation points. We call these areas southern satellites since at the topological transition they merge with the southern area.  
The southern area shrinks to zero as the phase approaches $\phi=\pi/6$ ($S_6$-symmetry). Four further areas of multiplicity $m=6$ (cyan) housing an extra minimum-, saddle point pair are located opposite to the yellow ones. 

The topological transition occurs at $\phi_c=\pi/9$ where the three southern satellites join with the corresponding southern area. Simultaneously the northern satellites join with the northern area. 
In each satellite one ${\cal B}_0$ bifurcation point merges with one ${\cal B}_0$ bifurcation point from the polar area. Thus the two polar fence segments of a satellite are both unified with two fence segments of the polar region. This results in a new topology with only two polar areas for the $C_6$-like case. Both areas are surrounded by a fence with twelve segments that are separated by a sequence of bifurcation points alternating between ${\cal B}_{0}$ and ${\cal B}_{-}$ (${\cal B}_{+}$). 

Due to the inversion symmetry $U^*({\mathbf H}_{ext})= -U^*( -{\mathbf H}_{ext})$ the transport of diamagnetic particles on ${\cal M}_+$ is the same as those of the transport of paramagnetic particles on ${\cal M}_-$ at the inverted external magnetic field. In Fig. \ref{figM3s}b we depict the topology of the stationary manifold for five different phases $\phi$. The true stationary manifold is embedded in a four dimensional curved phase space and we can only show its topology by deforming it until it finally is embedded into three dimensions. The deformation partially breaks the three-fold $C_3$-symmetry, however, the inversion symmetry shows up as a up-down mirror symmetry of the manifolds, accompanied by an inversion of the sign of the index of the submanifolds.  

In the $S_6$-like case there is a (hidden) preimage on $\cal M$ of the southern excess area of $\cal C$ that is entirely surrounded by ${\cal M}_0$ areas and therefore disconnected from the rest of ${\cal M}_{-}$.  We call this region ${\cal M}_{-}^{isolated}$ and it lies opposite to the visible ${\cal M}_{+}^{isolated}$ region  in Fig. \ref{figM3s}b.
This isolated area is surrounded by fences and does not contain pseudo fences. Therefore, all paths of steepest descend can only lead away from it since return points lie on pseudo fences. For this reason the isolated area ${\cal M}_{-}^{isolated}$  might be emptied once of a colloid but can never be refilled. Since we are interested in the motion occurring by the periodic repetition of modulation loops this area and hence its projection into $\cal C$ is completely irrelevant. After the topological transition to the $C_6$-like case the formerly irrelevant polar areas on $\cal C$ incorporate the three corresponding satellites. Hence ${\cal M}_{-}^{isolated}$ is no longer disconnected from the rest of ${\cal M}_-$ and becomes relevant for the motion.  

\begin{figure}[t]
	\includegraphics[width=\columnwidth]{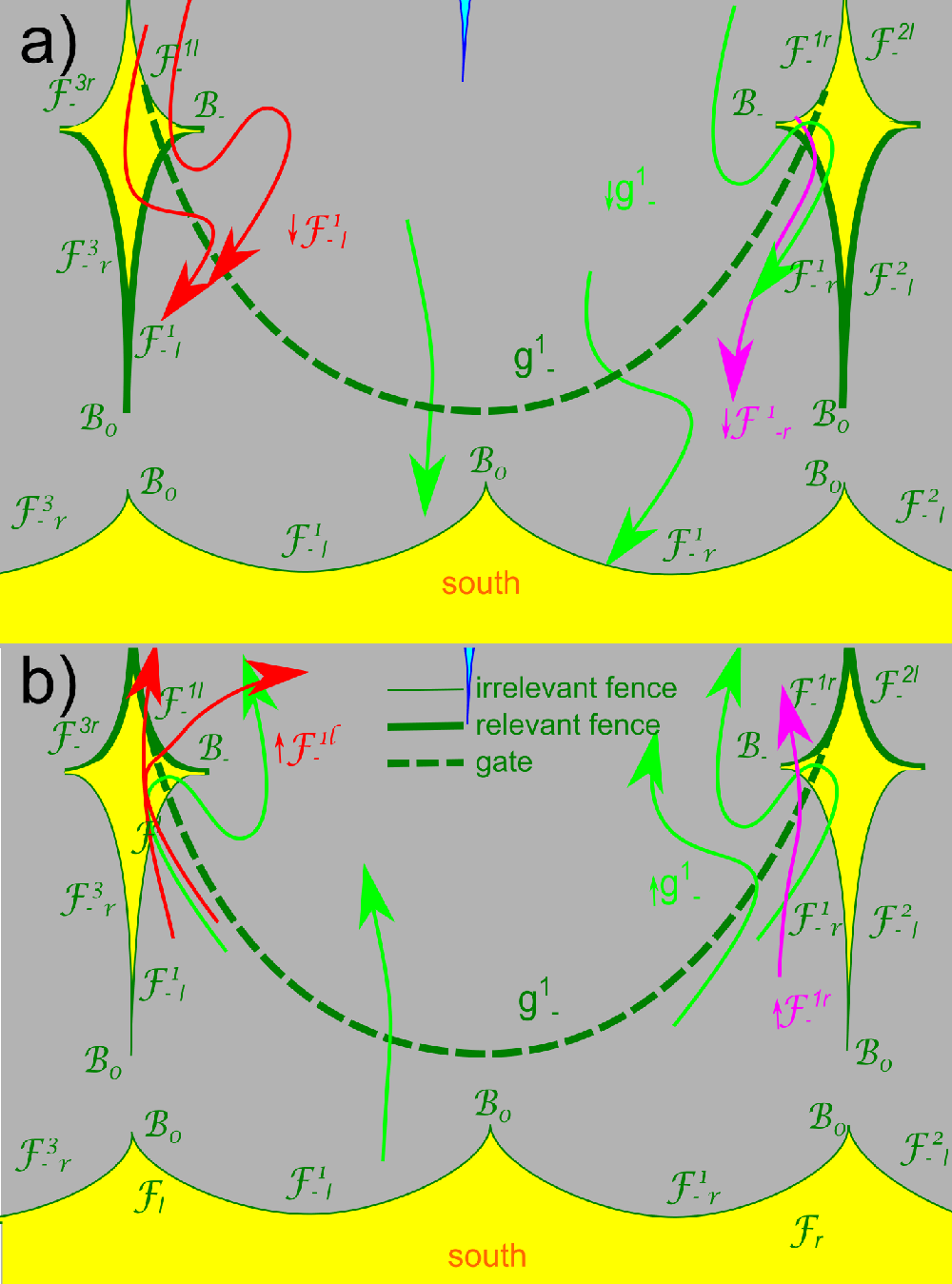}
	\caption{Paths in the southern hemisphere of $\cal C$ relevant for the loops in the $S_6$-like case. Paths of type $  \downarrow \kern -4px  g^1_-$  and $  \uparrow \kern -4px  g^1_-$ are shown in green, $  \downarrow \kern -4px {\cal F}^{1}_{-l}$ and  $  \uparrow \kern -4px {\cal F}^{1l}_{-}$ in red and  $  \downarrow \kern -4px {\cal F}^{1}_{-r}$  and $  \uparrow \kern -4px {\cal F}^{1r}_{-}$ in purple.
		The fences of the satellites are enumerated according to the gate closest to them. The index $l$ (r) indicates that the fences are left(right) of the corresponding gate and the position of the index (subscript or superscript) indicates the location of  the fence segment in the satellite (up or down). The fence segments of the irrelevant polar fence share the names with those segments of the satellites with which they will join beyond the topological transition. 
		a) South traveling paths for which the lower fences (highlighted) are relevant. b) North traveling paths for which the upper fence segments (highlighted) are relevant. \changes{See appendix \ref{definitions} for definitions and terminology.} }
	\label{paths}	
\end{figure}

For any $\phi$ the stationary manifold $\cal M$ is a genus seven surface and there are thus 14 different winding numbers. In the $S_6$-like case only two linear independent winding numbers correspond to loops ${\cal L}_{\cal M}\subset {\cal M}_{-}$ that are lying entirely in ${\cal M}_{-}$. Therefore there are only two ways of nontrivial adiabatic transport modes. When  ${\cal M}_{-}^{isolated}$ joins with the other part of ${\cal M}_{-}$ ($\phi=\pi/9$) two additional windings around holes of ${\cal M}_{-}$ occur allowing two new  transport routes through the formerly isolated region of $\cal A$.

Fig. \ref{figM3s}c shows the projection of the lower half of the stationary manifold  into action space $\cal A$. The projection of the upper half exactly matches the lower projection, however, with the colors of the upper half replacing those of the lower half. Fig. \ref{figM3s}d shows possible magnetization patterns that generate the universal potentials $U^*$. We also show the three-fold symmetric points ${\mathbf x}_{{\cal A},1}$,
	${\mathbf x}_{{\cal A},2}$, and ${\mathbf x}_{{\cal A},3}$ within the
	pattern. Their connections form a $12$-, $23$-, and $31$-network which are
	the three kinds of high symmetry lines of the lattice.

\subsection{Modulation loops in the $S_6$-like case}\label{theoryS6}

As in the four-fold symmetric case, in the three-fold case two neighboring allowed regions in $\cal A$ only touch each other at a single point, the gate. Hence modulation loops in $\cal C$ causing adiabatic transport in $\cal A$ have to pass through the grand circles of the gates in $\cal C$.

In the three-fold symmetric case there are six gates $g^i,g_i,i=1,2,3$ of two different types $g^i$ and $g_i$. All gates in $\cal M$ are closed curves dissected twice by ${\cal F}_+$ and twice by ${\cal F}_-$ (the gates on $\cal M$ are shown in more detailed images of $\cal M$ in the appendix \ref{appendixpicturesM} of this work). Hence, for the projection of each gate into $\cal C$ there is one minimum gate segment $g_+$ (blue in Fig. \ref{figM3s}a) projected from ${\cal M}_+$, one {maximum} segment $g_-$ (green) projected from ${\cal M}_-$, and two saddle point gate segments $g_0$ (red). 

 Whenever we cross a gate segment of type $g^i_-$ or $g_{i,-}$ in the $m=4$ (gray) region of $\cal C$ the unique maximum in $\cal A$ adiabatically passes from one allowed area through the gate $g^i_-$ or $g_{i,-}$ in $\cal A$ to the allowed area on the other side. For the $S_6$-like case the maximum segments $g_{i,-}$ of the three gates $g_i$, $i=1,2,3$ lie entirely in the irrelevant  southern excess region of $\cal C$ and are hence unimportant for transport. For the $C_6$-like case all six gates cross both polar excess regions. Therefore all gates become important for transport. Eventually if we have  $C_6$-symmetry (at $\phi=0$) the difference in character between both types of gates $g^i$ and $g_i$ completely vanishes.
Gates cross each other in $\cal C$ but in $\cal M$ they do not cross.
Only when we have a $S_6$-symmetry ($\phi=\pi/6$) the three gates $g_i$ of the isolated allowed region merge such that they touch each other in $\cal M$ and are all projected into the one monkey saddle point in $\cal A$. Otherwise the gates are separated curves on $\cal M$ much in the same way as in the four-fold case.

For the $S_6$-like case we can characterize fundamental modulation loops ${\cal L}_{\cal C}={  \downarrow \kern -4px s}  {\uparrow \kern -4px s'}$ in $\cal C$ by two loop segments. One is a south heading path $\downarrow \kern -4px s$ and  the other is a north heading path $ \uparrow \kern -4px s'$. 
There are three possible types of south traveling paths. It is either of type $  \downarrow \kern -4px  g^i_-$, of type $  \downarrow \kern -4px {\cal F}^{i}_{-l}$, or of type $  \downarrow \kern -4px {\cal F}^{i}_{-r}$ with $i=1,2,3$ in all cases.

Each gate segment $g^i_-$ has two ${\cal B}_-$ bifurcation points close to it. 
A path of type $  \downarrow \kern -4px g^i_-$  is a path that moves south between these two bifurcation points. It might thereby completely stay in the gray $m=4$ area or eventually enter a southern satellite (yellow) and exit it again via the same southern fence segment. Examples of all types of paths are shown in Fig. \ref{paths}a.
A path of type $  \downarrow \kern -4px {\cal F}^{i}_{-l}$ passes left of the two bifurcation points. It thereby has to enter the  $m=6$ satellite to the left of gate $g^i_{-}$ through one of the two upper fence segments. The path exits the satellite via the lower right fence segment that is also crossed by the corresponding gate segment $g^i_{-}$. A path of type $  \downarrow \kern -4px {\cal F}^{i}_{-r}$ is the equivalent path that passes right of the two bifurcation points and enters the satellite to the right of gate $g^i_{-}$. 
Since the paths  $  \downarrow \kern -4px {\cal F}^{i}_{-l}$ and  $  \downarrow \kern -4px {\cal F}^{i}_{-r}$ are fence crossing paths they induce ratchet motion and therefore they do not necessarily have to cross the gate. The paths $  \downarrow \kern -4px {\cal F}^{i}_{-l}$ and  $  \downarrow \kern -4px {\cal F}^{i}_{-r}$ are topologically protected by the path $  \downarrow \kern -4px g^i$ through the neighboring gate.

 We complete the fundamental loop with a north traveling path of type 
 $  \uparrow \kern -4px g^i_-$,  $  \uparrow \kern -4px {\cal F}^{il}_-$ or $  \uparrow \kern -4px {\cal F}^{ir}_-$.
A path of type $  \uparrow \kern -4px {\cal F}^{il}_-$ is a north traveling path that
passes left of the two bifurcation points. It enters the satellite left of gate $g^i_{-}$ and exits it via the upper  right fence segment attached to the gate segment $g^i_{-}$ (for examples see Fig. \ref{paths}b). 

\begin{figure}[t]
	\includegraphics[width=\columnwidth]{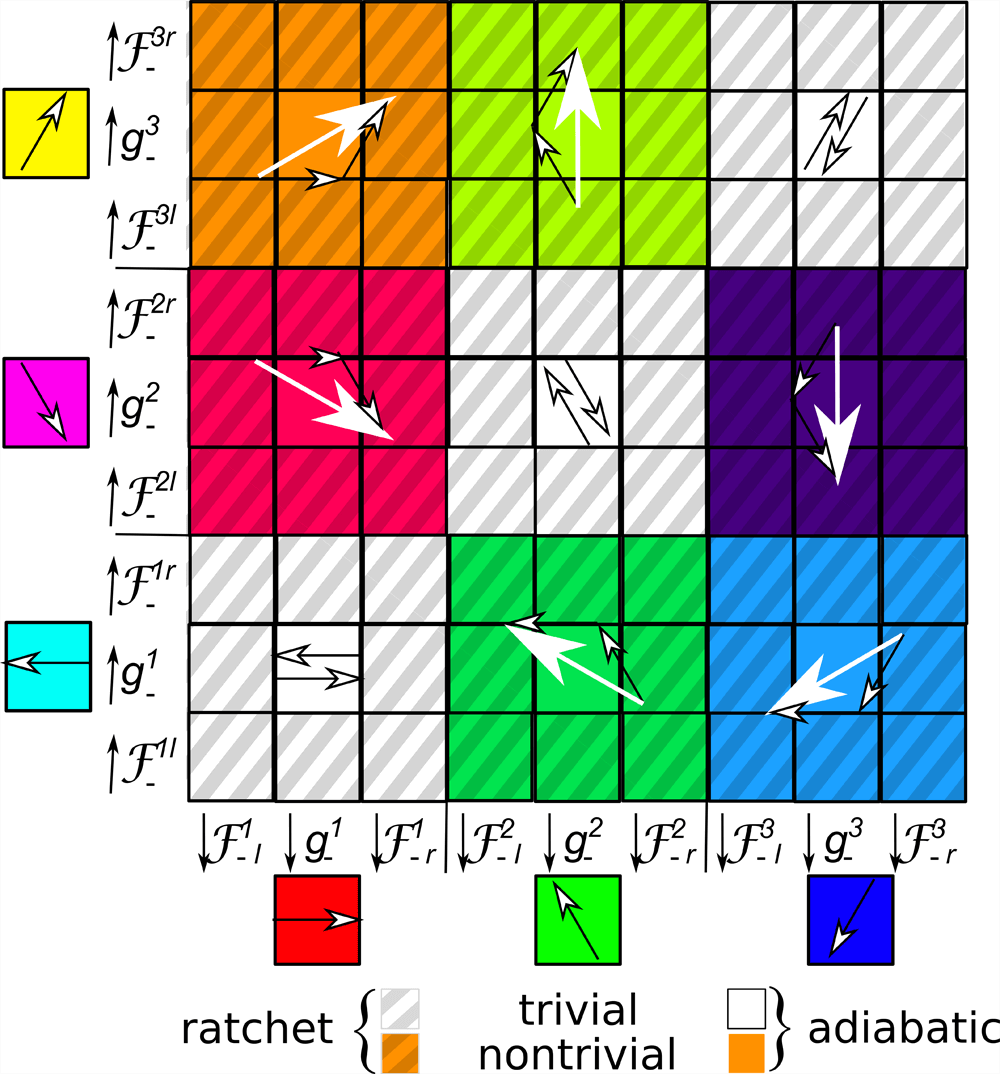}
	\caption{Phase diagram of the transport of paramagnetic colloids for $S_6$-like case. All paths (small arrows) occur on the $31$-network. \changes {The terminology of the paths is explained in section \ref{theoryS6}}}
	\label{figphasediagramS6}
\end{figure}

In Fig. \ref{figphasediagramS6} we depict the phase diagram of the transport induced by the fundamental loops ${\cal L}_{\cal C}={  \downarrow \kern -4px s}  {\uparrow \kern -4px s'}$  for the $S_6$-like case. Loops for which both paths are of type $g$ are adiabatic, while loops containing at least one path of type $\cal F$ are ratchets. Note that the transport direction is independent of how we enter an $m=6$ satellite region. We therefore do not specify the point of entry in the phase diagram. The entry determines whether a ratchet loop is a time reversal or non time reversal loop. If the entry and the exit are attached to a different gate segment the modulation loop is predicted to cause a non-time reversal ratchet. In contrast,
loops where paths enter and exit the satellites through the fence segments attached to the same gate are time reversal ratchet loops. 

\begin{figure}[t]
	\includegraphics[width=\columnwidth]{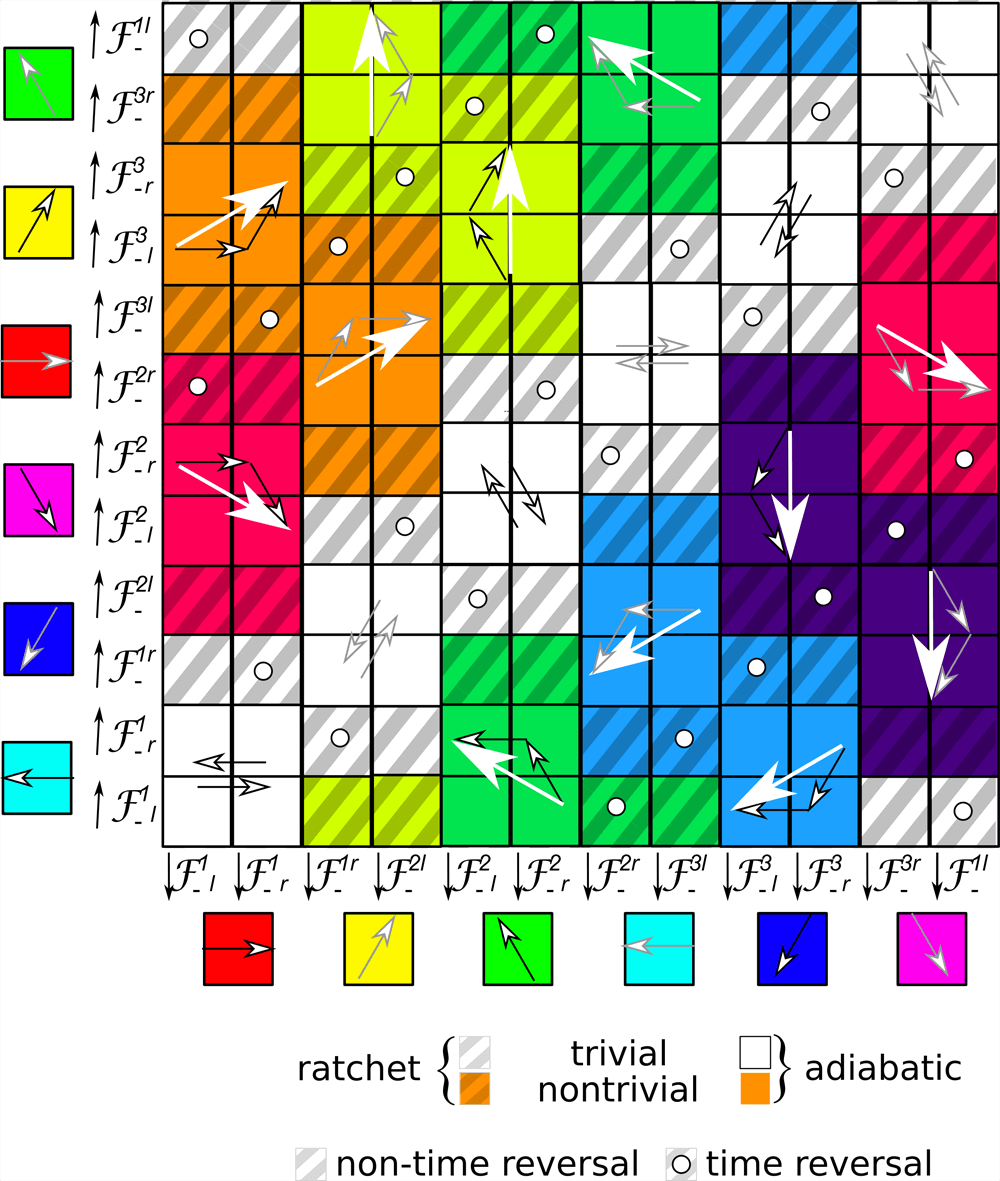}
	\caption{Phase diagram of the transport for $C_6$-like case. Ratchets are topologically protected by the adiabatic loop sharing the same south traveling path. Paths occur on the $31$-network (black) or on the $12$-network (gray arrows). The choice of network depends on the south heading path  $  \downarrow \kern -4px {\cal F}^{i}_{-}$. \changes {The terminology of the paths is explained in sections \ref{theoryS6} and \ref{theoryC6}}}
	\label{figphasediagramC3}
\end{figure}

\subsection{Modulation loops in the $C_6$-like case}\label{theoryC6}
The $C_6$-like case is easier than the $S_6$-like case. There is one single southern fence. Non trivial transport of paramagnetic particles occurs for modulation loops that cross the southern fence. 
Fundamental loops ${\cal L}_{\cal C}={  \downarrow \kern -4px {s}}  {\uparrow \kern -4px {s}'}$ can be characterized by the south traveling path
$  {\downarrow \kern -4px {s}}$ through fence segment ${s}$ and the path $ { \uparrow \kern -4px {s}'}$ traveling north through fence segment ${s}'$. We abbreviate the fence segments for the $C_6$-like case with the names of the segments for the $S_6$-like case from which they developed. 
The type of transport as well as the direction can also be explained by the bifurcation points the modulation loop encloses. The exact way the gates are crossed is still important. The gates, however, lie in such a way that crossing a fence segment dictates which gate the loop must pass. Hence, the fence segments passed by the loop fully determine the transport direction. 
Fig. 	\ref{figphasediagramC3} depicts the phase diagram of the transport directions of the $C_6$-like case. It is a checker board of adiabatic and ratchet loops.
Despite the topological transition the clustering of colors and therefore directions 
is quite similar to the phase diagram of the $S_6$-like case (Fig. \ref{figphasediagramS6}).
 Note that in contrast to the $S_6$ situation we use the same fence segments for both directions of the modulation loops.
 
Due to the symmetry of the universal potential $U^*$ diamagnetic transport can be achieved in the same way by simply reversing the field $\mathbf{H_{ext}}\rightarrow \mathbf{-H_{ext}}$. In contrast to the four-fold case the transport in all three-fold cases is more versatile. Paramagnetic and diamagnetic colloids are no longer fixed to the same transport direction but can be transported fully independently, because ${\cal F}_-$ and ${\cal F}_+$ are well separated in $\cal C$.
\begin{figure}
	\includegraphics[width=1\columnwidth]{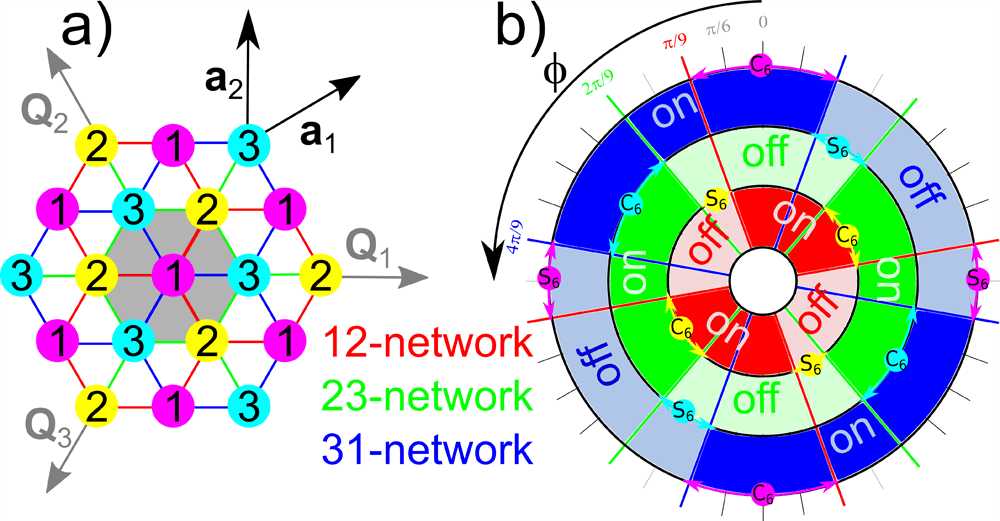}
	\caption{a) Threefold unit cell with the three possible symmetry points $\mathbf{x}_1$ (purple), $\mathbf{x}_2$ (yellow) and $\mathbf{x}_3$ (cyan). There are three networks along which transport is possible, the 12-network (red lines), the 31-network (blue) and the 23-network (green). 
		b) State of each network as a function of the phase of the pattern. Activated (on) networks have full colors while deactivated (off)  networks have light colors. Phases $\phi$, where one of the symmetry points acquire higher $S_6$ or $C_6$ symmetry are marked by circles of the color of the high symmetry point. Topological transitions between $S_6$ and $C_6$ symmetries are also marked with colored thick lines. The state of a network can only change at the topological transition. \changes{See appendix \ref{definitions} for definitions and terminology.}}
	\label{fignetwork}
\end{figure}

\subsection{Three and six fold symmetry}
Let us reconsider the symmetry of the three-fold lattice. As we have seen there are three
points $\mathbf{ x}_{\cal A}^1={\mathbf 0}$, 
${\mathbf x}_{\cal A}^2=({\mathbf a}_1+{\mathbf a}_2)/3$ and  ${\mathbf x}_{\cal A}^3=-({\mathbf a}_1+{\mathbf a}_2)/3$ in the unit cell of $\cal A$ with three-fold symmetry (see Fig. \ref{figpattern}c). As we vary $\phi$ one of these points acquires a higher $C_6$ symmetry at $\phi=n\pi/3$, with $n=1,2,3,..$. The higher symmetry permutes amongst the three points. Similarly, one of the points acquires a $S_6$ symmetry for $\phi=\pi/6
+n\pi/3$. 
 Connections between two different points ${\mathbf x}_{\cal A}^i$ and ${\mathbf x}_{\cal A}^j$ define a $ij$-network that might enable transport between two unit cells. There are three possible networks: the $12$-network, the $23$-network, and the $31$-network (see Fig. \ref{fignetwork}a).

For a polar orientation of the external field at least one of the three points is a minimum and at least one is a maximum. At the  $S_6$ symmetry point the potential has a monkey saddle for a polar external field orientation and a normal saddle point otherwise. In any case the $S_6$ point lies in the forbidden region. Hence the $S_6$ symmetry shuts off all connections to the point with $S_6$ symmetry. Only the network between the remaining two symmetry points can be used for transport via appropriate modulation loops. In contrast when the pattern acquires $C_6$ symmetry the point with $C_6$ symmetry is connected to both other symmetry points via two networks. The network between the lower $C_3$ symmetry points is shut off.

As we vary $\phi$ from $0$ to $2\pi$ each network is switched on and off twice. For any $\phi$ at least one network is on and at least one network is off. The exact number of active networks depends on whether $\phi$ is in the neighborhood of a $C_6$ or a $S_6$ symmetry. 
In Fig. \ref{fignetwork}b we plot the symmetry of the three points and the state of the three networks as a function of $\phi$. Note the close relationship to an antiferromagnetic equilibrium Ising system in a triangular lattice \cite{Ramirez}. Both systems are geometrically frustrated, with not all possible connections between sites being turned on. 

\subsection{Experiments on the $S_6$-like symmetry}
Three fold symmetric patterns with lattice constant $a=7\,\mu\textrm{m}$ have been created in the same way as the four-fold patterns. Here again lithographic edge effects of the patterning process render white regions larger than the black regions such that the average magnetization of the film is non-zero. This breaks the $S_6$-symmetry and shifts the phase $\phi<\phi_{mask}$ of the patterns away from the phase $\phi_{mask}$ of the lithographic mask toward the $C_6$-like symmetric direction. 

\begin{figure}
	\includegraphics[width=\columnwidth]{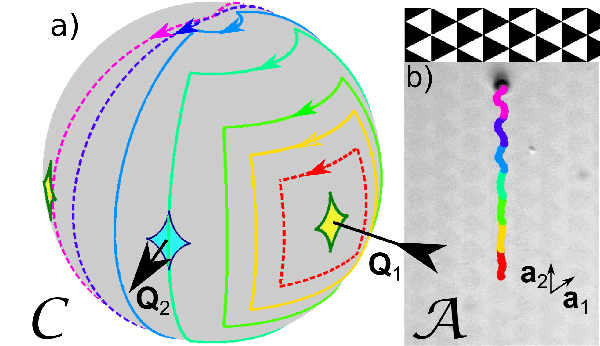}
	\caption{a) Different modulation loops in $\cal C$ encircling the satellite  around ${-\mathbf Q}_1$. The loops fall into the three classes  ${\cal L}_{\cal C} =  \downarrow \kern -4px {\cal F}^2_{-r} { \uparrow \kern -4px {\cal F}^{3l}_{-}}$ (dashed red),  ${\cal L}_{\cal C} =  {\downarrow \kern -4px {g}^2} { \uparrow \kern -4px {g}^3}$ (solid yellow, green, light green, and blue) and ${\cal L}_{\cal C} =  {\downarrow \kern -4px {\cal F}^2_{-l}} { \uparrow \kern -4px {\cal F}^{3r}_-}$ (dashed purple and magenta), where dashed lines are indicating modulation loops that induce ratchets. b) Corresponding experimental trajectories of a paramagnetic colloidal particle on top of the $S_6$-pattern. The (dashed) ratchet loops fall into the same homotopy class as the (solid) adiabatic loops and therefore the travel direction (along ${\mathbf a}_2$ ) is topologically protected.  Passing blue fences is irrelevant for the motion of the paramagnets.
		Note that some of the experimentally observed ratchet loops do not pass through the theoretical green fences of control space. 
		The background in b) is the reflection microscopy image of the underlaying lithographic magnetic  pattern. 
		A video clip of the motion of the paramagnetic colloidal particle is provided in \cite{CLIPS}.} 
	\label{figS6caroloop}
\end{figure}
\begin{figure}
	\includegraphics[width=\columnwidth]{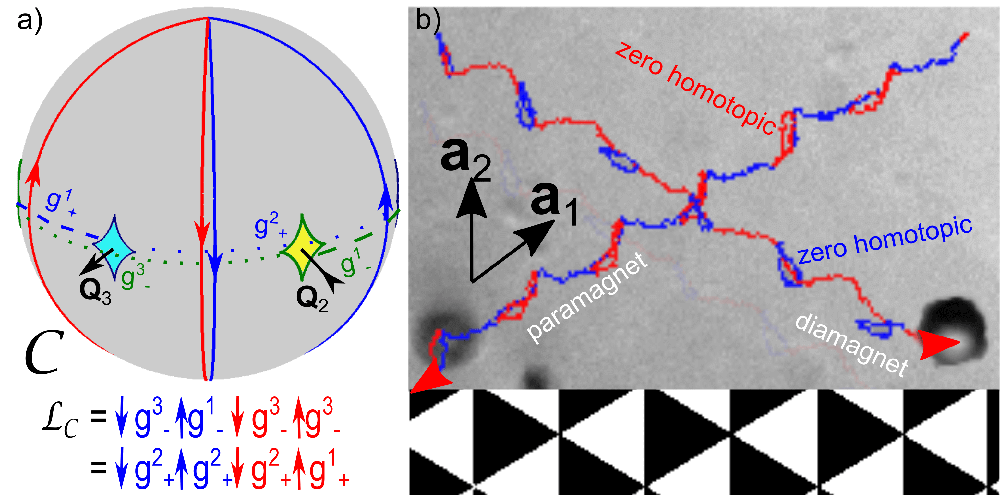}
	\caption{a) Control space $\cal C$ with the applied modulation double loop ${\cal L_C}=
		{  \downarrow \kern -4px {g}^3_-}
		{\uparrow \kern -4px {g}^1_-}
		{  \downarrow \kern -4px {g}^3_-}  {\uparrow \kern -4px {g}^3_-}
		= 
		{  \downarrow \kern -4px {g}^2_+}  
		{\uparrow \kern -4px {g}^2_+}
		{  \downarrow \kern -4px {g}^2_+}  {\uparrow \kern -4px {g}^1_+}
		$ 
		consisting of  two joint fundamental  modulation loops.  b)  Experimental trajectories of a paramagnetic and a diamagnetic colloidal particle in action space $\cal A$ caused by this loop. The result is the transport of paramagnetic and diamagnetic particles in directions differing by an angle of $2\pi/3$. While the first (blue) fundamental loop  transports the paramagnetic particles it is zero homotopic for the diamagnetic particles and vice versa for the second (red) loop. 
		The background is the reflection microscopy image of the lithographic magnetic  pattern. A video clip of the motion of the paramagnetic colloidal particle is provided in \cite{CLIPS}.} 
	\label{figS6independent}
\end{figure}

To show the topological protection of the transport directions in the $S_6$-like case we apply different fundamental modulation loops that all fall in the classes ${\cal L}_{\cal C} = {\downarrow \kern -4px {\cal F}^2_{-r}} { \uparrow \kern -4px {\cal F}^{3l}_{-}} ,  {\downarrow \kern -4px {g}^2} { \uparrow \kern -4px {g}^3}$, or $ {\downarrow \kern -4px {\cal F}^2_{-l}} { \uparrow \kern -4px {\cal F}^{3r}_-}$, 
but have different proximity to the satellite centered at ${-\mathbf Q}_1$ in $\cal C$. In Fig. \ref{figS6caroloop} we plot the corresponding trajectories of paramagnetic particles on a $S_6$-like pattern. All loops induce transport in the $\mathbf{a}_2$ direction, which is 
 in accordance with the predictions of section \ref{theoryS6}. It does not matter which particular modulation loop within the same homotopy class we choose, the global result after completing the loop is the transport of the paramagnetic particle by one unit vector ${\mathbf a}_2$.  Modulation loops closer to the encircled satellite have a straighter trajectory than loops passing the equator far from it (see Fig. \ref{figS6caroloop}). For small as well as for large modulation loops passing the equator close to one of the southern (green) satellites, we observe the transition from adiabatic toward ratchet motion (dashed modulation loops in Fig. \ref{figS6caroloop}a). Therefore,
 ratchet loops are observed in a larger region  than expected from the theoretically predicted positions of the  ${\cal B}_-$ bifurcation points and the fences of the satellites. However their occurrence is topologically equivalent to the theoretical model. Note that passing the blue fences is irrelevant for the motion of paramagnetic particles.  The difference between the adiabatic and ratchet motion will be 
 shown in detail in section \ref{expC6}. 

In a second step we immersed the paramagnetic particles into a ferrofluid on top of the pattern and added effectively diamagnetic particles. We subjected both types of particles to a double loop 
${\cal L_C}={\cal L_C}^1{\cal L_C}^2
$ 
consisting of two fundamental loops  
${\cal L_C}^1=
{  \downarrow \kern -4px {g}^3_-}
{\uparrow \kern -4px {g}^1_-}
=
{  \downarrow \kern -4px {g}^2_+}  
{\uparrow \kern -4px {g}^2_+}
$, and ${\cal L_C}^2=
{  \downarrow \kern -4px {g}^3_-}  
{\uparrow \kern -4px {g}^3_-}
=
{  \downarrow \kern -4px {g}^2_+}  
{\uparrow \kern -4px {g}^1_+}
$ (Fig. \ref{figS6independent}a).  The first loop ${\cal L_C}^1$ (blue) transports the paramagnetic particles by the unit vector $-{\mathbf a}_1$. ${\cal L_C}^1$ is zero homotopic for the diamagnets since it is only crossing the same minimum segment $g^2_+$ twice. The second fundamental loop ${\cal L_C}^2$ (red) is zero homotopic for the paramagnets and transports the diamagnets in the different ${\mathbf a}_1-{\mathbf a}_2$ direction. The resulting trajectories of paramagnetic and diamagnetic particles to the double loop ${\cal L_C}$ are shown in Fig. \ref{figS6independent}b. The double loop ${\cal L_C}$ is an example of a combination of two modulation loops that induces transport of paramagnetic and diamagnetic particles in two independent arbitrary directions on top of a $S_6$-like pattern.
\begin{figure}[t]
	\includegraphics[width=\columnwidth]{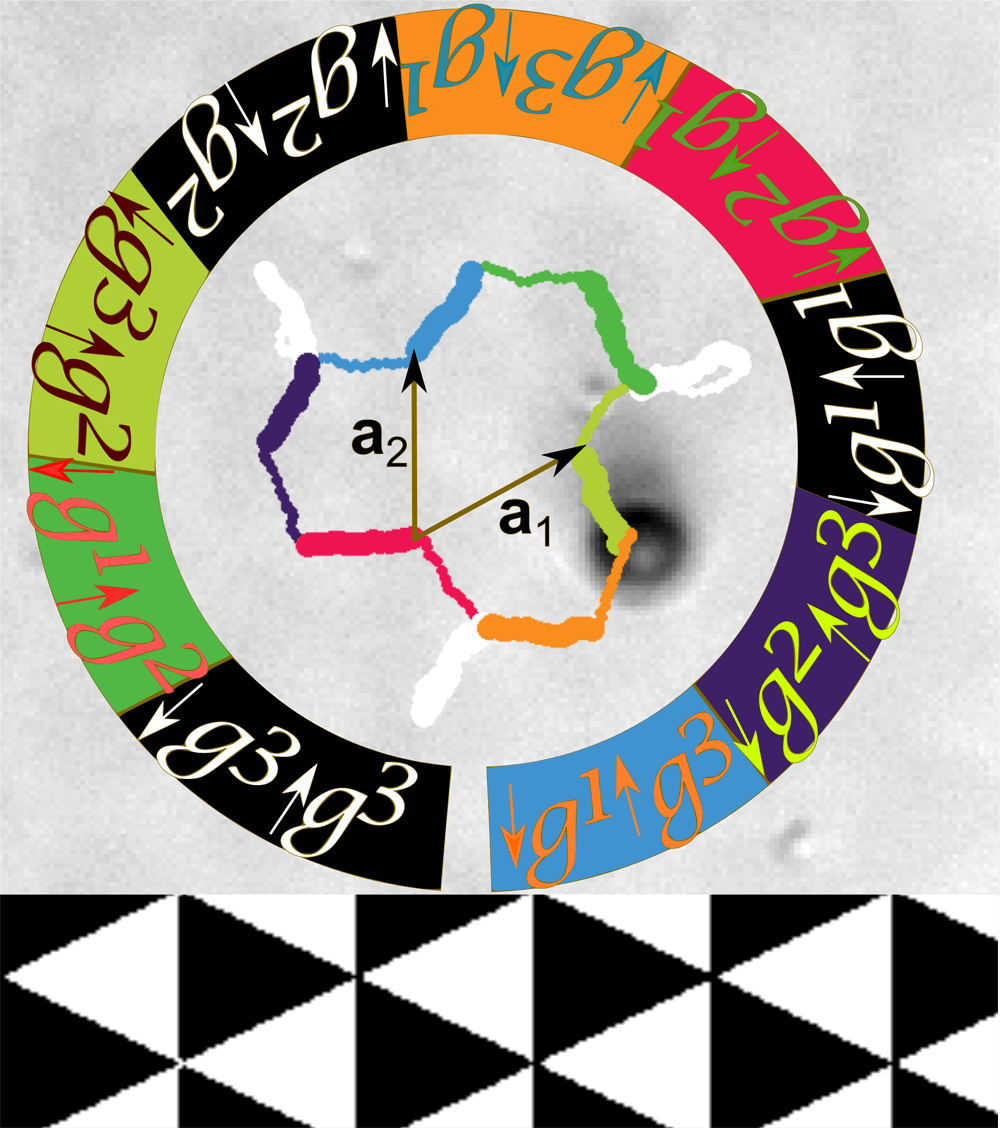}
	\caption{Experimental trajectory of a paramagnetic colloidal particle  on top of a $S_6$ pattern caused by a modulation poly-loop in $\cal C$ consisting of a sequence of all fundamental modulation loops. 
		The fundamental loops are colored according to the loops in the phase diagram in Fig.
		\ref{figphasediagramS6}. South traveling segments are marked as thick lines. North traveling segments are marked as thin lines.
		A video clip of the motion of the paramagnetic colloidal particle is provided in \cite{CLIPS}} 
	\label{figsuperloopS6}
\end{figure}

\begin{figure}[t]
	\includegraphics[width=\columnwidth]{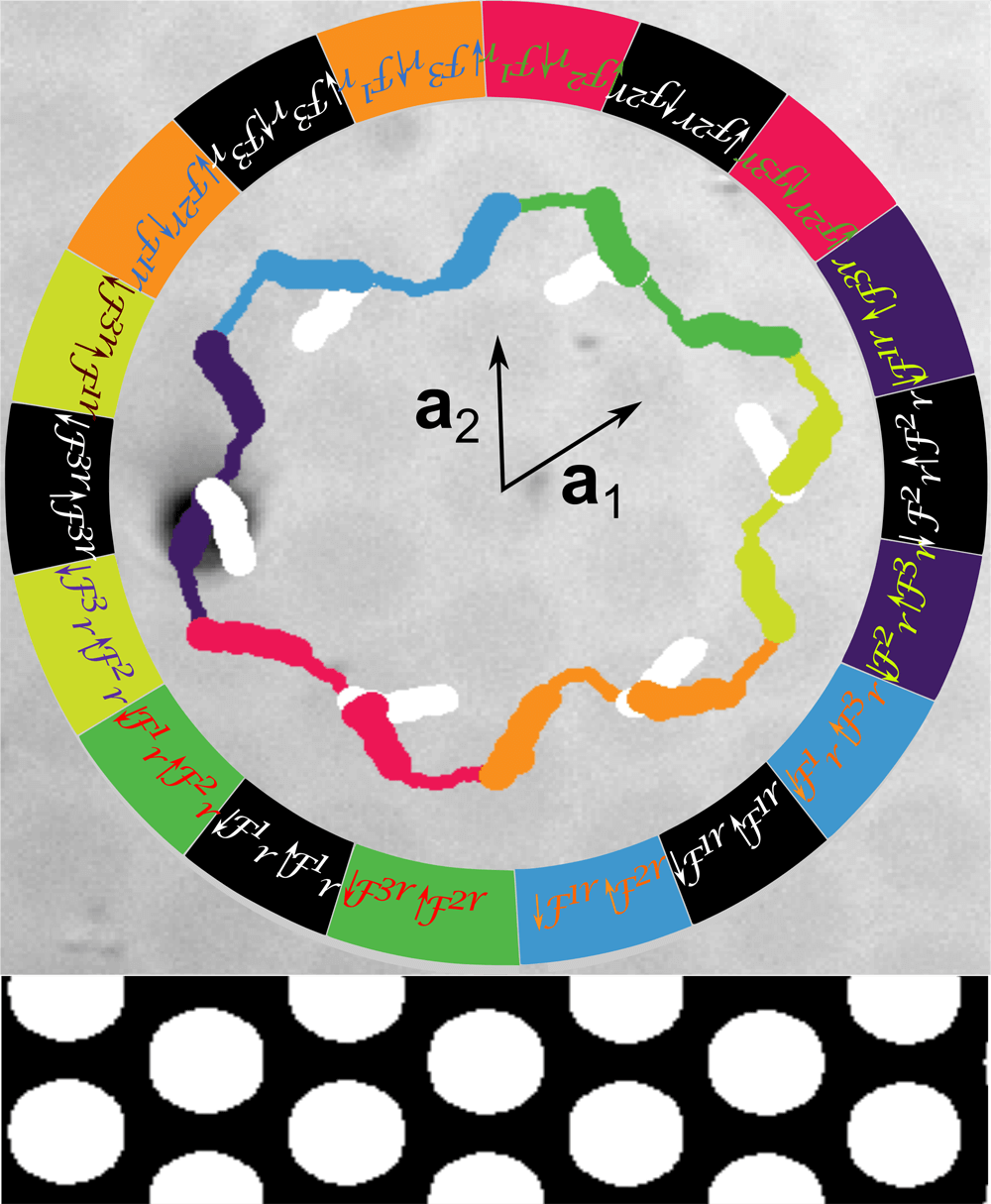}
	\caption{Experimental trajectory of a paramagnetic colloidal particle on top of a $C_6$-pattern. The colloidal particle is subjected to a modulation poly-loop in $\cal C$ which is a combination of all adiabatic right fence segment crossing fundamental modulation loops. The single fundamental loops are colored according to the loops in the phase diagram in Fig.
		\ref{figphasediagramC3}. South traveling segments are again marked as thick lines while north traveling segments are thin lines.
		Similar to the theory the circular bubble domains have positive magnetization. However the reflection microscopy image in the background has an inverted contrast such that the bubbles are dark.
		A video clip of the motion of the paramagnetic colloidal particle is provided in \cite{CLIPS}.} 
	\label{figsuperloopC6}
\end{figure}
\begin{figure}[t]
	\includegraphics[width=\columnwidth]{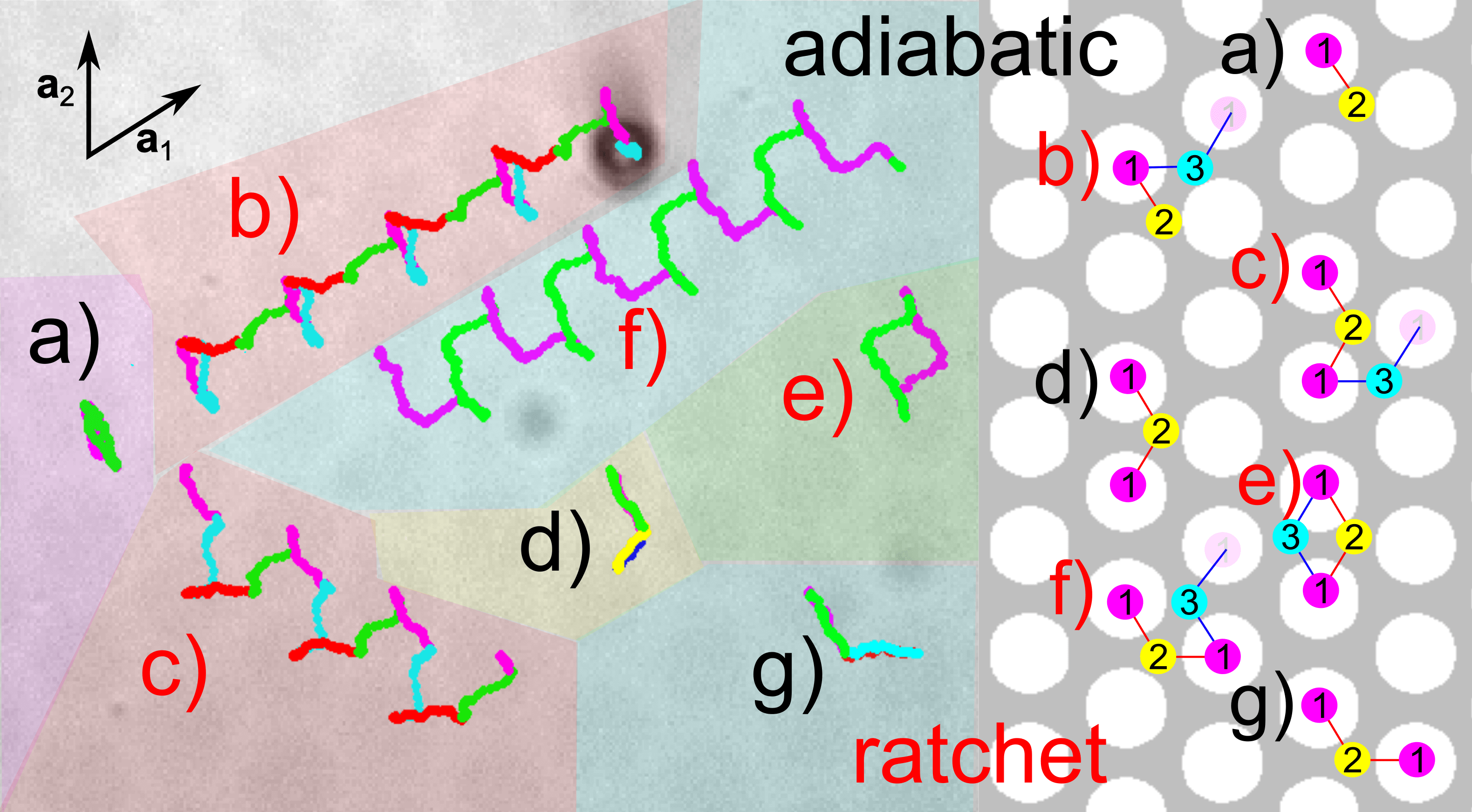}	
	\caption{Experimental trajectories of paramagnetic colloidal particles in action space $\cal A$ above a $C_6$-symmetric pattern. The trajectories are caused by various zero homotopic palindrome modulation loops ${\cal L_C}={  \downarrow \kern -4px {{\cal F}^{3r}}}  {\uparrow \kern -4px {s}}
		{  \downarrow \kern -4px {s}}  {\uparrow \kern -4px {{\cal F}^{3r}}}$
		with a) $s={\cal F}^{1l}$ b) $s={\cal F}^{1}_{l}$, c) $s={\cal F}^{1}_{r}$, d) $s={\cal F}^{2l}$, e) $s={\cal F}^{2}_{l}$, f) $s={\cal F}^{2}_{r}$, and g) $s={\cal F}^{2r}$. The paths in $\cal A$ are colored according to the four paths of the modulation loop as indicated by the squares in the phase diagram Fig. \ref{figphasediagramC3}. In the cases a), d) and g) the motion is adiabatic and the colloidal path in $\cal A$ consists of two forward paths that coincide with the backward path. The case e) corresponds to a time reversible ratchet with a zero homotopic path in $\cal A$. However  the colloid is moving on different forward and backward paths that belong to two different networks indicated to the right. The other cases b),c), and f) are non time reversible ratchets where the zero homotopic modulation loops in $\cal C$ induce non-zero homotopic (open) paths of the colloids in $\cal A$. The predicted paths between the high symmetry points for all loops are shown to the right.  
		Video clips of the motion of the paramagnetic colloidal particle caused by the loops in b), d) and e), are provided in \cite{CLIPS}.
	}
	\label{figadiabaticratchet}
\end{figure}

The experimental trajectories not only are in accordance with the theory for the previous loops, but for all possible fundamental loops.
To experimentally show this we applied a poly-loop   
for paramagnetic particles 
that combines all the fundamental loops of the phase diagram of Fig. \ref{figphasediagramS6}. In Fig. \ref{figsuperloopS6} we plot the experimental trajectory of paramagnetic particles with the fundamental sections colored with the color of the corresponding theoretical fundamental loop of Fig.  \ref{figphasediagramS6}.  All fundamental loops transport into the theoretically predicted directions. 
In conclusion the experimental response of the particles on a $S_6$-like pattern to all shown modulation loops is in topological agreement with the theoretical predictions. The only phenomenon that we could not observe in our experiments is a non time reversal ratchet. The reasons for this are discussed in section \ref{discussion}.

\subsection{Experiments on the $C_6$-like symmetry}\label{expC6}

The experimental trajectories of the adiabatic modulation loops of the $C_6$-like case are also in accordance with the theory. Fig. \ref{figsuperloopC6} shows the trajectory of a paramagnetic particle subject to an adiabatic poly-loop
that consists of all different adiabatic right fence segment crossing fundamental loops of the phase diagram in Fig. \ref{figphasediagramC3} combined. We plot the trajectories of the particles in the color of the corresponding fundamental loops of the phase diagram. All adiabatic loops transport into the directions predicted by the theory.

In contrast to the universal two-fold and four-fold symmetric patterns the three and sixfold symmetric patterns not only support adiabatic motion but also ratchet type motion can be observed.
To visualize the characteristics of the different types of motion we use palindrome modulation loops ${\cal L_C}={\tilde{\cal{L}}_{\cal{C}}}{\tilde{\cal{L}}_{\cal{C}}}^{-1}={  \downarrow \kern -4px {{{\cal F}^{3r}}} } {\uparrow \kern -4px {s}}
{  \downarrow \kern -4px {s}}  {\uparrow \kern -4px {{{\cal F}^{3r}}}}={\cal L_C}^{-1}$ .
They consist of a loop ${\tilde{\cal{L}}_{\cal{C}}}={  \downarrow \kern -4px {{\cal F}^{3r}}}  {\uparrow \kern -4px {s}}$ that is first played in the forward direction and afterwards played again but this time reversed, i. e., in the backward direction. 
While the first path ${  \downarrow \kern -4px {{{\cal F}^{3r}}} }$ of  ${\tilde{\cal{L}}_{\cal{C}}}$ is kept the same, the second path ${\uparrow \kern -4px {s}}$ varies along the eleventh column of the phase diagram (Fig. \ref{figphasediagramC3}). We start with a) $s={\cal F}^{1l}$ which makes ${\cal L_C}$ an adiabatic zero homotopic loop and then trace the transition towards adiabatic transport d) ($s={\cal F}^{2l}$) via two different non time reversal ratchets b) ($s={\cal F}_l^1$) and c) ($s={\cal F}_r^1$). Afterwards we show the crossover toward another adiabatic transport direction g) ($s={\cal F}^{2r}$), this time by passing a time reversal ratchet e) ($s={\cal F}_l^2$) and another non time reversal ratchet f) ($s={\cal F}_r^2$). Trajectories of these motions are shown in Fig. \ref{figadiabaticratchet}.
 
 Obviously, if the induced motion is adiabatic the colloidal particle is tracing some path in $\cal A$ during the forward motion, and then returns to the initial position by tracing the exact same path in the backward direction. Three such adiabatic  paths (a,d and g) are shown in Fig.  \ref{figadiabaticratchet}. All adiabatic paths are caused by modulation loops making use of only upper type fence crossings and cause motion on the $12$-network only.
In contrast the irreversible nature of ratchet jumps causes the colloidal particles to move on a different path in $\cal A$ during the forward and backward modulation loop. The reason for this is that the forward loop ${\tilde{\cal{L}}_{\cal{C}}}$ uses a south traveling path crossing an upper type fence ${\cal F}^{3r}$ and a north traveling path crossing a lower type fence. When ${\tilde{\cal{L}}_{\cal{C}}}$ is played forward the colloid travels the first half adiabatically from ${\mathbf x}_{{\cal A},1}$ toward ${\mathbf x}_{{\cal A},2}$ since the modulation path enters the southern excess region and upper type fence crossings support motion on the $12$-network. The second half of ${\tilde{\cal{L}}_{\cal{C}}}$  must bring the particle back to ${\mathbf x}_{{\cal A},1}$. However, adiabatic motion with lower type fence 
crossing paths is possible only on the $31$-network and our particle is currently at ${\mathbf x}_{{\cal A},2}$ that is not part of this network. Hence the particle performs a ratchet jump back toward ${\mathbf x}_{{\cal A},1}$. When ${\tilde{\cal{L}}_{\cal{C}}}$ is played backward the particle adiabatically moves from ${\mathbf x}_{{\cal A},1}$ toward ${\mathbf x}_{{\cal A},3}$ and jumps back via a ratchet jump. The full palindrome loop hence visits the high symmetry points in the sequence:  ${\mathbf x}_{{\cal A},1}$,${\mathbf x}_{{\cal A},2}$,${\mathbf x}_{{\cal A},1}$,${\mathbf x}_{{\cal A},3}$,${\mathbf x}_{{\cal A},1}$.
For time reversal ratchets the colloidal particle returns to its initial position after the full modulation loop $\cal{L_C}$, however by using a backward path in $\cal A$ different from the forward path. Such a time reversible ratchet path is shown in Fig.  \ref{figadiabaticratchet}e. In general palindrome modulation loops cause non-time reversal ratchet motion. The particle does not return to its initial position after a complete modulation loop but is transported  by one unit vector. 
Three non-time reversible ratchet paths of this type are shown in Fig.  (\ref{figadiabaticratchet}b, c, and f). 

\begin{figure}[t]
	\includegraphics[width=\columnwidth]{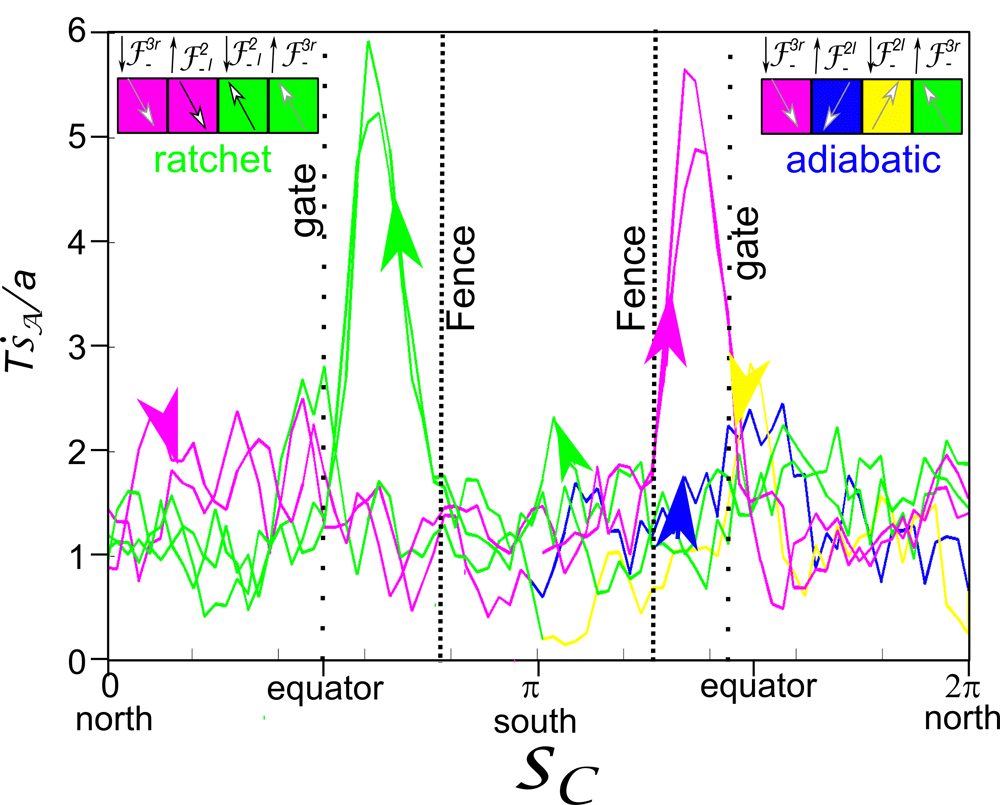}
	\caption{Speed $\dot{s}_{\cal A}$ in $\cal A$ of the colloidal motion induced by the palindrome modulation loop d) and e) of Fig. \ref{figadiabaticratchet}. The speed is normalized by the lattice constant $a$ and the period $T$ of one sub loop. It is plotted against the normalized path length $s_{\cal C}$ of the modulation in $\cal C$ ranging from $0$ to $2 \pi$ for the forward modulation and from $2 \pi$ to $0$ for the backward modulation.
	Ratchet jumps in the ratchet modulation loop (maximum speed) occur in the second half of the forward (magenta maximum) and backward path (green maximum) when the modulation loop leaves the southern excess region in $\cal C$ via the fence. 
	Also the adiabatic speed profile (magenta/blue/yellow/green modulation, d in Fig. \ref{figadiabaticratchet}) exhibits maxima when the modulation crosses the gates. But they are clearly smaller then the maxima of the ratchet jumps.
	} 
	\label{figvelocityadiabaticratchet}
\end{figure}

The characteristics of the adiabatic and ratchet motion can also be inferred without looking at the differences between the forward and backward paths in $\cal A$. We measure the speed $\dot{ s}_{\cal A}$ of the colloids in $\cal A$ versus the normalized path length $s_{\cal C}$ of the  modulation loop. We parametrize the forward modulation loop ${\tilde{\cal{L}}_{\cal{C}}}$ from $0 $ to $2\pi$ and the backward loop ${\tilde{\cal{L}}_{\cal{C}}}^{-1}$ from $2\pi$ to $0$ such that the path length $s_{\cal C}$ in Fig. \ref{figvelocityadiabaticratchet} runs back and forth between $0$ and $2\pi$.  Ratchet loops can be distinguished from adiabatic loops by the ratchet jumps that have a significantly higher speed than the adiabatic motion. These jumps occur during the second half (magenta) of the forward and the second half (green) of the backward modulation when the modulation hits the fences and leaves the southern excess region in $\cal C$. There are also smaller maxima in the speed of the adiabatic motion when the beads pass the gates. The increased gate speed is a result of the way that curves which are passing the gates in $\cal M$ are projected into $\cal A$ and $\cal C$. The projections  are causing a maximum in the conversion of the speed in action space versus the speed in control space at the gate. In our special case the gates seem to be located less polar than the fences, which contradicts the theoretical predictions for the $C_6$-symmetric case but is in accordance with theoretical predictions for weakly broken $C_6$-symmetry.

We are hence able to independently characterize the type of motion and the particular path taken by the colloids. Both the experimentally determined types of motion as well as the directions are in perfect agreement with the theoretically predicted phase diagram (Fig. \ref{figphasediagramC3}) for the $C_6$-like case.

For the $S_6$-like case we also observe adiabatic and ratchet motion in topological agreement to the theory. However, we did not succeed in finding palindrome loops causing non-time reversible ratchets as predicted by the theory and simulations. Instead, we observe that loops, which are supposed to induce non time reversal ratchets, cause the coexistence of time reversible ratchets with different directions above different unit cells. The directions thereby correspond to either the theoretically predicted forward or backward direction.

\subsection{Experiments on the $S_6$-$C_6$-topological transition}

\begin{figure}
	\includegraphics[width=\columnwidth]{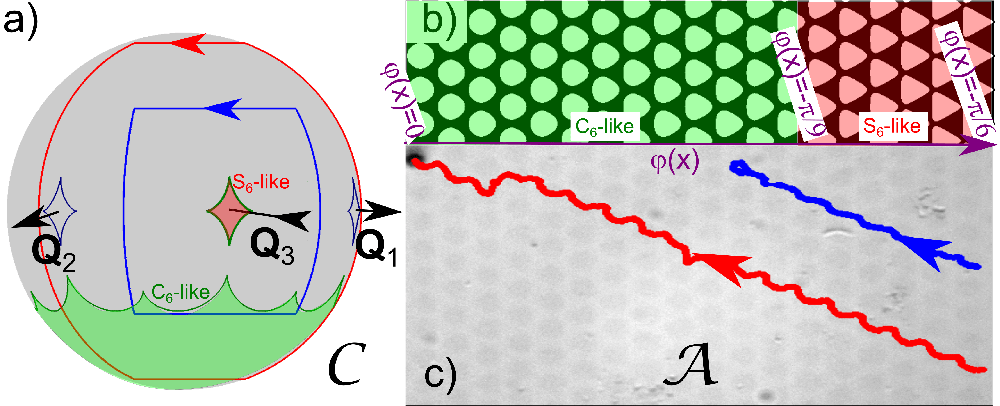}
	\caption{Motion of colloids in a phase gradient pattern. a) Control space with two  modulation loops (blue and red) circulating around ${-\mathbf Q}_3$. We have plotted the relevant excess satellite regions of the $S_6$-symmetric case (red area) and the excess region of the $C_6$-symmetric case (green area). b) Scheme of the slowly varying phase pattern. The pattern is $C_6$-like to the left and $S_6$-like to the right. 
    c) Experimental trajectories of paramagnetic particles induced by the two modulation loops.
	 Both loops encircle the $S_6$-symmetric satellite excess region and thus induce transport on the $S_6$-like pattern. The blue modulation loop barely touches the $C_6$-like fence in $\cal C$ which destroys the motion of the corresponding particle when it reaches $C_6$-like territory. 
The red loop in contrast  fully enters the southern $C_6$-symmetric excess region in control space and leaves it only once. Therefore the red trajectory persists well in to the $C_6$-like territory.
		A video clip of the motion of the paramagnetic colloidal particle is provided in \cite{CLIPS}.
		} 
	\label{figphasegradientmotion}
\end{figure}
To illustrate the $S_6$-$C_6$-topological transition we produced lithographic patterns with a slowly varying pattern phase $\phi(x)$. This continuously converts a $C_6$ pattern into a $S_6$ pattern within a spacial range of approximately 20 unit cells. In Fig. \ref{figphasegradientmotion} we show the motion of paramagnetic particles on such a phase gradient pattern induced by two different modulation loops (blue and red) encircling the ${-\mathbf Q}_3$ point. Both loops induce transport on the $S_6$-like pattern. However as the phase of the pattern declines towards zero (the phase of the $C_6$-pattern) the encircled satellite excess region of control space moves out of the blue loop such that the motion ceases beyond the critical phase $\phi<\phi_c=\pi/9$. The blue loop then touches the southern fence of the $C_6$-symmetric pattern, which is no longer sufficient  to induce transport on the $C_6$-like pattern. The red loop fully crosses the southern fences of the $C_6$-symmetric pattern. Therefore the motion of the particle persists as it enters $C_6$-like territory in action space $\cal A$. The direction of transport is thereby topologically protected over the transition 

\begin{figure}[t]
	\includegraphics[width=\columnwidth]{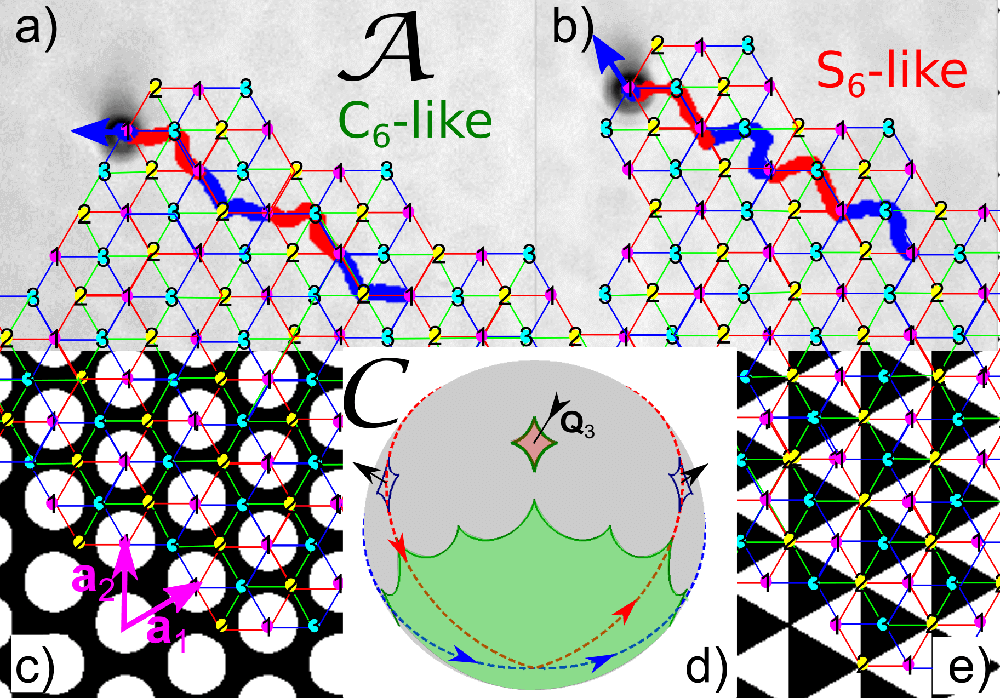}
	\caption{Motion of paramagnetic colloids on a) a $C_6$ symmetric pattern (scheme in c)) and b) a $S_6$ symmetric pattern (scheme in e)). The particles are subject to a modulation double loop ${\cal L}_{\cal C}={\cal L}_{\cal C}^{12}{\cal L}_{\cal C}^{31}$ with ${\cal L}_{\cal C}^{12}=
		{  \downarrow \kern -4px {\cal F}^{2}_{-r}}  {\uparrow \kern -4px {\cal F}^{1}_{-r}}$ (blue loop) and
		${\cal L}_{\cal C}^{31}={  \downarrow \kern -4px {\cal F}^{2r}_{-}}  {\uparrow \kern -4px {\cal F}^{3r}_{-}}$ (red loop) crossing different segments of the $C_6$-fence. d) Control space with the combined modulation loop consisting of two fundamental loops (blue and red) which are both encircling the $S_6$-symmetric excess region. We have plotted the relevant excess regions of the $S_6$-symmetric ($\phi=-\pi/6$) case (red area) and the $C_6$-symmetric case (green area). The induced motion on a $C_6$-symmetric pattern is shown in a). Both sub-loops induce motion on different networks resulting in  a trajectory that alternately uses the $12$ and the $31$ network. 
		 In the $S_6$-symmetric pattern only the  $31$-network is active. Therefore
		the induced motion shown in b) has to use the $31$-network during both sub-loops.
	}
	\label{fignetworkswap}
\end{figure}

Upon the transition between $S_6$ and $C_6$ also the state of networks  available for transport changes. While in the $C_6$-like pattern the $12$- and the $31$-networks are active the first one is switched off in a $S_6$-like pattern and only the $31$-network is available for transport (see Fig. \ref{fignetwork}). To experimentally demonstrate this  we apply a double loop of the type 
${\cal L}_{\cal C}={\cal L}_{\cal C}^{12}{\cal L}_{\cal C}^{31}$ with ${\cal L}_{\cal C}^{12}=
{  \downarrow \kern -4px {\cal F}^{2}_{-r}}  {\uparrow \kern -4px {\cal F}^{1}_{-r}}$  a fundamental loop passing through the lower fence segments (blue loop) and
${\cal L}_{\cal C}^{31}={  \downarrow \kern -4px {\cal F}^{2r}_{-}}  {\uparrow \kern -4px {\cal F}^{3r}_{-}}$ (red loop)
a fundamental loop passing through upper fence segments of 
the $C_6$-symmetric case as shown in Fig. \ref{fignetworkswap}d. For the $C_6$-like patterns the theory predicts an alternating use of the $12$-network and the 
$31$-network. The overall transport direction is the same for both fundamental loops. The same double loop converts into a ${\cal L}_{\cal C}=
{  \downarrow \kern -4px {g}^{2}_{-}}  {\uparrow \kern -4px {g}^{1}_{-}}
{  \downarrow \kern -4px {g}^{2}_{-}}  {\uparrow \kern -4px {g}^{1}_{-}}$ loop for the $S_6$-like case where transport is only possible on the $31$-network. In Fig. \ref{fignetworkswap} a) and b) we show the motion subject to this modulation loop on the $C_6$-like and the $S_6$-like patterns, respectively. Clearly the motion of the paramagnetic particle on the $C_6$-like pattern makes use of the $12$- and the $31$-network. We observe an alternating transport over these two networks. On the $S_6$-like pattern  transport happens via the $31$-network only. The motion  is again topologically protected in the direction, i.e. the modulation that before enforced the use the other network now also has to use the $31$-network into the same direction.

\section{Discussion}\label{discussion}

We have seen that most of the theoretically predicted features are experimentally robust. This ensures that colloids elevated only a few microns above the pattern behave pretty much the same way as predicted for universal potentials. The few deviations  of experiment and theory can mostly be attributed to non-universal proximity effects. These arise from larger reciprocal lattice vectors contributing to the colloidal potential. We have shown, however, that higher reciprocal lattice vectors change the position of certain transport direction transitions, but not the topology of the problem as long as their influence is not too strong. Experimental proofs for proximity effects have been shown at different elevations for the two-fold symmetric problem. These effects will of course also play a role on lattices of higher symmetry \changes{and for non-symmetric magnetic lattices where such symmetry is broken by higher reciprocal lattice vector contributions.} For the higher symmetric patterns we did not discuss these effects in detail and minimized them by performing experiments at sufficient elevation above the pattern. However, they are still visible in some experimental features. In the four-fold symmetric experiments for example the fence point is not a point but a finite area. Modulation loops must wind around this larger area instead of winding around the theoretical point and hence modulation loops can not be chosen arbitrarily small to cause adiabatic transport.  

\changes{The Bravais lattice of any periodic pattern has inversion and thus $C_2$ symmetry. Filling the unit cell of such a Bravais lattice with a magnetization pattern that has no net magnetic moment will generate a Fourier series that has contributions from Fourier coefficients at the non zero reciprocal lattice vectors. The contributions from the shortest reciprocal lattice vectors will always have one of the universal rotation symmetries. The symmetry can be broken by higher order reciprocal lattice vectors. The magnetic field contribution to a reciprocal lattice vector decays in the $z$-direction with the magnitude of the reciprocal lattice vectors, which is the reason why every transport at sufficient elevation of the order of the period will have exactly the characteristics of one of the patterns described in this paper. The transport remains topologically protected also for the symmetry broken case when the breaking of the symmetry is not too strong. There will be a topological transition to a non-transport regime for any type of pattern if one places the colloids close enough to the pattern. There might be other topological transport modes for symmetry broken patterns at intermediate elevation. These however are not universal as they will depend on all details of the pattern, field strength etc.}

\begin{figure}
	\includegraphics[width=1\columnwidth]{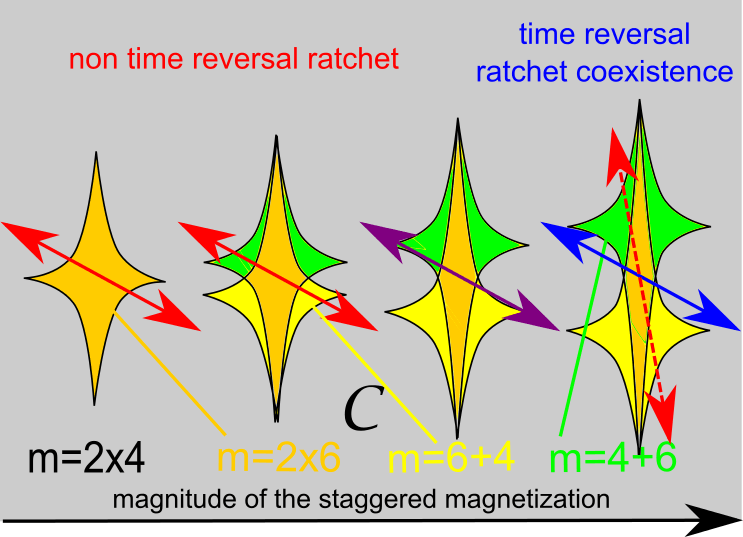}
	\caption{Splitting of a satellite when switching on a staggered magnetization that doubles the period, doubles the fence and doubles the multiplicity $m$. The net magnetization of the two new different half unit cells shifts one half of the satellite fence to the north and the other half to the south. As a result a path that initially passed the fences on opposite sides of the satellites (red arrow) now cuts the northern (southern) half of the double satellites on the  neighboring southern (northern) fence segments (blue arrow). 
	Instead of a non time reversal ratchet this produces time reversible ratchets with different directions in one and the other new half unit cell.}
	\label{noise}
\end{figure}

A difference between experiment and theory that cannot be explained with non universal proximity effects is the absence of non time reversible ratchets in the three-fold symmetric $S_6$-like case. There instead of non time reversible ratchets we observed the coexistence of time reversible ratchets of different direction above different unit cells of the pattern. We attribute those effects to the noise of the magnetic patterns. Presumably the net magnetization of each unit cell does not vanish as required by equation (\ref{zeromagnetization}), but acquires values that might differ from one unit cell to the next. A non vanishing magnetization acts like an additional external field in the z-direction and therefore shifts the satellites to the north or to the south. We may see the effect of magnetization noise for the simple example of an additional staggered magnetization alternating between positive and negative values in neighboring unit cells. The staggered magnetization doubles the unit cell and therefore also doubles the length of the fence. Each satellite becomes a double satellite around which the fence circles twice. When we increase the magnitude of the staggered magnetization one half of the double satellite moves north while the other half moves south (see Fig. \ref{noise}). 
Let us consider a modulation loop segment (red) that passes the unsplit double satellite on opposing segments. We expect this loop to induce a non time reversal ratchet. When the satellite splits the modulation loop segment will eventually pass the upper half of the double satellite south of the two ${\cal B}_-$ bifurcation points and the lower half north of the other two ${\cal B}_-$ bifurcation points. This, however, will now cause time reversible ratchets into different directions on one and the other half of the larger unit cell. This is exactly what we observe in the experiments, however, of course not in the simple staggered way predicted by our simplified period doubled theory.    

In Fig. \ref{figvelocityadiabaticratchet} we measured the speed of adiabatically moving colloidal particles. Gates in control space  can then be identified by the location of the maximum speed in $\cal A$. The experiments measured the positions of the extremum segments of the gates to lie in the $m=4$ region, while the theory of the $C_6$ symmetric case predicts that they lie inside of the excess area in $\cal C$. We have already mentioned that there might be a mismatch between the phase of the lithographic mask and the phase of the actual pattern. Indeed the lithographic writing process  presumably produces a magnetization pattern with a phase that differs from the desired phase. A phase that is slightly different from the $C_6$-symmetry would allow for gates in the tropics of ${\cal C}$ and hence explain the observed deviation.
 Since the exact path of the gate on $\cal M$ is a feature that is not topologically robust it is conceivable that either the phase shift or proximity effects might cause this discrepancy.

To achieve adiabatic transport our modulation loops in control space must be modulated at an angular frequency $\omega_{ext}$ that is significantly smaller than the intrinsic angular frequency $\omega_{int}\propto e^{-Qz} M$. For the lithographic magnetic patterns this restricts our modulation frequencies to $\omega_{ext}\approx 0.1$ Hz. For useful applications one would have to improve the saturation magnetization or the thickness of the lithographic patterns to increase the modulation frequency. The garnet films we used for the two-fold stripe pattern as well as for experiments on $C_6$-like patterns in \cite{Loehr} allowed for the use of up to two orders of magnitude higher modulation frequencies. The closer the particles are to the pattern the faster we might modulate the field, however, the less universal will be the behavior of the transport. An elevation of roughly half the lattice constant seems to be a good compromise that does not yet change the topology of the transport.   

We describe our ratchet as a deterministic ratchet, i.e. thermal diffusion of particles only happens during very short and therefore irrelevant times when the colloidal particles sit right on the fence. This short diffusion will not lead to a broadening distribution of transport directions as long as we avoid the ${\cal B}_0$ points. When using modulation loops passing close to a ${\cal B}_0$ point the particles may access the two alternative paths of steepest descend also in the surroundings of this point. Thermal effects broaden the fences. A transition to a thermal ratchet will occur for temperatures where the broadened fences overlap. Some of the topological properties might persist even then and thus also explain the omni-directional transport observed in such thermal ratchets \cite{Arzola}.

Comparing our system with topological crystalline insulators \cite{Fu,Hsieh,Dziawa,Slager,Liu,Miert} we
	note that the gates in our system are the analogues to the Dirac-cones in
	the quantum systems. Gates are lying on high symmetry points in the
	lattices with even $C_4$, and $C_6$ symmetries, while they lie on the
	$ij$-network for the three-fold symmetric lattices. The situation is
	comparable to the position of Dirac-cones lying on high symmetry points
	and lines in the first Brillouin zone of the lattices of different
	symmetry. As in topological crystalline insulators their number and
	robustness varies based on the symmetry of the lattice.

Comparing our driven system with Floquet topological quantum systems \cite{Kitagawa,Rudner} we note that
time dependent interactions of Floquet topological insulators usually must wind around the north-south axis to cause topologically non trivial behavior. This is because the unperturbed time independent Hamilton operator is diagonalized with respect to the z-component of the spin respectively pseudo spin operator. Different time dependent driving, such as THz-oscillating magnetic fields, stress modulation, or modest in plane electric field modulations \cite{Lindner} are experimental ways to achieve non trivial behavior.
Only perturbations that have non commuting contributions of non-diagonalized spin components will couple the different bands and cause non trivial dynamics.
Floquet topological insulators so far have been investigated mainly with
	respect to time reversal symmetry and particle hole symmetry protecting
	the topology. We are not aware of a crystalline Floquet
	topological insulator, which would be the quantum system in closest
	analogy to our system. Due to the lattice symmetry in our
colloidal system we have a variety of different axes around which
the perturbing external field may be wound. The reason for this is the multi-fold lattice symmetry that causes multiple stable points in the absence of a perturbing external field. In contrast to the quantum systems we have a richer variety of driving loops that can wind around alternative points of control space.

We should also mention that the dynamics of our colloidal system occurs in direct space not in reciprocal space. Direct space is an affine lattice having no natural origin. Each unit cell is equivalent to any other unit cell. Floquet topological quantum systems operate in reciprocal space where we can distinguish the first Brillouin zone from all the higher order Brillouin zones. For example in a hexagonal lattice the $\Gamma$-point in reciprocal space plays a different role than the $K$-points, while in our affine three-fold lattice all high symmetry points are equivalent
and cause lattice symmetries to have different effects in our colloidal system than in quantum systems.

Finally our system is dissipative causing irreversible relaxation processes to contribute to the dynamics. These irreversible processes can be rendered unimportant only on the stationary manifold via the adiabatic driving, but not on the paths of steepest decent. This is causing the non-time reversible ratchet processes that have no analogue in the topological quantum systems. 

\section{conclusions}\label{conclusion}
\begin{figure}
	\includegraphics[width=1\columnwidth]{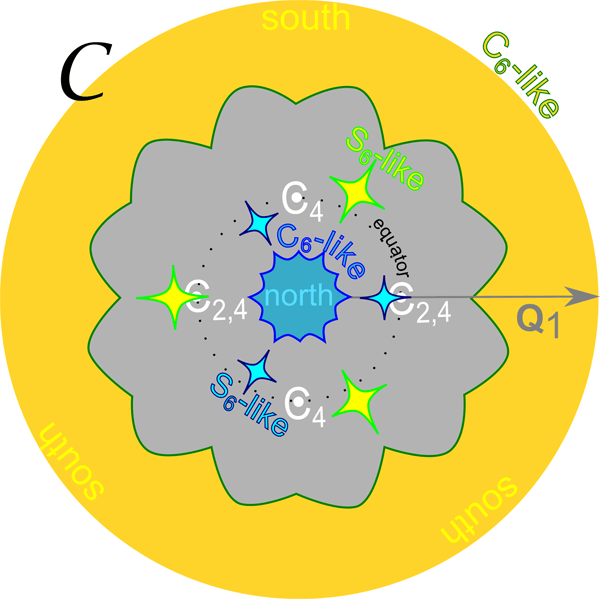}
	\caption{Stereographic projection of control space with all relevant objects for the lattices of different symmetry. White circles are relevant for both paramagnets and diamagnets while green fences are relevant for the paramagnets only and blue fences are relevant for diamagnets only.}
	\label{figsummary}
\end{figure}
Paramagnetic and diamagnetic colloids above a magnetic pattern can be transported by modulating the potential with time dependent homogeneous external fields. If such modulation loops wind around specific points \changes{ (fence points for the two- and four-fold symmetries, bifurcation points for the three- and six-fold symmetry) or pass through fence segments (three- and six-fold symmetry)} in control space the topologically trivial modulation can be translated into non trivial motion of colloids.  A summary of the relevant points and segments is shown as stereographic projection of control space for lattices of $C_2,C_4, S_6$-like, and $C_6$-like symmetry in Fig. \ref{figsummary}. 
It shows the deep connection between symmetry and topology since all objects are completely different for the various symmetries. \changes{The lattice symmetry determines the transport modes, which are possible along the primitive lattice vectors. }

Modulation loops can be sorted into topologically equivalent classes, according to their winding around those points and/or by the sequence of segment crossings.  
All modulation loops belonging to the same class induce motion in the same direction, which  makes the transport very robust against perturbations. Noise in the pattern only affects the less robust features of the transport while it doesn't alter its topological class.

On top of $C_2$- and the $C_4$-symmetric patterns para- and diamagnets are adiabatically transported into the same direction. In contrast above 3-fold and 6-fold symmetric patterns both types of particles can be transported into independent directions and the motion happens either adiabatically or via irreversible ratchets.

 Classes of modulation loops causing transport modes into one direction cluster around the adiabatic paths. Ratchet modulation loops are topological protected by their neighboring adiabatic loops and hence transport into the same direction. 
The whole variety of possible transport is described by a set of topological invariants, which are winding numbers around the holes of the stationary surfaces $\cal M$. 

\changes{The robustness of the topological transport can be used to transport a collection of colloids with a broad distribution of properties, such as size-polydispersity without dispersion. This is a clear advantage over other collective transport methods such as thermal ratchets, external gradients and active motion. The possibility of independent motion of paramagnets and diamagnets facilitates other applications such as guiding chemical reactions and assembly\cite{Loehr}.}

\section{acknowledgments}
J. B and A. T. acknowledge support by a Ghana MOE - DAAD joined fellowship and a University of Kassel PhD fellowship respectively. 

\section{appendix}\label{appendix}
\subsection{Three fold symmetric stationary manifolds} \label{appendixpicturesM}
\begin{figure*}
	\begin{center}
	\includegraphics[width=1.55\columnwidth]{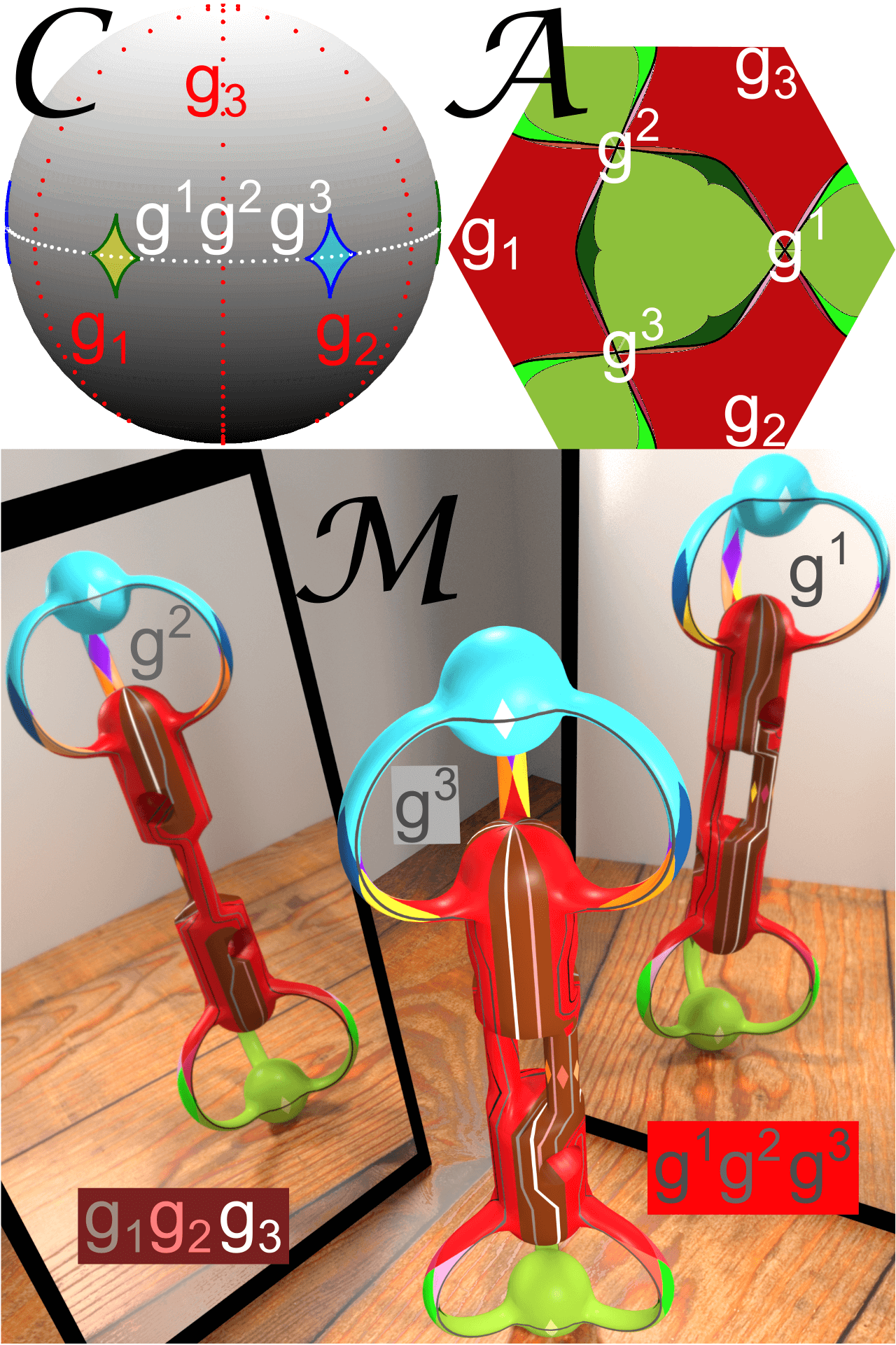}
	\caption{Universal topology of $\cal C$, $\cal A$ and $\cal M$ for a pattern with $S_6$ symmetry ( $\phi =\pi/6$). We have marked the six gates $g_1,g_2,g_3,g^1,g^2,g^3$ that are projected into the six gate points in $\cal A$. On $\cal M$ the upper gates $g^1,g^2,g^3$ travel on the handles while the lower gates $g_1,g_2,g_3$ pass through polar regions that will become isolated in the $S_6$-like case.\vspace{2 cm}  }
	\label{figM3phi5}\end{center}
\end{figure*}

\begin{figure*}	\begin{center}
	\includegraphics[width=1.55\columnwidth]{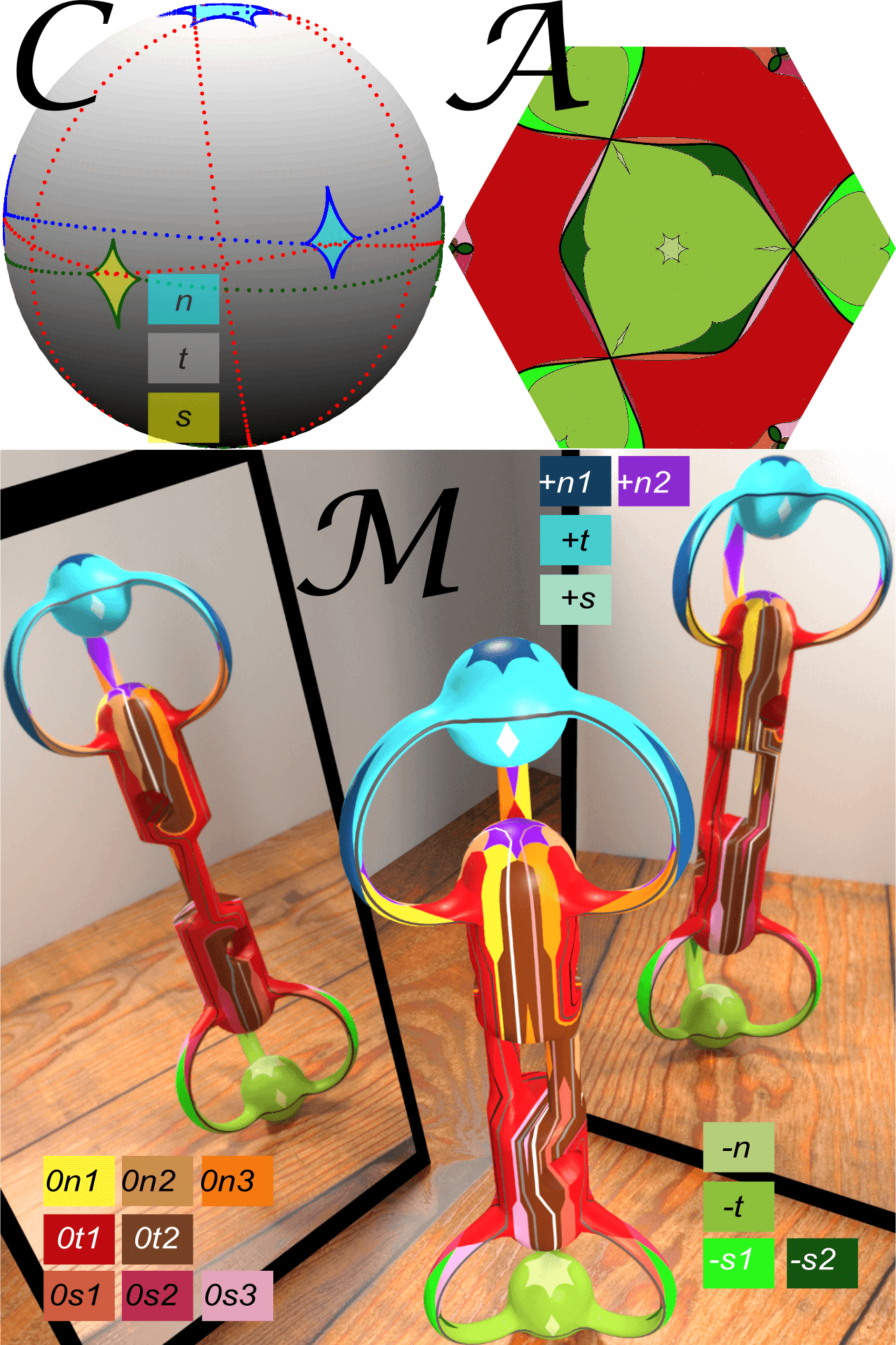}
	\caption{Universal topology of $\cal C$, $\cal A$ and $\cal M$ for a pattern with $S_6$-like symmetry ( $\phi =5\pi/36$)  together with color codes for the areas of $\cal C$ and the shared color codes of $\cal M$ and $\cal A$. The coloring of the gates is the same as in Fig. \ref{figM3phi5}. \vspace{1 cm}}
	\label{figM3phi4}\end{center}
\end{figure*}

\begin{figure*}	\begin{center}
	\includegraphics[width=1.55\columnwidth]{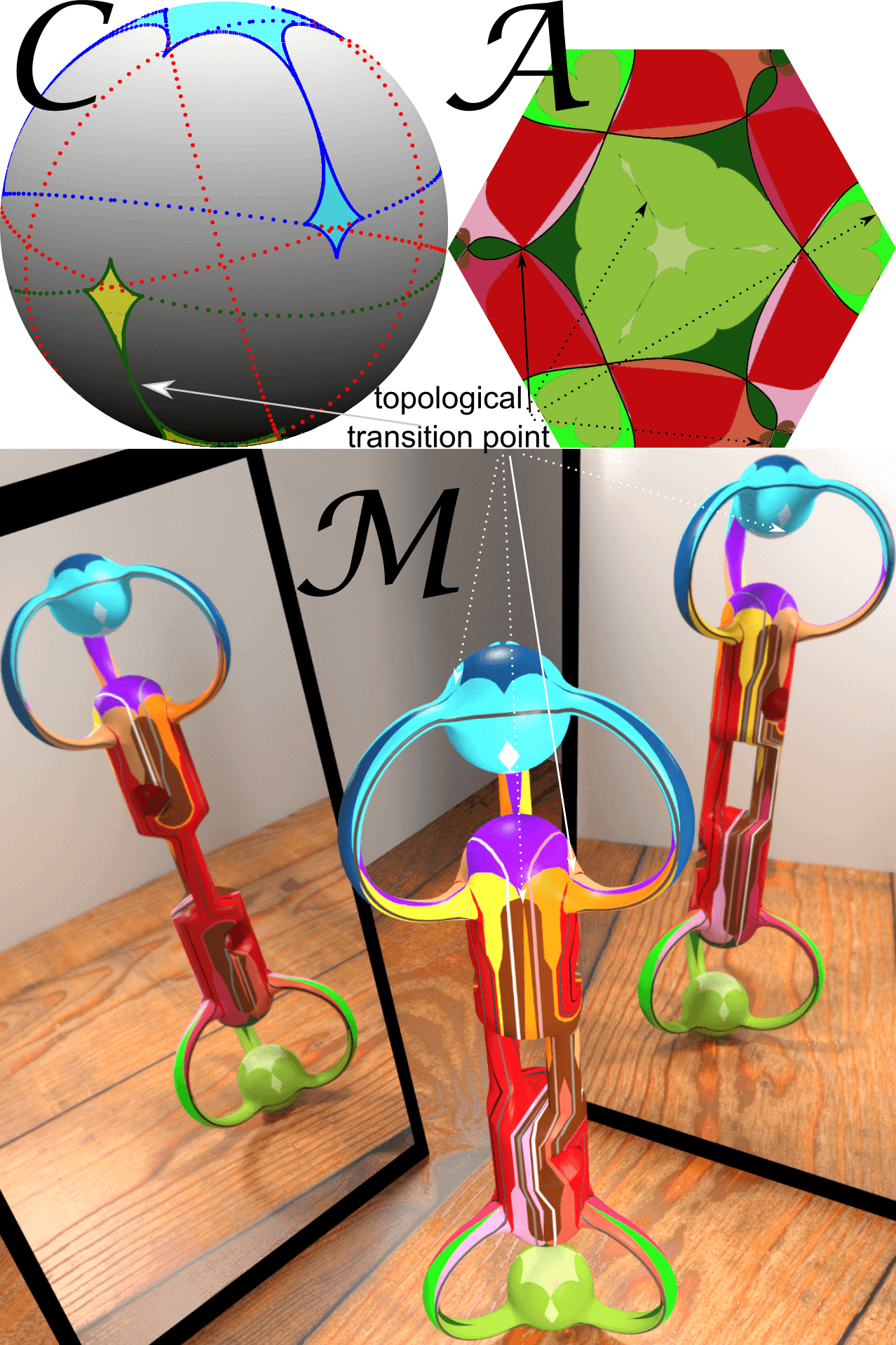}
	\caption{Universal topology of $\cal C$, $\cal A$ and $\cal M$ for a pattern at the transition from $S_6$-like to $C_6$-like symmetry ($\phi_c =\pi/9$) with gates colored  similar to Fig. \ref{figM3phi5}. Two ${\cal B}_0$ (pseudo) bifurcation points from two (pseudo) fences annihilate at the topological transition points at the solid (dashed) arrows where the satellites merge with the polar excess areas.\vspace{1 cm}}
	\label{figM3phi3}\end{center}
\end{figure*}

\begin{figure*}	\begin{center}
	\includegraphics[width=1.55\columnwidth]{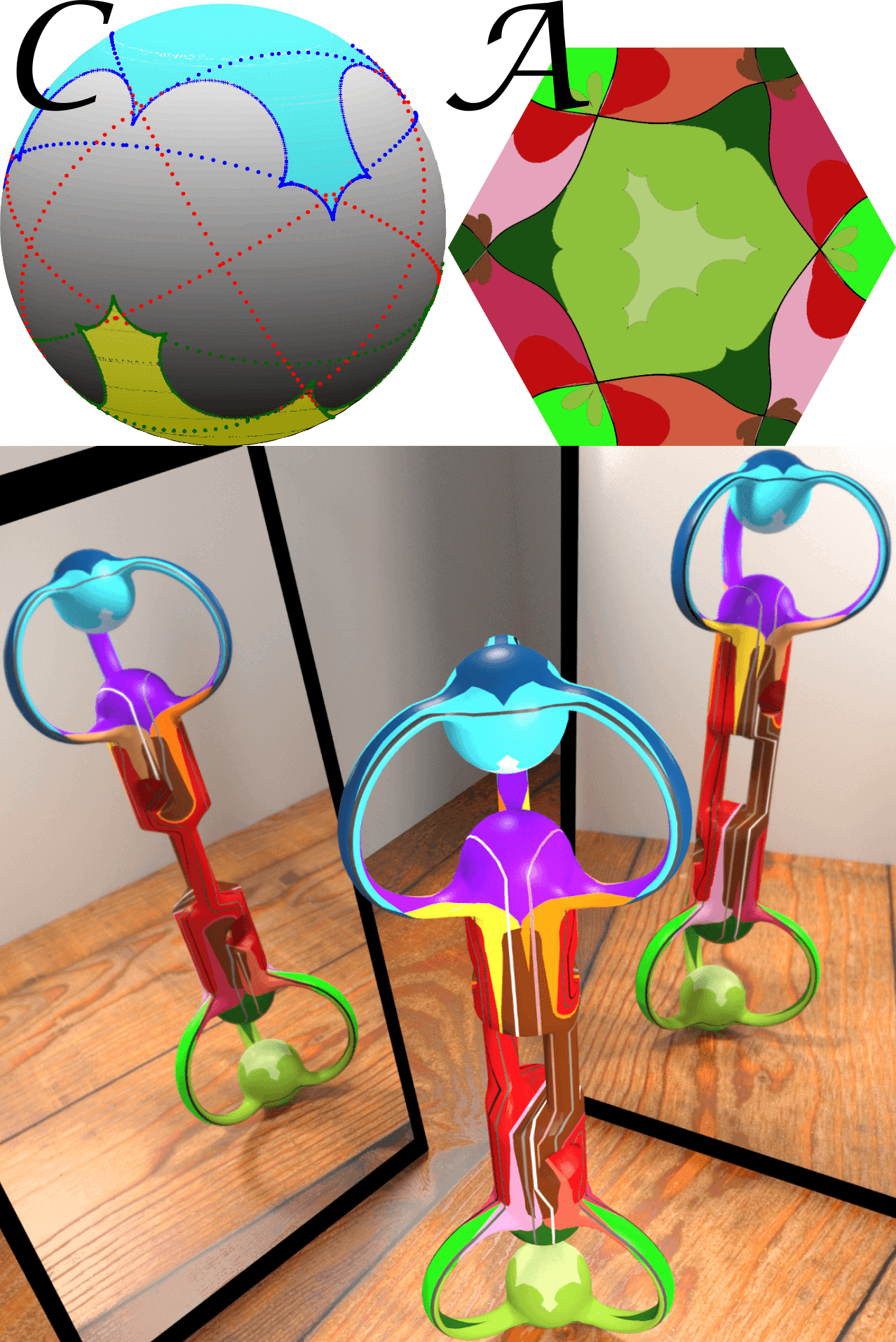}
	\caption{Universal topology of $\cal C$, $\cal A$ and $\cal M$ for a pattern with $C_6$-like symmetry ( $\phi =\pi/18$).  \vspace{2 cm}.}
	\label{figM3phi2}\end{center}
\end{figure*}

\begin{figure*}	\begin{center}
	\includegraphics[width=1.55\columnwidth]{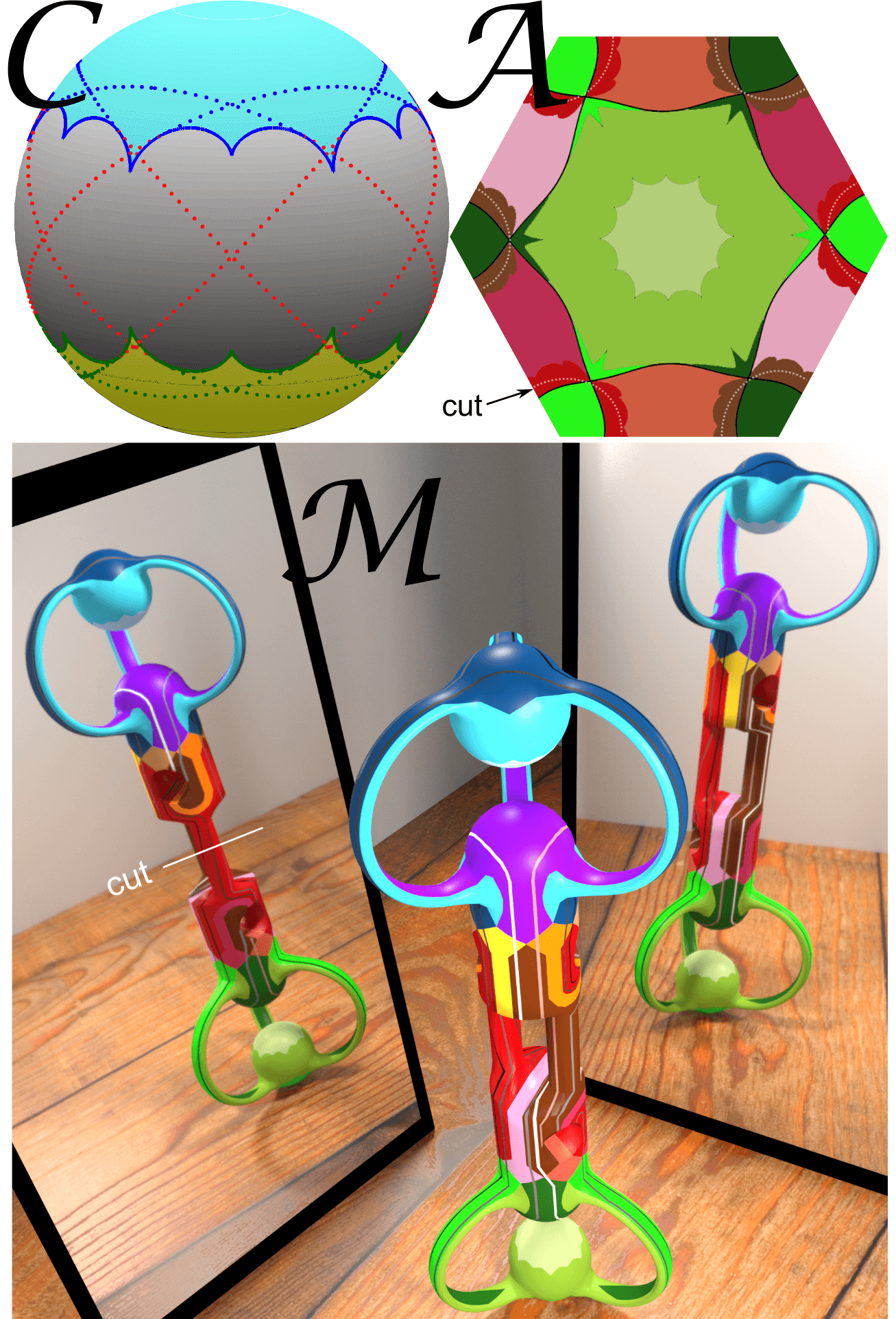}
	\caption{Universal topology of $\cal C$, $\cal A$ and $\cal M$ for a pattern with $C_6$ symmetry ($\phi =0$). We have marked the cut in $\cal A$ that is the projection of the boundary between the projected lower half of $\cal M$ and the upper half. \vspace{1 cm}}
	\label{figM3phi1}\end{center}
\end{figure*}

 In Fig. \ref{figM3phi5} - Fig. \ref{figM3phi1} we give a high resolution view of $\cal C$, $\cal A$, and $\cal M$ of the three-fold symmetric patterns at five different values of $\phi$, where we explain specific details in one of the figures each. These details apply to all different phases if not stated otherwise. The positions of the six gates in each space is explained in Fig. \ref{figM3phi5} and remains the same throughout the rest of the figures. In Fig. \ref{figM3phi4} we show the color coding of the areas in $\cal C$ as well as the color coding shared between $\cal M$ and $\cal A$. The poles of $\cal C$ have $2\times 6$ preimages in $\cal M$ that all lie on the central axis of $\cal M$ either on a pole of a hemispherical cap or at the apex or base of the three central holes. When projecting $\cal M$ into $\cal A$ the poles on the hemispheres fall onto the three-fold symmetric points of $\cal A$, while the saddle point poles of ${\cal M}_0$ in the three central holes are expelled in the surroundings of ${\mathbf x}_{{\cal A},2}$. The topological transition happens in Fig. \ref{figM3phi5}. Two ${\cal B}_0$ bifurcation points (pseudo bifurcation points) one from a satellite and one from a polar fence (polar pseudo fence) annihilate when the satellite excess area coalesces with the polar excess area at the ends of the full (dashed) arrows. Since only the lower half of $\cal M$ is projected into $\cal A$ there occur two cuts in the brown and red tropical regions of ${\cal M}_0$. The cut in $\cal M$ and its projection into $\cal A$ is shown in Fig. \ref{figM3phi1}. The cut in $\cal A$ circles twice around ${\mathbf x}_{{\cal A},2}$ and around ${\mathbf x}_{{\cal A},3}$ and twists each of the six times it passes a gate thereby alternating between the lower half lying inside and outside the cut. The cuts in the other figures are topologically equivalent to those in Fig. \ref{figM3phi1}. The projection of areas in $\cal M$ into $\cal C$ preserves the orientation of half the areas and switches sign for the others. Each time one passes a pseudo fence that is connected to a bifurcation point one switches the orientation of the projection in $\cal C$. The orientation of the projection from $\cal M$ into $\cal A$ switches sign when we pass from one side of the gate to the other side. The southern excess region south of the gates $g^i_-$ ($g_{i,-}$) in $\cal C$ switches orientation when its preimages in ${\cal M}_-$ are mapped into the bright (dark) green regions around ${\mathbf x}_{{\cal A},3}$ (around ${\mathbf x}_{{\cal A},2}$) in $\cal A$.

\vfill

\subsection{Lithographic magnetic structures}\label{appendixpatterning}
Magnetic patterns with the desired symmetry
 have been created by $10$ keV He-ion bombardment induced magnetic patterning of magnetic multilayer structures with perpendicular magnetic anisotropy \cite{KET2010, EKH2015} using a home-built ion source for 5-30 keV He ions \cite {Lengemann}. First, the layer system ${\textrm Ti^{4 \textrm{nm}}/Au^{60 \textrm{nm}}/[Co^{0.7 \textrm{nm}}/Au^{1 \textrm{nm}}]_5}$ with $M_s$ of 1420 kA/m was fabricated by DC magnetron sputter deposition on a silicon substrate \cite{TGK2011, MPA2013}. The sample's magnetic properties were characterized by polar magneto-optical Kerr effect magnetometry, possessing an initial coercive field of $19.5\pm 0.5 \,\textrm{kA/m}$. The magnetic domain structure was introduced by a local change of the sample's coercive field via $10\,\textrm{keV}$ He ion bombardment through a shadow mask with an ion fluency of $1\times 10^{15} \textrm{Ions/cm}^2$. Here, the geometry of the mask coincides with the desired four-fold symmetric, three-fold symmetric, or phase gradient pattern with a period length of $7\, \mu \textrm{m}$ (Fig. \ref{figpattern}). The mask locally prevents the He ions to penetrate into the layer system \cite{EKH2015}. In the uncovered areas, however, ion bombardment leads to a decrease of the perpendicular magnetic anisotropy and hence, the coercive field, primarily due to defect creation at the interfaces of the [Co/Au] multilayer structure \cite{UKK2010, CBF1998}. In preliminary experiments, the decrease of the coercive field was characterized via polar magneto-optical Kerr effect magnetometry and determined to be $6.5\pm 0.5 \,\textrm{kA/m}$. The shadow mask was prepared via UV lithography on top of the sample. For this purpose, the sample was first spin coated with a photo-resist layer of AZ nLOF 2070 (MicroChemicals, AZ nLOF 2070 diluted with AZ EBR, ratio 4:1) with an average layer thickness of $2 \,\mu\textrm{m}$ as determined from atomic force microscopy measurements. The lithographic structure was introduced by UV exposure through a structured chromium hard mask and subsequent development in AZ 826 MIF (MicroChemicals) to remove the unexposed parts of the resist. After ion bombardment without external magnetic fields applied during the process, the sample was first treated with 1-Methyl-2-pyrrolidone  for 24 h at $80^{\circ}\textrm{C}$, than ultrasonicated for 1 minute and finally cleaned with acetone and isopropanol. Due to the thickness $t=3.5 \,\textrm{nm}$ of the magnetic layer, which is small in comparison to the wavelength of our structures ($tQ<1$), the pattern magnetic field on top of the lithographic pattern is attenuated to $H^p=M_s\cdot t \cdot Q$ in comparison to the value $H^p=M_s$ of a thick ($tQ>1$) garnet film.

\changes{
\subsection{Definitions}\label{definitions}}

{\bf Action space:} the plane $z=\textrm{const}$, where the colloidal particles move. Due to the periodicity different unit cells can be identified with each other which folds action space into a torus.

{\bf Adiabatic motion:} A motion enslaved  by the external modulation, possible when one preimage in $\cal M$ of a modulation loop in $\cal C$ lies in ${\cal M}_-$.

{\bf Allowed regions:} projection of the minimum/maximum sections of $\cal M$ into $\cal A$.

{\bf Bifurcation points:} Bifurcation points on $\cal M$ and on $\cal A$ are crossings of fences with pseudo fences. In $\cal C$ the bifurcation points are cusps of the fence. Bifurcation points exist for the three- and six-fold pattern not for the two- and four-fold pattern.

{\bf Control space:} the endpoints of the external magnetic field of constant magnitude, a sphere.

{\bf Equator:} The boundary between the two hemispheres in control space excluding fence points. The equators in $\cal M$ are the preimages of the equator in $\cal C$ of the projection from $\cal M$ onto $\cal C$. The equators are relevant for the two-fold pattern, where there are no gates. 

{\bf Excess area:} A connected set of points in $\cal C$ with higher multiplicity.

{\bf Fence:} The fence in $\cal M$ is the boundary between minima (or maxima) and the saddle points on $\cal M$. We use the same names for its projection into control and action space. Fences on $\cal M$ and on the torus $\cal A$ are closed lines. Fences on $\cal C$ are points for the two- and four-fold symmetric pattern and lines for the three- and six-fold symmetric pattern.

{\bf Forbidden regions:} projection of the saddle point regions  of $\cal M$ into $\cal A$. Allowed and forbidden regions are disjunct areas in $\cal A$ for all but the two-fold patterns.

{\bf Gates:} A gate in $\cal A$ is a crossing point of two fences in $\cal A$. Gates exist for the three-, four-, and six-fold pattern not for the two-fold pattern. The preimage in $\cal M$ of a gate in $\cal A$ of the projection from ${\cal C}\otimes {\cal A}$ onto $\cal A$ is the gate (a closed line) on $\cal M$. The projection of the gate in $\cal M$ onto $\cal C$ is the gate in $\cal C$. A gate in $\cal C$ is a grand circle.

{\bf Irrelevant fence:} A fence that has nor ${\cal B}_+$ and no ${\cal B}_-$ bifurcation points.

{\bf Lemniscate:} A preimage in $\cal M$ of a modulation loop in $\cal C$ that is not a set of loops in $\cal M$. 

{\bf Modulation loop:} A loop in $\cal C$

{\bf Multiplicity:} The multiplicity of a point $\boldmath{H}_{ext}\in\cal C$ is the number of preimages $(\boldmath{H}_{ext},\boldmath{x}_{\cal A})\in {\cal M}\subset {\cal C}\otimes {\cal A}$ mapped from $\cal M$ onto $\boldmath{H}_{ext}\in\cal C$ by the projection onto control space. 

{\bf Non-time reversible ratchet:} A ratchet motion that follows an open path when playing a palindrome modulation loop.

{\bf Northern hemisphere:} The northern hemisphere are simply connected regions on $\cal C$ and on $\cal M$ with $H_{z,ext}>0$. A similar definition holds for the southern hemisphere.

{\bf Palindrome modulation loop:} A loop in $\cal C$ consisting of two loops that are the inverse of each other.

{\bf Path:} A path is a directed segment of a modulation loop.

{\bf Phase space}  the (multiply connected) product space of control space and action space and thus the product of a sphere and a torus.

{\bf Pseudo bifurcation points:}. Pseudo bifurcation points in $\cal M$ are preimages of the bifurcation points in $\cal C$ that are not bifurcation points.  Pseudo bifurcation points exist in three- and six-fold symmetric patterns. Pseudo bifurcation points in $\cal A$ are the projection of the pseudo bifurcation points in $\cal M$. Pseudo bifurcation points in $\cal M$ and in $\cal A$ are located at cusps of the pseudo fences.

{\bf Pseudo fence:}. A line in $\cal M$ different from the fence in $\cal M$ that is projected onto the fence in $\cal C$. Pseudo fences are closed lines in $\cal M$ and $\cal A$ that exist for the three- and six-fold symmetric pattern not the universal two- and four-fold symmetric pattern.  

{\bf Ratchet motion:}  A motion where the adiabatic motion is interrupted by jumps following the intrinsic dynamics.

{\bf Reduced control space} The cut of control space with the space spanned by the single reciprocal lattice vector $\mathbf Q_1$ of the two-fold pattern and the normal vector $\mathbf n$. 

{\bf Satellites:} Excess areas for the $S_6$-like pattern that merge with their polar parent excess area upon the topological transition to a $C_6$-like pattern. 

{\bf Stationary manifold} a two dimensional manifold in phase space, where the action gradient of the colloidal potential vanishes. 

{\bf Time reversibel ratchet:} A ratchet motion that follows a closed path when playing a palindrome modulation loop.

{\bf 12-network:} The three-fold symmmetric pattern has three different points per unit cell with three-fold rotation symmetry. The straight lines between the first two points define the 12-network. Similar definitions hold for the 23-network and the 31-network.

\vfill
\pagebreak\pagebreak
\vfill
\pagebreak\pagebreak

\end{document}